\documentclass[journal,10pt]{IEEEtran}

\makeatletter
\def\ps@headings{%
\def\@oddhead{\mbox{}\scriptsize\rightmark \hfil \thepage}%
\def\@evenhead{\scriptsize\thepage \hfil \leftmark\mbox{}}%
\def\@oddfoot{}%
\def\@evenfoot{}}
\makeatother
\usepackage{fancyhdr}

\usepackage{setspace}

\usepackage{bm}
\usepackage{amssymb}
\usepackage{amsmath}
\usepackage{amsthm}

\newtheorem{proposition}{Proposition}

\newtheorem{assumption}{Assumption}

\newtheorem{observation}{Observation}
\usepackage{graphicx}

\newcommand{\minitab}[2][l]{\begin{tabular}{#1}#2\end{tabular}}

\usepackage{thmtools}
\declaretheoremstyle[headfont=\bf\slshape]{mystyle}
\declaretheorem[style=mystyle]{problem}

\usepackage{dsfont}

\usepackage{cite}

\usepackage{subfigure}

\usepackage{caption}

\usepackage{algorithm}
\usepackage{algorithmic}
\usepackage{multirow}
\usepackage{array}

\usepackage{comment}

\usepackage{stfloats}

\usepackage{url}

\usepackage{color}

\begin{document}

\title{Public Wi-Fi Monetization via Advertising}

\author{Haoran Yu, Man Hon Cheung, Lin Gao, \IEEEmembership{Senior Member, IEEE}, and Jianwei Huang, \IEEEmembership{Fellow, IEEE}\vspace{-0.6cm}
\thanks{Manuscript received July 11, 2016; revised January 4, 2017; accepted February 13, 2017. 
This work is supported by the General Research Funds (Project Number CUHK 14202814 and 14219016) established under the University Grant Committee of the Hong Kong Special Administrative Region, China. 
This paper was presented in part at IEEE INFOCOM, San Francisco, CA, USA, April 2016 \cite{yu2016wifiad}. \emph{(Corresponding author: Jianwei Huang.)}}

\thanks{H. Yu, M. H. Cheung, and J. Huang are with the Department of Information Engineering, the Chinese University of Hong Kong, Hong Kong, China (\{hryu, mhcheung, jwhuang\}@ie.cuhk.edu.hk).} 
\thanks{L. Gao is with the School of Electronic and Information Engineering, Harbin Institute of Technology, Shenzhen, China (gaolin@hitsz.edu.cn).

}

}


\maketitle

\thispagestyle{empty}
\setcounter{page}{1}

\begin{abstract}
The {{proliferation of public Wi-Fi hotspots has brought new business potentials for Wi-Fi networks, which carry a significant amount of global mobile data traffic today.}} 
In this paper, we propose a {{novel}} \emph{Wi-Fi monetization} model for {venue owners (VOs) deploying public Wi-Fi hotspots, where the VOs can} generate revenue by providing two different Wi-Fi access schemes for mobile users (MUs): (i) the \emph{premium access}, in which MUs directly pay VOs for their Wi-Fi usage, and (ii) the \emph{advertising sponsored access}, in which MUs watch advertisements in exchange of the free usage of Wi-Fi. 
VOs sell their ad spaces to advertisers (ADs) via an ad platform, and share {{the ADs' payments}} with the ad platform.
We formulate the economic interactions among the ad platform, VOs, MUs, and ADs as a three-stage Stackelberg game. 
{In Stage I, the ad platform announces its advertising revenue sharing policy. In Stage II, VOs determine the Wi-Fi prices (for MUs) and advertising prices (for ADs). In Stage III, MUs make access choices and ADs purchase advertising spaces.} 
{We analyze the sub-game perfect equilibrium (SPE) of the proposed game systematically, and our analysis shows the following useful observations. First, the ad platform's advertising revenue sharing policy in Stage I will affect only the VOs' Wi-Fi prices but not the VOs' advertising prices in Stage II. 
Second, both the VOs' Wi-Fi prices and advertising prices are non-decreasing in the advertising concentration level and non-increasing in the MU visiting frequency.} 
{{Numerical results {further} show that the VOs are capable of generating large revenues through mainly providing one type of Wi-Fi access (the premium access or advertising sponsored access), depending on their advertising concentration levels and MU visiting frequencies.
}}
\end{abstract}

\begin{IEEEkeywords}
Wi-Fi pricing, Wi-Fi advertising, Stackelberg game, revenue sharing.
\end{IEEEkeywords}

\IEEEpeerreviewmaketitle

\vspace{-0.2cm}
\section{Introduction}

\subsection{Motivations}

Global mobile traffic grows unprecedentedly, and is expected to reach $30.6$ exabytes per month by 2020 \cite{Cisco2016}.
Facilitated by the recent technology development, data offloading has become one of the main approaches to accommodate the mobile traffic explosion.
According to the forecast of Cisco, $55$\% of the global mobile traffic will be offloaded to Wi-Fi and small cell networks by 2020 \cite{Cisco2016}. ~~~~~~~~~~~~~~~~~~~~~~~~~~~~~~~~~~~~~~~~

According to the report of Wireless Broadband Alliance \cite{WBAWiFi2014}, $50$\% of worldwide commercial Wi-Fi hotspots are owned by different \emph{venues}, such as cafes, restaurants, hotels, and airports.\footnote{Specifically, ``Retails'', and ``Cafes \& Restaurants'' are the venues with the largest number of hotspots ($4.5$ and $3.3$ million globally in 2015, respectively), followed by ``Hotels'', ``Municipalities'', and ``Airports'' \cite{WBAWiFi2014}.}
The  {venue owners (VOs)} build public Wi-Fi for the access of  {mobile users (MUs)}, in order to enhance   MUs' experiences and meanwhile provide   location-based services (\emph{e.g.}, shopping guides, navigation, billing) to benefit the VOs' own business \cite{WBA2015LBS}.

To compensate for the Wi-Fi deployment and operational costs, VOs have been actively considering \emph{monetizing} their hotspots.
One conventional business model is that VOs directly charge MUs for their Wi-Fi usage.
However, as most MUs prefer free Wi-Fi access, it is suggested that VOs should come up with new business models to create extra revenue streams \cite{WBAWiFi2014}.
Wi-Fi \emph{advertising}, where VOs obtain revenue from  {advertisers (ADs)} by broadcasting ADs' advertisements on their hotspots, has emerged as a promising monetization approach.
It is especially attractive to ADs, as the accurate localization of Wi-Fi allows ADs to make location-aware advertising.
Furthermore, with MUs' basic information collected by the hotspots,{\footnote{For instance, when MUs login the public hotspots with their social network accounts, SOCIFI collects customers' information, such as age and gender \cite{socifi}.}} ADs can efficiently find their targeted customers and deliver the personalized contents to them.

Nowadays, several companies, including SOCIFI (collaborated with Cisco) \cite{socifi} and Boingo \cite{boingo}, are providing the following types of technical supports for VOs and ADs on Wi-Fi advertising.
First, they offer the devices and softwares which enable VOs to display selected advertisements on the Wi-Fi login pages and collect the statistics information (\emph{e.g.}, number of visitors and click-through rates).
Second, they manage the \emph{ad platforms}, where VOs and ADs trade the ad spaces.
Once ADs purchase the ad spaces, VOs and ad platforms share ADs' payments based on the sharing policy designed by ad platforms.
Although Wi-Fi advertising has been emerging in practice, its influence on entities like VOs and MUs, as well as the detailed pricing and revenue sharing policies, has not been carefully studied in the existing {{literature}}. This motivates our study in this work.

\vspace{-0.2cm}
\subsection{Contributions}

\begin{figure}[t]
  \centering
  \includegraphics[scale=0.4]{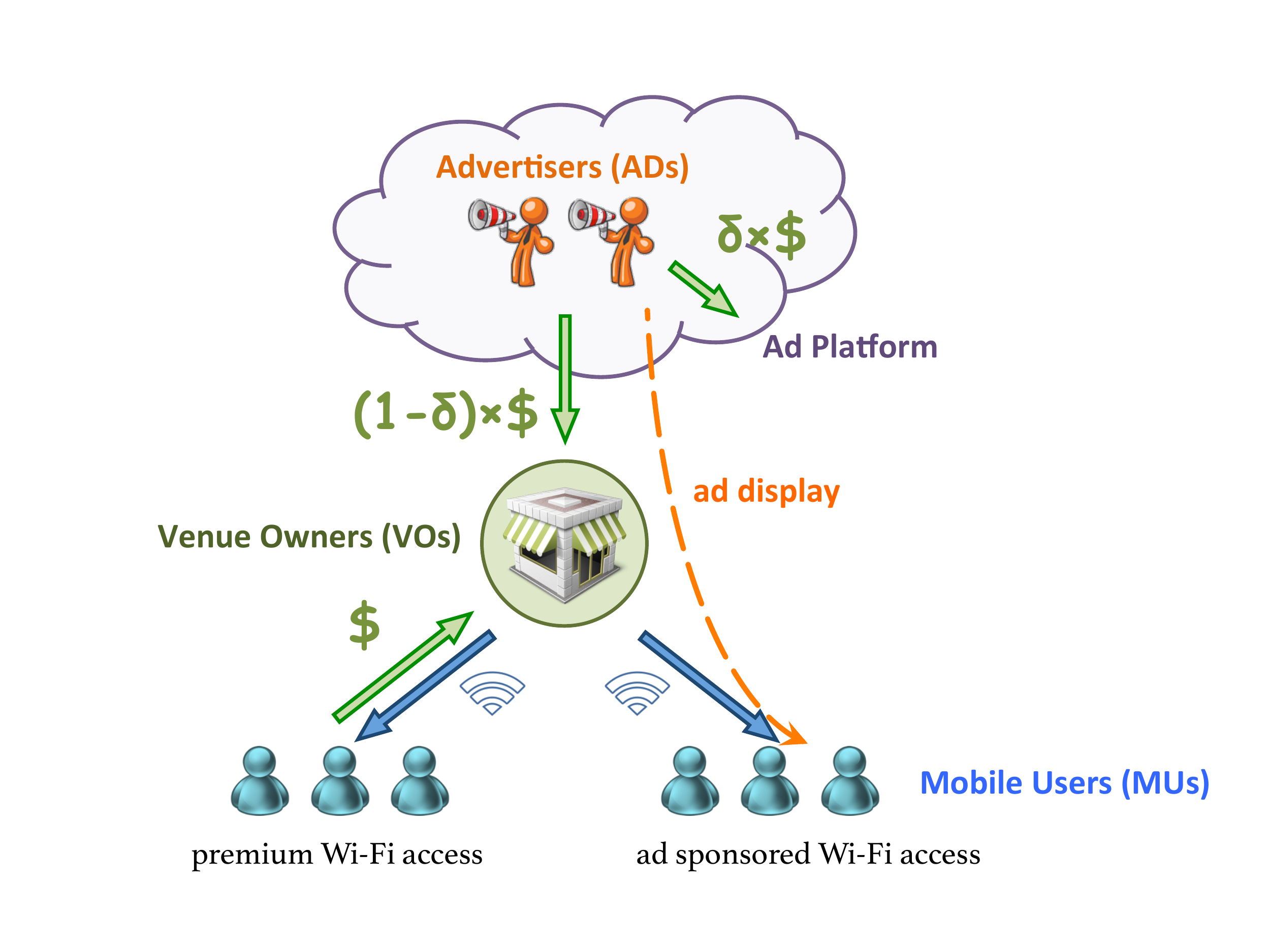}
  \caption{Public Wi-Fi monetization: (a) the \emph{premium access}: VOs directly charge MUs; (b) the \emph{advertising sponsored access}: VOs sell the ad spaces to ADs via the ad platform, and ADs broadcast their advertisements to MUs.}
  \label{fig:ecosystem}
  \vspace{-0.35cm}
\end{figure}

We consider a general Wi-Fi monetization model, where VOs monetize their hotspots by providing two types of Wi-Fi access: \emph{premium access} and  \emph{advertising sponsored access}.
With the premium access, MUs pay VOs according to certain pricing schemes.
With the advertising sponsored access, MUs are required to watch the advertisements, after which MUs use Wi-Fi for free during a certain period.\footnote{As an example, SOCIFI technically supports the premium access as well as the advertising sponsored access for its subscribed VOs \cite{socifi}.}
Depending on the VOs' pricing schemes, MUs with different valuations on Wi-Fi access will choose different types of access.
When MUs choose the advertising sponsored access, VOs sell the corresponding ad spaces to ADs through participating in the ad platform. 
{{Based on the ad platform's sharing policy $\delta$, the ad platform and VOs obtain $\delta$ and $1-\delta$ fractions of the ADs' payments, respectively.}} 
Fig.~\ref{fig:ecosystem} illustrates the Wi-Fi monetization ecosystem.

In this work, we will study such a Wi-Fi monetization system in two parts.

\subsubsection{Modeling and Equilibrium Characterization}
In the first part of our work, we model the economic interactions among different decision makers as a \emph{three-stage Stackelberg game}, and study the game equilibrium systematically.
Specifically, in Stage I, the ad platform designs an advertising revenue sharing policy for each VO, which indicates the fraction of advertising revenue that a VO needs to share with the ad platform.
In Stage II, each VO decides and announces its Wi-Fi price to MUs for the premium access, and its advertising price to ADs.
In Stage III, MUs choose the access types (premium or advertising sponsored access), and ADs decide the number of ad spaces to purchase from the VO.~~~~~~~~~~~~~~~~

We analyze the {sub-game perfect equilibrium (SPE)} of the proposed Stackelberg game systematically.
Our analysis {{shows}} that:
(a) the VO's advertising price (to ADs) in Stage II is independent of the ad platform's advertising revenue sharing policy in Stage I, as a VO always charges the advertising price to maximize the total advertising revenue;
(b) the VO's Wi-Fi price (to MUs) in Stage II is set based on the ad platform's sharing policy in Stage I, since a VO will increase the Wi-Fi price to push more MUs to the advertising sponsored access if the VO can {{obtain}} more advertising revenue.

\subsubsection{Sensitivity Analysis}
In the second part of our work, we define an \emph{equilibrium indicator}, the value of which determines the equilibrium outcomes, such as the ad platform's sharing policy and the VO's Wi-Fi price. Intuitively, the equilibrium indicator describes the VO's relative benefit in providing the premium access over the advertising sponsored access. 
We show that when the equilibrium indicator is small, the VO charges the highest Wi-Fi price to push all MUs to the advertising sponsored access. 
On the other hand, when the equilibrium is large, the ad platform sets the highest advertising revenue sharing ratio, and the VO mainly generates its revenue from the premium access. 

Furthermore, we investigate the influences of (a) the \emph{advertising concentration level}, which measures the degree of asymmetry in ADs' popularity, and (b) the \emph{visiting frequency}, which reflects the average time that MUs visit the venue.
Our analysis shows that these two parameters have the opposite impacts on a VO's pricing strategies: (a) both the VO's Wi-Fi price and advertising price are non-decreasing when the popularity among ADs becomes more asymmetric, and (b) both prices are non-increasing when MUs visit the VO more often.

The key contributions of this work are as follows:

\begin{itemize}
\item \emph{Novel Wi-Fi Monetization Model}: To the best of our knowledge, this is the first work studying the advertising sponsored public Wi-Fi hotspots.
We consider a general Wi-Fi monetization model with both the premium access and the advertising sponsored access,
which enable a VO to segment the market based on MUs' valuations, and maximize the VO's revenue.

\item \emph{Wi-Fi Monetization Ecosystem Analysis}: We study a Wi-Fi monetization ecosystem consisting of the ad platform, VOs, MUs, and ADs, and analyze the equilibrium via a three-stage Stackelberg game. 
We show that a VO's advertising price is independent of the ad platform's sharing policy, and a single term called equilibrium indicator determines the VO's Wi-Fi price and the ad platform's sharing policy.

\item \emph{{{Analysis of Parameters' Impacts}}}: We study the impacts of the advertising concentration level and the visiting frequency, and show that they have the opposite impacts on the equilibrium outcomes. {{For example, a VO's Wi-Fi price and advertising price are non-decreasing in the advertising concentration level and non-increasing in the visiting frequency.}}

\item \emph{Performance Evaluations}: Numerical results show that the VOs are able to generate large total revenues by mainly offering one type of Wi-Fi access (the premium access or advertising sponsored access), depending on their advertising concentration levels and visiting frequencies. 
\end{itemize}

\vspace{-0.3cm}
\subsection{Related Work}
Several recent works have studied the business models related to Wi-Fi networks.
{{Duan \emph{et al.} in \cite{duan2015pricing} and Musacchio \emph{et al.} in \cite{musacchio2006wifi} studied the pricing schemes of Wi-Fi owners.}}
Yu \emph{et al.} in \cite{yu2016cooperative} analyzed the optimal strategies for network operators and VOs to deploy public Wi-Fi networks cooperatively.
Gao \emph{et al.} in \cite{gao2014bargaining} and Iosifidis \emph{et al.} in \cite{iosifidis2015double} investigated the Wi-Fi capacity trading problem, where cellular network operators lease third-party Wi-Fi to offload their traffic. 
Some other recent works \cite{manshaei2008wireless,afrasiabi2015choice,ma2016economic,ma2016contract,linicc2017} proposed and analyzed a novel crowdsourced Wi-Fi network, where Wi-Fi owners collaborate and share their Wi-Fi access points with each other. 
Different from these works, we study the monetization of public Wi-Fi through the Wi-Fi advertising,
and focus on the economic interactions among different entities in the entire Wi-Fi ecosystem.~~~~

A closely related work on Wi-Fi advertising is \cite{onlineoffline}, where Bergemann \emph{et al.} considered an advertising market with ADs having different market shares. 
The differences between \cite{onlineoffline} and our work are as follows.
First, in \cite{onlineoffline}, an MU is only interested in one AD's product, while in our model, an MU can be interested in multiple ADs'  products.
Second, in \cite{onlineoffline}, the authors analyzed the market with an infinite number of ADs, while in our work, we first analyze the problem with a finite number of ADs, and then consider the limiting asymptotic case with an infinite number of ADs.
Moreover, in \cite{targetingad,geometricad}, the authors explored the influence of targeting on the advertising market.
However, none of the works \cite{onlineoffline,targetingad,geometricad} considered the ad platform and the associated advertising revenue sharing, which is a key focus of our study.



\vspace{-0.2cm}
\section{System Model}\label{sec:systemmodel}

In this section, we define the strategies of four types of decision makers in the Wi-Fi monetization ecosystem: the ad platform, VOs, ADs, and MUs.
We formulate their interactions as a three-stage Stackelberg game.

\vspace{-0.3cm}
\subsection{Ad Platform}\label{subsec:APL}

The ad platform plays two major roles in the ecosystem.
First, it offers the platform for VOs to locate ADs and sell their ad spaces to ADs.
Second, it offers the necessary technical supports for VOs to display advertisements on their Wi-Fi hotspots.{\footnote{For example, SOCIFI Media Network is the ad platform managed by SOCIFI \cite{socifi}.
SOCIFI Media Network collects visitors' data, provides the statistics, such as the click-through rates, and supports the ad display in different formats (\emph{e.g.}, website, video, message).}}
To compensate for its operational cost, the ad platform can share {{a fraction of the advertising revenue when the VOs sell ad spaces to ADs}}.{\footnote{As stated in \cite{socifi}, there is no cost for VOs to register SOCIFI Media Network, which earns profits from {sharing the advertising revenue} with VOs.}}

{\textbf{Revenue Sharing Ratio $\delta$}}: 
{{We first consider the VO-specific revenue sharing case, where the ad platform can set different advertising revenue sharing ratios for different VOs. 
In this case, we can focus on the interaction between the ad platform and a particular VO without loss of generality, as different VOs are decoupled.}} 
Let $\delta$ denote the ad platform's revenue sharing policy for the VO, which corresponds to the fraction of the advertising revenue that {{the ad platform obtains when the VO sells the ad spaces to ADs.}} 
When the ad platform takes away all the advertising revenue (\emph{i.e.}, $\delta=1$), the VO will not be interested in providing the advertising sponsored access, and the ad platform cannot obtain any revenue. Hence, we assume that the ad platform can only choose $\delta$ from interval $\left[0,1-\epsilon\right]$, where $\epsilon$ is a positive number close to zero. {{{{Mathematically, all results in this paper hold for any $\epsilon\in\left(0,\frac{1}{3}\right)$.}}}} 
{{The corresponding analysis for this VO-specific revenue sharing case is given in Sections \ref{sec:systemmodel} to \ref{sec:impact}.}} 

{{In Section \ref{sec:simulation}, we will further discuss the uniform revenue sharing case, where the ad platform sets a uniform $\delta_U\in\left[0,1-\epsilon\right]$ for all VOs due to the fairness consideration.}}


\vspace{-0.3cm}
\subsection{VO's Pricing Decision}

The VO provides two types of Wi-Fi access for MUs: the premium access and the advertising sponsored access.

\begin{figure}[t]
  \centering
  \includegraphics[scale=0.4]{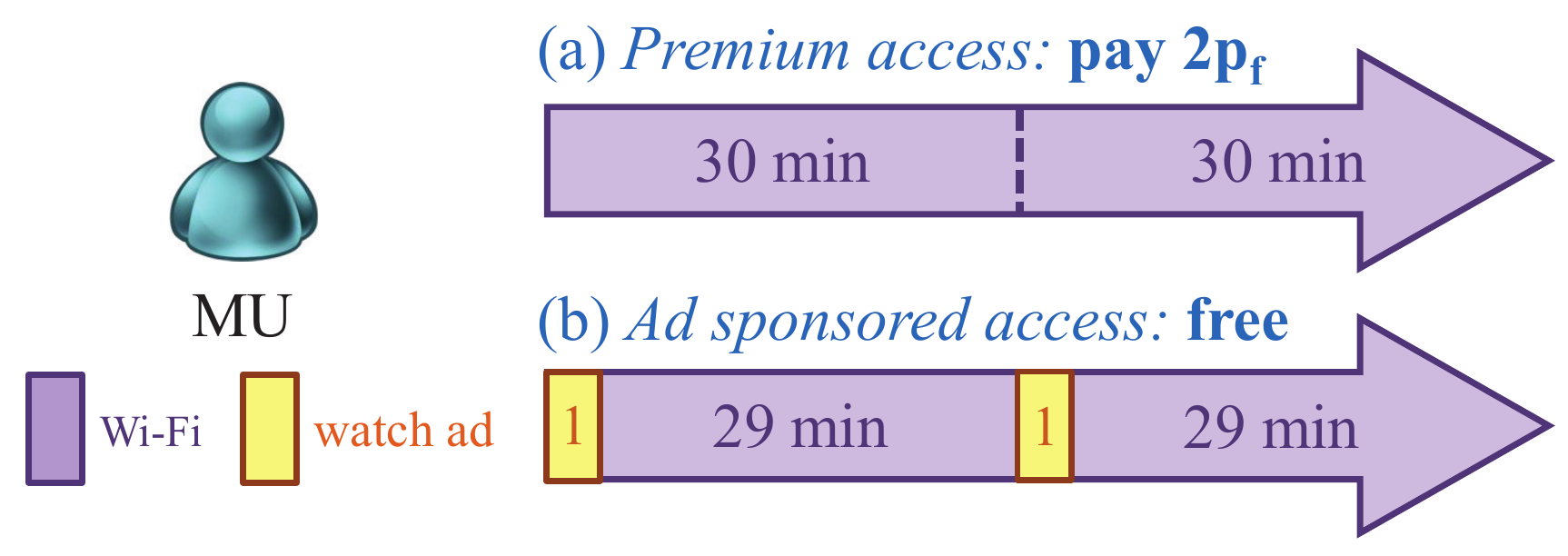}
  \caption{Illustration of Wi-Fi Access.}
  \label{timesegment}
  \vspace{-5mm}
\end{figure}

{\textbf{Wi-Fi Price $p_f$}}: We assume that the VO charges the premium access based on a time segment structure: Each time segment has a fixed length, and the VO charges $p_{f}$ per time segment.
Fig.~\ref{timesegment} illustrates such an example, where the length of one time segment is $30$ minutes.
If an MU chooses the premium access for two segments, it pays $2 p_{f}$, and can use the Wi-Fi for $60$ minutes.

\textbf{Advertising Price $p_a$}:
The MU can also use the Wi-Fi for free by choosing the advertising sponsored access.
In this case, the MU has to watch an advertisement at the beginning of each time segment.{\footnote{{In Boingo's example, an MU can watch an advertisement in order to access Boingo's Wi-Fi networks around $30$ minutes \cite{boingo}.}}}
To guarantee the fairness among the MUs who choose the advertising sponsored access, we assume that all advertisements have the same displaying time.
Let $p_{a}$ denote the advertising price for ADs (for showing one advertisement).
In the example of Fig.~\ref{timesegment}, the ad display time is $1$ minute.
If an MU chooses the advertising sponsored access for two segments,
it needs to watch 2 minutes of advertisements in total, and can use the Wi-Fi for the remaining $58$ minutes free of charge.
Meanwhile, {{the total payment of ADs is $2p_{a}$, which will be shared by the ad platform and the VO according to the revenue sharing ratio mentioned before.}}

{{In our model, we assume that the advertising price $p_a$ is set by the VO, and this setting has been adopted by the companies, such as SOCIFI \cite{socifi}. 
It is important to note that in the case where the advertising price $p_a$ is set by the ad platform, our analysis and results will remain unchanged. This is because the VO and ad platform will choose the same advertising price, which maximizes the total advertising revenue. Therefore, our results and conclusions also apply to the scenario where the ad platform determines the advertising price and sells the ad spaces to the ADs on behalf of the VO.}}


\vspace{-3mm}

\subsection{MUs' Access Choices}\label{subsec:MUaccess}

\textbf{MU's Payoff and Decision}:
We consider the operations in a fixed relatively long time period (\emph{e.g.}, one week).{\footnote{The time length of the period is chosen such that all the system parameters introduced in this paper can be well approximated by constants.}}
Let $N>0$ denote the number of MUs visiting the VO during the period.
We use $\theta\in\left[0,\theta_{\max}\right]$ ($\theta_{\max}>0$) to describe a particular MU's valuation on the Wi-Fi connection.
We assume that $\theta$ follows the uniform distribution.{\footnote{The uniform distribution has been widely used to model MUs' valuations on the wireless service \cite{shetty2009economics,duan2015pricing}. The consideration of other distributions does not change the main conclusions in this paper.}}

Let $d \in \left\{0,1\right\}$ denote an MU's access choice, with $d=0$ denoting the advertising sponsored access, and $d=1$ denoting the premium access. We normalize the length of each segment to $1$, and define the payoff of a type-$\theta$ MU in \emph{one} time segment as
\begin{align}
{\Pi^{\rm MU}}\left( {{\theta},d,p_f} \right) = \left\{ {\begin{array}{*{20}{l}}
{{\theta}\left( {1 - {\beta}} \right),}&{{\rm if~}d = 0,}\\
{{\theta-p_f},}&{{\rm if~}d = 1,}
\end{array}} \right.\label{MUpayoff}
\end{align}
where $\beta\in\left(0,1\right]$ is the utility reduction factor, and term $1-\beta$ describes the discount of the MU's utility due to the inconvenience of watching advertisements.{\footnote{In Fig. \ref{timesegment}'s example, the time segment length is $30$ minutes. If an MU chooses the advertising sponsored access and its utility is equivalent to the case where it directly uses Wi-Fi for $20$ minutes (which we call {\emph{equivalent Wi-Fi usage time}}) without watching advertisements, parameter $\beta$ is computed as $1 - \frac{{20}}{{30}} = \frac{1}{3}$. 
}} 
For simplicity, we assume that $\beta$ is MU-independent. 
When $d=0$, {{the MU's equivalent Wi-Fi usage time during each time segment is $1-\beta$}}; when $d=1$, the MU pays $p_{f}$ to use the Wi-Fi during the whole segment. {{Note that we model the inconvenience of watching advertisements as a multiplicative cost and the payment for the premium access as an additive cost. As we will show in Section \ref{subsec:stageIII:a}, the results obtained under our model are consistent with the reality, where the MUs with high valuations on the Wi-Fi connection choose the premium access and the MUs with low valuations choose the advertising sponsored access.}}

Each MU will choose an access type that maximizes its payoff.
Let ${\varphi _{f}}\left( {{p_{f}}} \right),{\varphi _{a}}\left( {{p_{f}}} \right)\in\left[0,1\right]$ denote the fractions of MUs choosing the premium access and the advertising sponsored access under price $p_{f}$, respectively.

{\textbf{MU Visiting Frequency $\lambda$}:} We further assume that the number of time segments that an MU demands at the venue within the considered period (say one week) is a random variable ${{K}}$, which takes the value from set $\left\{0,1,2,\ldots\right\}$ and follows the Poisson distribution with parameter $\lambda>0$.{\footnote{Poisson distribution has been widely used to model the distribution of the number of events that occur within a time period \cite{kingman1993poisson}. It is a good initial approximation before we get more measurement data that allow us to build a more elaborated model of MUs' behaviors.}} We assume that all MUs visiting the venue have a homogenous parameter $\lambda$. 
Since $\lambda = {\mathbb E}\left\{{{K}}\right\}$, $\lambda$ reflects MU visiting frequency at the venue, and a larger $\lambda$ implies that MUs visit the venue more often.

Since the current Wi-Fi technology already achieves a large throughput, we assume that the capacity of the VO's Wi-Fi is not a bottleneck and can be considered as unlimited.{\footnote{A similar assumption on the unlimited Wi-Fi capacity has been made in reference \cite{duan2015pricing}. Next we briefly discuss the problem with a limited Wi-Fi capacity. First, if the capacity is limited but MUs who choose $d=0$ and $d=1$ experience the same congestion level, then essentially it will not change our analysis. Second, if MUs with $d=0$ and $d=1$ experience different congestion levels but the difference in the congestion level is a constant, then the congestion difference can be easily factorized in our model, and also does not change the results. Third, if MUs with $d=0$ and $d=1$ experience different congestion levels and the difference is not a constant, the analysis will be more complicated, and we plan to investigate this in our future work.}}

\vspace{-2mm}

\subsection{ADs' Advertising Model}\label{subsec:ADmodel}
There are $M$ ADs who seek to display advertisements at the venue.{\footnote{{{In a more general situation, the ADs may simply send their requests of displaying advertisements to the ad platform without specifying the specific venues for the ad display. In this case, the ad platform needs to determine the distribution of the ADs' advertisements over different venues by jointly considering the VOs' characteristics. We leave the study of this general situation as our future work.}}}} We assume that MUs have intrinsic interests on different ADs' products.
An MU will purchase a particular AD's product, if and only if it is interested in that AD's product, and has seen the AD's advertisement at least once.
This assumption reflects the \emph{complementary perspective} of advertising \cite{economicofad}, and has been widely used in the advertising literature \cite{onlineoffline,targetingad,geometricad}. Intuitively, this assumption means that the advertising does not change the consumers' preferences, but becomes a necessary condition to generate a purchase.{\footnote{Besides the \emph{complementary perspective}, reference \cite{economicofad} also mentioned the \emph{persuasive perspective}, where the advertising alters consumers' preferences. We will study the persuasive perspective in our future work.}}

{\textbf{AD's Popularity $\sigma$}}:
We define the popularity of an AD as the percentage of MUs who are interested in the AD's product. Each AD's popularity at the venue is described by its type $\sigma$, which is uniformly distributed in $\left[0,\sigma_{\max}\right]$. We assume the popularity of a type-$\sigma$ AD is
\begin{align}
s\left({\sigma}\right) \triangleq {\gamma}{e^{ - {\gamma}{\sigma}}},\label{equ:popularity}
\end{align}
where $\gamma\in\left(0,1\right]$ {{is a system parameter}}.{\footnote{Reference \cite{onlineoffline} used a similar exponential function to model the market share of a particular AD. However, reference \cite{onlineoffline} considered a model with an infinite number of ADs, and directly made assumptions on an AD's market share. In our work, we model a finite number of ADs, and use a randomly distributed parameter $\sigma$ to describe an AD's popularity. In Section \ref{sec:stageII}, we first analyze the VO's optimal pricing for a finite number of ADs, and then focus on the limiting asymptotic case with an infinite number of ADs. Therefore, compared with \cite{onlineoffline}, our model and analysis are different and more reasonable.} We can show that $s\left({\sigma}\right)$ is decreasing in the type index $\sigma$ and $s\left({\sigma}\right)\le1$.
The parameter $\gamma$ measures the advertising \emph{concentration level} at the venue, which is defined as the asymmetry of the popularities of ADs with different type $\sigma$. A large $\gamma$ implies a high advertising concentration level, since those ADs with small values of $\sigma$ have much higher popularities than other ADs.

\begin{figure}[t]
  \centering
  \includegraphics[scale=0.42]{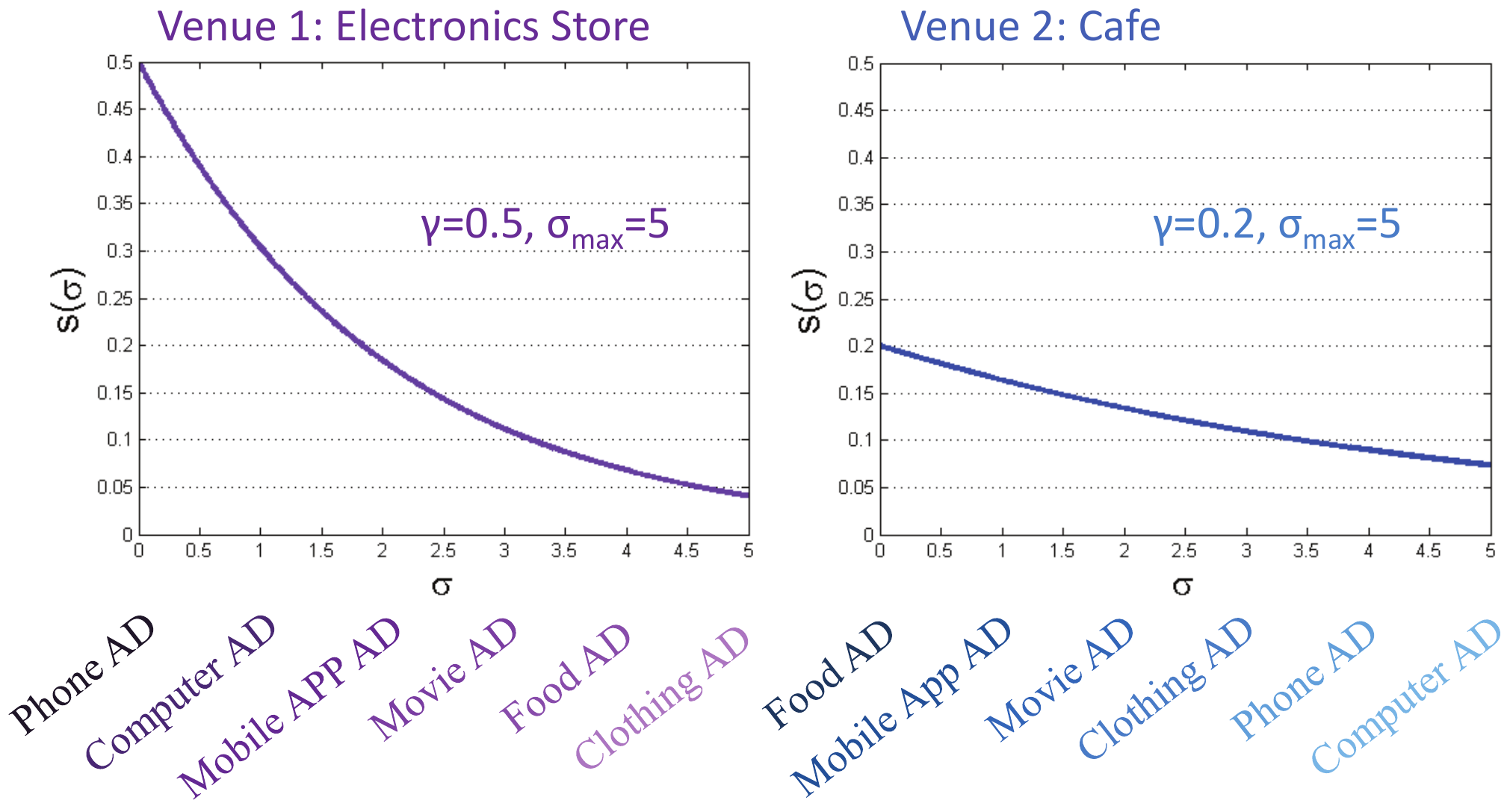}
  \caption{Comparison of Venues with Different $\gamma$.}
  \label{fig:concentration}
  \vspace{-4mm}
\end{figure}

In Fig.~\ref{fig:concentration}, we show different types of ADs' popularities at an electronics store and a cafe, respectively.
Since the electronics store is more specialized and most visitors have interests on the electronics products, the phone AD and computer AD are much more popular than other types of ADs. Hence, the concentration level $\gamma$ of the electronics store is high. On the contrary, the cafe is less specialized and has a lower concentration level than the electronics store. ~~~~~~~~~~~~~~~~~~~~~~~~~~~~~~~~~~

{\textbf{Advertisement Display:}} Next we introduce the advertisement displaying setting. Recall that the number of time segments demanded by an MU is Poisson distributed with an average of $\lambda$ (segments/MU), and the proportion of MUs choosing the advertising sponsored access is ${\varphi _{a}}\left( {{p_{f}}} \right)$.
Hence, the expected number of ad spaces that the VO has during the entire time period is ${\lambda}{N}{\varphi _{a}}\left( {{p_{f}}} \right)$. Let $m$ be the number of advertisements that an AD decides to display at the venue during the entire time period. 
If an MU chooses the advertising sponsored access, then the VO shows an advertisement from this particular {{AD}} to the MU with the following probability at the beginning of every time segment:{\footnote{As shown in the later analysis, the VO will set $p_a$ large enough so that the total number of displayed advertisements does not exceed ${\lambda}{N}{\varphi _{a}}\left( {{p_{f}}} \right)$. Hence, the summation of (\ref{equ:probability}) over all ADs will not be greater than $1$.}}
\begin{align}
\chi \left(m,p_f\right) \triangleq \frac{{{m}}}{{{\lambda}{N}{\varphi _{a}}\left( {{p_{f}}} \right)}}. \label{equ:probability}
\end{align}
{{Note that if the VO does not sell out all the ad spaces, the VO will fill the unsold ad spaces with the VO's own business promotions. This is to guarantee the fairness among the MUs choosing the advertising sponsored access. Specifically, if the VO does not fill the unsold ad spaces with its own business promotions, some MUs choosing the advertising sponsored access may not watch any advertisements or promotions, which leads to fairness issues among the MUs choosing the advertising sponsored access.}}

{\textbf{AD's Payoff:}} Next we study a {{type-$\sigma$ AD's}} payoff. {{We name the considered type-$\sigma$ AD as the \emph{tagged AD}. We use $\nu\left(m,p_f\right)$ to denote the probability of seeing the \emph{tagged} {{AD's}} advertisements {\emph{at least once}} for an MU choosing the advertising sponsored access. Next we compute  $\nu\left(m,p_f\right)$.

Recall that the number of time segments that an MU demands is the discrete random variable $K$, which follows the Poisson distribution with parameter $\lambda$. Hence, the probability for an MU choosing the advertising sponsored access to demand ${{K=k}}$ time segments is $\frac{{{e^{ - \lambda}}{\lambda^k}}}{{k!}}$. Assuming that the MU demands $k$ time segments, hence the conditional probability that the MU does not see the \emph{tagged} {{AD's}} advertisements during these $k$ time segments is ${\left( {1 - \chi \left(m,p_f\right) } \right)^k}$. Therefore, considering all possibilities of the discrete random variable $K$, we have
\begin{align}
\nu\left(m,p_f\right)=1 - \sum\limits_{k = 0}^\infty  {\left( {\frac{{{e^{ - \lambda}}{\lambda ^k}}}{{k!}}{{\left( {1 -\chi \left(m,p_f\right) } \right)}^k}} \right)}.\label{equ:nu}
\end{align}
Based on the Maclaurin expansion of the exponential function, we can simplify (\ref{equ:nu}) as
\begin{align}
\nu\left(m,p_f\right)=1 - {e^{ - \frac{{{m}}}{{{N}{\varphi _{a}}\left( {{p_{f}}} \right)}}}},\label{equ:accuracy}
\end{align}
which is an increasing and concave function of $m$. We can see that $\nu\left(m,p_f\right)$ in (\ref{equ:accuracy}) is independent of $\lambda$. This is because $\lambda$ has two opposite influences on $\nu\left(m,p_f\right)$. First, when $\lambda$ increases, the probability that an MU demands a large number of time segments increases, which potentially increases $\nu\left(m,p_f\right)$. Second, when $\lambda$ increases, the total number of ad spaces $\lambda N \varphi_a\left(p_f\right)$ increases. Based on (\ref{equ:probability}), this leads to the decrease of $\chi \left(m,p_f\right)$, which reduces $\nu\left(m,p_f\right)$. Because the two opposite influences cancel out, $\nu\left(m,p_f\right)$ is independent of $\lambda$.}} 

Recall that an MU will purchase the {{AD's}} product, if and only if the MU is interested in the {{AD's}} product and has seen the AD's advertisement at least once. We use $\Pi^{\rm AD}\left(\sigma,m,p_f,p_a\right)$ to denote a {{type-$\sigma$ AD's}} payoff (\emph{i.e.}, revenue minus payment):
\begin{align}
\Pi^{\rm AD}\!\left({\sigma},m,p_f,p_a\right)\!=\!{a}{N}{{\varphi _{a}}\left( {{p_{f}}} \right)}{s\left({{\sigma}}\right)} \nu\left(m,p_f\right) \!- \!{p_{a}}{m}.\label{equ:ADpayoff}
\end{align}
The parameter $a>0$ is the profit that an AD generates when an MU purchases its product,{\footnote{Since our work focuses on studying the heterogeneity of ADs' popularities, we assume $a$ is homogeneous for all ADs at the venue.}} ${N}{{\varphi _{a}}\left( {{p_{f}}} \right)}$ is the expected number of MUs choosing the advertising sponsored access, $s\left({\sigma}\right)$ is the {{type-$\sigma$ AD's popularity}}, {{$\nu\left(m,p_f\right)$ is the probability of seeing the {{AD's}} advertisements {{at least once}} for an MU choosing the advertising sponsored access,}} and $p_{a}$ is the VO's advertising price.

\vspace{-0.2cm}

\subsection{Three-Stage Stackelberg Game}

\begin{figure}[t]
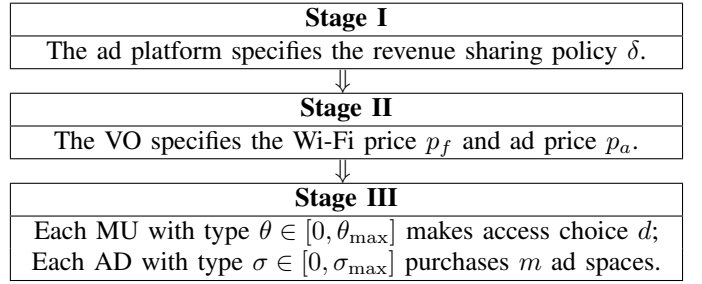

 \centering
\begin{tabular}{|p{8.5cm}<{\centering}|}
\hline
{{\bf{Stage I}}}\\
\hline
{The ad platform specifies the revenue sharing policy $\delta$.}\\
\hline
\end{tabular}
\centerline{$\Downarrow$}
\begin{tabular}{|p{8.5cm}<{\centering}|}
\hline
{\bf{Stage II}}\\
\hline
{The VO specifies the Wi-Fi price $p_{f}$ and ad price $p_{a}$.}\\
\hline
\end{tabular}
\centerline{$\Downarrow$}
\begin{tabular}{|p{8.5cm}<{\centering}|}
\hline
{\bf{Stage III}}\\
\hline
{Each MU with type $\theta\in\left[0,\theta_{\max}\right]$ makes access choice $d$;}\\
{Each AD with type {{$\sigma\in\left[0,\sigma_{\max}\right]$}} purchases $m$ ad spaces.}\\
\hline
\end{tabular}
\caption{Three-Stage Stackelberg Game.}\label{fig:threestage}
\vspace{-4mm}
\end{figure}

We formulate the interactions among the ad platform, the VO, MUs, and ADs by a three-stage Stackelberg game, as illustrated in Fig. \ref{fig:threestage}. From Section \ref{sec:stageIII} to Section \ref{sec:stageI}, we analyze the three-stage game by backward induction.

{{For convenience, we summarize the key notations in Table \ref{table:notation}, including some notations to be discussed in Sections \ref{sec:stageIII}, \ref{sec:stageII}, and \ref{sec:stageI}.}}



{{
\begin{table}[t]
\centering
\caption{{{Key Notations}}}\label{table:notation}
\begin{tabular}{|p{2.7cm}|p{5.3cm}|}
\hline
\multicolumn{2}{|c|}{{\bf Decision Variables}}\\
\hline
{\minitab[c]{$\delta\in\left[0,1-\epsilon\right]$}} & {Ad platform's revenue sharing ratio}\\
\hline
{\minitab[c]{$p_f\in\left[0,\infty\right)$}} & {VO's Wi-Fi price}\\
\hline
{\minitab[c]{$p_a\in\left[0,\infty\right)$}} & {VO's advertising price}\\
\hline
{\minitab[c]{$d\in\left\{0,1\right\}$}} & {An MU's access choice}\\
\hline
{\minitab[c]{$m\in\left[0,\infty\right)$}} & {An AD's ad display choice}\\
\hline
\hline
\multicolumn{2}{|c|}{{\bf Parameters}}\\
\hline
{\minitab[c]{$N\in\left(0,\infty\right)$}} & {Expected Number of MUs}\\
\hline
{\minitab[c]{$\theta\in\left[0,\theta_{\max}\right]$}} & {An MU's Wi-Fi valuation (\emph{MU type})}\\
\hline
{\minitab[c]{$\beta\in\left(0,1\right]$}} & {Utility reduction due to ads}\\
\hline
{\minitab[c]{$\lambda\in\left(0,\infty\right)$}} & {MU visiting frequency}\\
\hline
{\minitab[c]{$M\in\left(0,\infty\right)$}} & {Expected Number of ADs}\\
\hline
{\minitab[c]{$\sigma\in\left[0,\sigma_{\max}\right]$}} & {An AD's popularity index (\emph{AD type})}\\
\hline
{\minitab[c]{$\gamma\in\left(0,1\right]$}} & {Advertising concentration level}\\
\hline
{\minitab[c]{$a\in\left(0,\infty\right)$}} & {ADs' unit profit of MUs' purchasing}\\
\hline
{\minitab[c]{$\eta\in\left[0,\infty\right)$}} & {Popularity of the advertising market}\\
\hline
{\minitab[c]{$\Omega\in\left(0,\infty\right)$}} & {Equilibrium indicator}\\
\hline
\hline
\multicolumn{2}{|c|}{{\bf Functions}}\\
\hline
{\minitab[c]{$\Pi^{\rm MU}\left(\theta,d,p_f\right)$}} & {A type-$\theta$ MU's payoff (one segment)}\\
\hline
{\minitab[c]{$\Pi^{\rm AD}\left(\sigma,m,p_f,p_a\right)$}} & {A type-$\sigma$ AD's payoff}\\
\hline
{\minitab[c]{$\Pi_a^{\rm VO}\left(p_f,p_a,\delta\right)$}} & {VO's revenue from sponsored access}\\
\hline
{\minitab[c]{$\Pi_f^{\rm VO}\left(p_f\right)$}} & {VO's revenue from premium access}\\
\hline
{\minitab[c]{$\Pi^{\rm APL}\left(\delta\right)$}} & {Ad platform's revenue}\\
\hline
{\minitab[c]{$\varphi_f\left(p_f\right)$}} & {Fraction of MUs in premium access}\\
\hline
{\minitab[c]{$\varphi_a\left(p_f\right)$}} & {Fraction of MUs in sponsored access}\\
\hline
{\minitab[c]{$s\left(\sigma\right)$}} & {A type-$\sigma$ AD's popularity}\\
\hline
{\minitab[c]{$\nu\left(m,p_f\right)$}} & {Probability of seeing the \emph{tagged} AD's ads at least once for an MU choosing the advertising sponsored access}\\
\hline
{\minitab[c]{$\theta_T\left(p_f\right)$}} & {Threshold MU type}\\
\hline
{\minitab[c]{$\sigma_T\left(p_a\right)$}} & {Threshold AD type}\\
\hline
{\minitab[c]{$Q\left(p_a,p_f\right)$}} & {Total number of the sold ad spaces}\\
\hline
\end{tabular}
\end{table}
}}

\vspace{-0.1cm}

\section{Stage III: MUs' Access and ADs' Advertising}\label{sec:stageIII}

In this section, we analyze MUs' optimal access strategies and ADs' optimal advertising strategies in Stage III. 
{{The MUs and ADs make their decisions by responding to the ad platform's revenue sharing policy $\delta$ in Stage I, and to the VO's pricing decisions $p_f$ and $p_a$ in Stage II.}}

\vspace{-0.2cm}
 
\subsection{MUs' Optimal Access}\label{subsec:stageIII:a}

Equation (\ref{MUpayoff}) characterizes an MU's payoff for one time segment. Since an MU's payoff for multiple time segments is simply the summation of its payoff from each time segment, an MU's access choice only depends on its type $\theta$ and is independent of the number of time segments it demands. Equation (\ref{MUpayoff}) suggests that a type-$\theta$ MU will choose ${d}=1$ if $\theta  - {p_f}\ge\theta \left( {1 - \beta } \right)$, and $d=0$ otherwise. Therefore,
the optimal access choice of a type-$\theta$ MU is
\begin{align}
{d^*}\left( {\theta,{p_f}} \right) = \left\{ {\begin{array}{*{20}{l}}
{1,}&{{\rm if~}\theta \ge \theta_T\left(p_f\right),}\\
{0,}&{{\rm if~}\theta < \theta_T\left(p_f\right),}
\end{array}} \right.\label{equ:MUoptimald}
\end{align}
where ${\theta _{T}}\left(p_f\right) \triangleq \min\left\{ {  {\frac{{{p_{f}}}}{{{\beta}}}},{\theta _{\max }}} \right\}$ is the \emph{threshold MU type}. 
Intuitively, MUs with high valuations on the Wi-Fi connection will pay for the premium access and use the Wi-Fi for the whole time segment, while MUs with low valuations will watch advertisements in order to access Wi-Fi for free.

Since $\theta$ follows the uniform distribution, under a price $p_f$, the fractions of MUs choosing different types of access are
\begin{align}
{\varphi _a}\left( {{p_f}} \right) = \frac{{{\theta _{T}\left(p_f\right)}}}{{{\theta _{\max }}}} {\rm ~and~} {\varphi _f}\left( {{p_f}} \right) = 1 - \frac{{{\theta _{T}\left(p_f\right)}}}{{{\theta _{\max }}}}.\label{equ:varphi}
\end{align}
 
\subsection{ADs' Optimal Advertising}\label{subsec:stageIII:b}

According to (\ref{equ:ADpayoff}), a {{type-$\sigma$ AD's}} optimal advertising problem is as follows.
\begin{problem}\label{problem:ADm}
The {{type-$\sigma$ AD}} determines the optimal number of ad displays that maximizes its payoff in (\ref{equ:ADpayoff}):
\begin{align}
& \max{~} {a}{N}{{{\varphi _{a}}\left( {{p_{f}}} \right)}}{s\left({{\sigma}}\right)} \nu\left(m,p_f\right) - {p_{a}}{m}\label{equ:ADopt:a}\\
& {\rm var.}{~~~~} m\ge0,\label{equ:ADopt:b}
\end{align}
where $s\left({\sigma}\right)$ is the {{type-$\sigma$ AD's}} popularity defined in (\ref{equ:popularity}).
\end{problem}

The {{type-$\sigma$ AD's}} optimal advertising strategy solving Problem \ref{problem:ADm} is:
\begin{align}
\nonumber
&~~ m^*\left( {{\sigma},{p_a},{p_f}} \right) =\\
& \left\{ {\begin{array}{*{20}{l}}
{\!\! N{\varphi _a}\!\left( {{p_f}} \right)\!\left( {\ln \left( {\frac{{a\gamma }}{{{p_a}}}} \right)\!-\!\gamma \sigma} \right)\!,}&{{\!\rm if~}0 \le \! \sigma \le \sigma_T\left(p_a\right),}\\
{0,\!}&{{\!\rm if~}  \sigma_T\left(p_a\right)\!<\!\sigma\!\le \!\sigma_{\max}.}
\end{array}} \right.\label{equ:ADoptsolution}
\end{align}
Here, ${\sigma_T}\left( {{p_a}} \right)$ is the \emph{threshold AD type}, indicating whether an AD places advertisements. It is defined as
\begin{align}
{\sigma_T}\left( {{p_a}} \right)\triangleq \min\left\{\frac{1}{\gamma }\ln \left( {\frac{{a\gamma }}{{{p_a}}}} \right),\sigma_{\max}\right\}.\label{equ:sigmaT}
\end{align}

We can show that $m^*\left({\sigma},p_a,p_f\right)$ is non-increasing in the AD's type $\sigma$. The reason is that an AD's popularity $s\left(\sigma\right)$ decreases with its type $\sigma$. Only for ADs with high popularities, the benefit of advertising can compensate for the cost of purchasing ad spaces from the VO.~~~~~~~~~~~~~~~~~~~~~~~~~~~~~~~~~~~~~~~~

Moreover, $m^*\left({\sigma},p_a,p_f\right)$ increases with the number of MUs choosing the advertising sponsored access, $N{\varphi _a}\left( {{p_f}} \right)$. 
It is somewhat counter-intuitive to notice that the threshold $\sigma_T\left( {{p_a}} \right)$ is independent of $N{\varphi _a}\left( {{p_f}} \right)$. When $N {\varphi _a}\left( {{p_f}} \right)$ increases, the number of MUs that both choose the advertising sponsored access and like the product from an AD with type $\sigma=\sigma_T\left( {{p_a}} \right)$ indeed increases. While expression (\ref{equ:accuracy}) implies that since there are more MUs, the probability for an MU to see the advertisements from the AD with type $\sigma=\sigma_T\left( {{p_a}} \right)$ decreases. As a result, the change of $N{\varphi _a}\left( {{p_f}} \right)$ does not affect the number of ADs who choose to display advertisements at the venue.

{{When $m^*\left({\sigma},p_a,p_f\right)$ is not an integer, the type-$\sigma$ AD can purchase the ad spaces in a randomized manner, and ensure that the expected number of purchased ad spaces equals $m^*\left({\sigma},p_a,p_f\right)$. The randomized implementation does not affect the ad platform's, VO's, and MUs' equilibrium strategies. It only reduces some ADs' payoffs, and our numerical results show that such an influence is minor. We provide the details about the randomized implementation and numerical experiments in the appendix.}}




\section{Stage II: VO's Wi-Fi and Advertising Pricing}\label{sec:stageII}

In this section, we study the VO's advertising pricing $p_a$ and Wi-Fi pricing $p_f$ in Stage II. {{The VO determines its pricing by responding to the ad platform's revenue sharing policy $\delta$ in Stage I, and anticipating the MUs' and ADs' strategies in Stage III.}}



\subsection{VO's Optimal Advertising Price}\label{subsec:adprice}

We first fix the VO's Wi-Fi price $p_f$ and optimize the VO's advertising price $p_a$. 
{{We will show that the VO's optimal advertising price $p_a^*$ is independent of $p_f$.}} 
In the next subsection, we will further optimize $p_f$.

Let $Q\left(p_a,p_f\right)$ denote the expected total number of ad spaces sold to all ADs. According to (\ref{equ:ADoptsolution}), if $p_a>a\gamma$, no AD will purchase the ad spaces and $Q\left(p_a,p_f\right)=0$; if $0\le p_a \le a \gamma$, we compute $Q\left(p_a,p_f\right)$ as follows:
\begin{align}
\nonumber
& Q\left(p_a,p_f\right)=M \int_0^{\sigma_T\left(p_a\right)}\frac{1}{\sigma_{\max}} m^*\left(\sigma,p_a,p_f\right) {d\sigma}\\
&  =\frac{MN{\varphi _a\left( {{p_f}} \right)}}{\sigma_{\max}}\left( {\ln \left( {\frac{{a\gamma }}{{{p_a}}}} \right){\sigma _T}\left( {{p_a}} \right) - \frac{\gamma }{2}\sigma _T^2\left( {{p_a}} \right)} \right),\label{equ:Qpa}
\end{align}
where $M$ is the number of ADs, and $\frac{1}{\sigma_{\max}}$ is the probability density function for an AD's type $\sigma$.

We define ${\Pi_a^{\rm VO}}\left( {{p_f},{p_a},\delta} \right)$ as the VO's expected advertising revenue, which can be computed as
\begin{align}
{\Pi_a^{\rm VO}}\!\left( {{p_f},{p_a},\delta } \right)\! =\! \left\{ {\begin{array}{*{20}{l}}
{\!\left({1\! -\! \delta } \right)\!{p_a}Q\!\left(p_a,p_f\right)\!,}&{{\rm if~}0\! \le {p_a}\! \le a\gamma,}\\
{0,}&{{\rm if~}{p_a} > a\gamma,}
\end{array}} \right.\label{equ:VO:Ra}
\end{align}
where $1-\delta$ denotes the fraction of advertising revenue {received} by the VO under the ad platform's policy. Based on (\ref{equ:VO:Ra}), we formulate the VO's advertising pricing problem as follows.
\begin{problem}\label{problem:VOadprice}
The VO determines the optimal advertising price by solving
\begin{align}
& \max {~~~} \left({1 - \delta } \right){p_a}Q\left(p_a,p_f\right) \label{equ:Raopt:a}\\
& {\rm s.t.~~~~~} Q\left(p_a,p_f\right)\le {\lambda}{N}{\varphi _{a}}\left( {{p_{f}}} \right), \label{equ:Raopt:b}\\
& {\rm var.~~~~~} 0\le p_a\le a \gamma,\label{equ:Raopt:c}
\end{align}
where constraint (\ref{equ:Raopt:b}) means that the VO can sell at most $\lambda N{\varphi _a}\left( {{p_f}} \right)$ ad spaces as discussed in Section \ref{subsec:ADmodel}.
\end{problem}
The solution to Problem \ref{problem:VOadprice} is summarized in the following proposition (the proofs of all propositions can be found in the appendix).
\begin{proposition}[Advertising price]\label{proposition:finite}
The VO's unique optimal advertising price $p_a^*$ is independent of {{the VO's Wi-Fi price $p_f$ and}} the ad platform's advertising revenue sharing policy $\delta$, and is given by
\begin{align}
\!\!\!\!{p_a^* \!= \!\left\{ {\begin{array}{*{20}{l}}
{\!\!a\gamma {e^{ - \sqrt {\frac{{2\lambda \gamma {\sigma _{\max }}}}{M}} }},}&{{\rm \!\!\!\!if~\!}\frac{\lambda }{M} \!\le \!\min\! \left\{ \!{\frac{{\gamma {\sigma _{\max }}}}{2},\!1,\!\frac{2}{{\gamma {\sigma _{\max }}}}} \!\right\}\!\!,}\\
{\!\!a\gamma {e^{ - \left( {\frac{{\gamma {\sigma _{\max }}}}{2} + \frac{\lambda }{M}} \right)}},}&{{\rm \!\!\!\!if~}\frac{{\gamma {\sigma _{\max }}}}{2} < \frac{\lambda }{M} \le 1,}\\
{\!\!a\gamma {e^{ - \left( {\frac{{\gamma {\sigma _{\max }}}}{2} + 1} \right)}},}&{{\rm \!\!\!\!if}\frac{{\gamma {\sigma _{\max }}}}{2} < 1 < \frac{\lambda }{M},}\\
{\!\!a\gamma {e^{ - 2}},}&{{\rm \!\!\!\!other~cases}.}
\end{array}} \right.}\label{equ:generaladprice}
\end{align}
\end{proposition}

We observe that the expression of $p_a^*$ is sensitive to the number of ADs $M$ and the parameter of ADs' popularity distribution $\sigma_{\max}$. 
To reduce the cases to be considered and have cleaner engineering insights, we will focus on a large advertising market asymptotics with the following assumption in the rest of the paper.{\footnote{Assumption \ref{assumption:infinite} is for the sake of presentations. Without Assumption \ref{assumption:infinite}, there will be seven different regimes (which are divided based on the relations among $\frac{\lambda }{M}$, $\frac{{\gamma {\sigma _{\max }}}}{2}$, $\frac{2}{{\gamma {\sigma _{\max }}}}\!$, and $1$) that we need to discuss (and we can solve), and we will not be able to include the full analysis here due to the space limit. The consideration of finite systems without Assumption \ref{assumption:infinite} does not change the main results in the later sections. In reference \cite{onlineoffline}, the authors directly modeled and analyzed the advertising market with an infinite number of ADs.}}
\begin{assumption}\label{assumption:infinite}
There are infinitely many ADs in the advertising market, i.e., $M\rightarrow \infty$, and the lowest popularity among all types of ADs is zero, i.e., ${\sigma_{\max}}\rightarrow\infty$.{\footnote{From (\ref{equ:popularity}), when ${\sigma_{\max}}\rightarrow\infty$, the popularity of a type-$\sigma_{\max}$ AD is $\lim_{\sigma_{\max}\rightarrow \infty} {s\left({\sigma_{\max}}\right)} =\lim_{\sigma_{\max}\rightarrow \infty}{{\gamma}{e^{ - {\gamma}{\sigma_{\max}}}}=0}$.}}
\end{assumption}
We define $p_a^{\infty}$ as the VO's optimal advertising price under Assumption \ref{assumption:infinite}. According to Proposition \ref{proposition:finite}, we {{show}} $p_a^{\infty}$ in the following proposition.{\footnote{In Section \ref{sec:simulation}, we show that even without Assumption \ref{assumption:infinite}, the $p_a^{\infty}$ derived in Proposition \ref{proposition:adprice} {{achieves a close-to-optimal advertising revenue}} for most parameter settings.}}
\begin{proposition}[Advertising price under Assumption \ref{assumption:infinite}]\label{proposition:adprice}
Under Assumption \ref{assumption:infinite}, the VO's unique optimal advertising price $p_a^{\infty}$ {{is independent of the VO's Wi-Fi price $p_f$ and the ad platform's advertising revenue sharing policy $\delta$, and is given by}}
\begin{align}
p_a^{\infty} = \left\{ {\begin{array}{*{20}{l}}
{a\gamma {e^{ - \sqrt {\frac{2\lambda \gamma}{\eta} } }},}&{{\rm if~}0< \lambda   \le \frac{2\eta}{\gamma},}\\
{a\gamma {e^{ - 2}},}&{{\rm if~}\lambda   > \frac{2\eta}{\gamma},}
\end{array}} \right.\label{equ:pa:infinite}
\end{align}
where $\eta\triangleq \lim_{M,{\sigma _{\max }} \to \infty } \frac{M}{{{\sigma _{\max }}}}$ and takes a value in $\left[0,\infty\right)$.
\end{proposition}
Next we explain the physical meaning of $\eta$. Under Assumption \ref{assumption:infinite}, if we randomly pick an MU, the expected number of ADs that the MU likes is computed as
\begin{align}
\mathop {\lim }\limits_{M,{\sigma _{\max }} \to \infty } M\int_0^{{\sigma _{\max }}} {\frac{{{s\left(\sigma \right)}}}{{{\sigma _{\max }}}}} d\sigma  = \mathop {\lim }\limits_{M,{\sigma _{\max }} \to \infty } \frac{M}{{{\sigma _{\max }}}}=\eta.
\end{align}
Hence, $\eta$ describes the popularity of the advertising market.

Next we discuss how the VO's advertising price $p_a^{\infty}$ changes with $\lambda\in\left(0,\frac{2\eta}{\gamma}\right]$ and $\lambda\in\left(\frac{2\eta}{\gamma},\infty\right)$, respectively.

\subsubsection{Small $\lambda\in\left(0,\frac{2\eta}{\gamma}\right]$}\label{subsubsec:smalllambda}
In this case, the advertising price $p_a^{\infty}$ decreases with $\lambda$. 
This is because MUs' small demand rate $\lambda$ leads to a limited number of ad spaces. When $\lambda$ increases, the VO has more ad spaces to sell, and will decrease $p_a^{\infty}$ to attract more ADs. We can verify that the number of sold ad spaces is $\lambda N{\varphi _a}\left( {{p_f}} \right)$, \emph{i.e.}, the VO always sells out all of the spaces. We call ADs that purchase the ad spaces as \emph{active} ADs, and use $\rho\left(p_a^{\infty}\right)$ to denote the expected number of active ADs. We can compute $\rho\left(p_a^{\infty}\right)$ as
\begin{align}
{{\rho\left(p_a^{\infty}\right)=}}M\frac{{{\sigma _T\left(p_a^{\infty}\right)}}}{{{\sigma _{\max }}}} = \sqrt {\frac{{2\lambda \eta }}{\gamma }},\label{equ:smalllambda:a}
\end{align}
where ${\sigma _T\left(p_a^{\infty}\right)}$ is defined in (\ref{equ:sigmaT}). 
Moreover, the VO's expected advertising revenue is
\begin{align}
{\Pi_a^{\rm VO}}\left( {{p_f},p_a^{\infty},\delta } \right) =\left( {1 - \delta } \right)aN{\varphi _a}\left( {{p_f}} \right)\lambda \gamma {e^{ - \sqrt {\frac{{2\lambda \gamma }}{\eta }} }}.\label{equ:smalllambda:b}
\end{align}
Both (\ref{equ:smalllambda:a}) and (\ref{equ:smalllambda:b}) increase with $\lambda$ when $\lambda$ is small.

\subsubsection{{{Large}} $\lambda\in\left(\frac{2\eta}{\gamma},\infty\right)$}\label{subsubsec:largelambda}
In this case, the advertising price $p_a^{\infty}$ is independent of $\lambda$. 
The reason is that the VO has sufficient ad spaces to sell, so it can directly set $p_a^{\infty}$ to maximize the objective function (\ref{equ:Raopt:a}) while guaranteeing the capacity constraint (\ref{equ:Raopt:b}) satisfied. We can verify that the number of sold ad spaces $Q\left(p_a^{\infty},p_f\right)$ is $\frac{2\eta}{\gamma }N{\varphi _a}\left( {{p_f}} \right)$, {{which is smaller than the capacity $\lambda N{\varphi _a}\left( {{p_f}} \right)$.}}{\footnote{In this case, the VO can fill the unsold ad spaces with its own business promotions to guarantee the fairness among MUs choosing the advertising sponsored access.}} Furthermore, the expected number of active ADs $\rho\left(p_a^{\infty}\right)$ is
\begin{align}
\rho\left(p_a^{\infty}\right)=M\frac{{{\sigma _T}\left( {p_a^{\infty}} \right)}}{{{\sigma _{\max }}}} = \frac{{2\eta }}{\gamma },\label{equ:largelambda:a}
\end{align}
and the VO's expected advertising revenue is
\begin{align}
{\Pi_a^{\rm VO}}\left( {{p_f},p_a^{\infty},\delta } \right) = 2\left( {1 - \delta } \right)aN{\varphi _a}\left( {{p_f}} \right)\eta {e^{ - 2}}.\label{equ:largelambda:b}
\end{align}
Both (\ref{equ:largelambda:a}) and (\ref{equ:largelambda:b}) are independent of  $\lambda$. 

Based on (\ref{equ:smalllambda:b}) and (\ref{equ:largelambda:b}), we summarize the VO's expected advertising revenue as
\begin{align}
{\Pi_a^{\rm VO}}\left( {{p_f},p_a^{\infty},\delta } \right)=\left( {1 - \delta } \right)a N {\varphi _a}\left( {{p_f}} \right) g\left(\lambda,\gamma,\eta\right), \label{equ:combineVOad}
\end{align}
where 
\begin{align}
g\left( {\lambda ,\gamma ,\eta } \right) \triangleq \left\{ {\begin{array}{*{20}{l}}
{\lambda \gamma {e^{ - \sqrt {\frac{{2\lambda \gamma }}{\eta }} }},}&{{\rm{if~}}\lambda\in\left(0,\frac{2\eta}{\gamma}\right],}\\
{2\eta {e^{ - 2}},}&{{\rm{if~}}\lambda\in\left(\frac{2\eta}{\gamma},\infty\right).}
\end{array}} \right.\label{equ:gfunction}
\end{align}

\subsection{VO's Optimal Wi-Fi Price}
Now we analyze the VO's optimal choice of Wi-Fi pricing $p_f$.
We define ${\Pi_f^{\rm VO}}\left( {{p_f}} \right)$ as the VO's revenue in providing the premium access with a given $p_f$.{\footnote{{Notice that the VO's revenue in providing the premium access is collected from the MUs, and hence is independent of the VO's advertising price.}}} Since there are $N \varphi_f\left(p_f\right)$ MUs choosing the premium access and the expected number of time segments demanded by an MU is $\lambda$, we have
\begin{align}
{\Pi_f^{\rm VO}}\left( {{p_f}} \right) = \lambda {p_f}N{\varphi _f}\left( {{p_f}} \right).\label{equ:VO:Rf}
\end{align}

Based on (\ref{equ:combineVOad}) and (\ref{equ:VO:Rf}), we find that $p_f$ affects the VO's revenue in providing both types of access. The VO's total revenue is computed as
\begin{align}
\nonumber
& {\Pi ^{\rm VO}}\left( {{p_f},\delta } \right) ={\Pi_f^{\rm VO}}\left( {{p_f}} \right) +{\Pi_a^{\rm VO}}\left( {{p_f},p_a^{\infty},\delta} \right) 
\\
&  =\lambda {p_f}N{\varphi _f}\left( {{p_f}} \right) +\left( {1 - \delta } \right)a N g\left(\lambda,\gamma,\eta\right) {\varphi _a}\left( {{p_f}} \right).\label{equ:VOtotalrevenue}
\end{align}
By checking $\varphi_f\left(p_f\right)$ and $\varphi_a\left(p_f\right)$ in (\ref{equ:varphi}), we can show that ${\Pi ^{\rm VO}}\left( {{p_f},\delta } \right)$ does not change with $p_f$ when $p_f\in\left[\beta \theta_{\max},\infty\right)$. This is because all MUs will choose the advertising sponsored access if $p_f\ge \beta\theta_{\max}$, and increasing $p_f$ in this range will no longer have an impact on ${\Pi ^{\rm VO}}\left( {{p_f},\delta } \right)$. Therefore, we only need to consider optimizing ${\Pi ^{\rm VO}}\left( {{p_f},\delta } \right)$ over $p_f\in\left[0,\beta \theta_{\max}\right]$. 
This leads to the following optimal Wi-Fi pricing problem.

\begin{problem}\label{problem:VOWiFiprice}
The VO determines the optimal Wi-Fi price to maximize its total revenue in (\ref{equ:VOtotalrevenue}):
\begin{align}
&\max \lambda {p_f}N{\varphi _f}\left( {{p_f}} \right) +\left( {1 - \delta } \right)a N g\left(\lambda,\gamma,\eta\right) {\varphi _a}\left( {{p_f}} \right)\label{equ:formula:WiFiprice:a}\\
& {\rm var.~~~~~} 0\le p_f\le \beta\theta_{\max}.\label{equ:formula:WiFiprice:b}
\end{align}
\end{problem}

Solving Problem \ref{problem:VOWiFiprice}, we obtain the VO's optimal Wi-Fi pricing in the following proposition.
\begin{proposition}[Optimal Wi-Fi price under $\delta$]\label{proposition:WiFiprice}
Given the ad platform's fixed sharing policy $\delta$, the VO's unique optimal Wi-Fi price $p_f^*\left(\delta\right)$ is given by
\begin{align}
p_f^*\left( \delta  \right)\! =\! \frac{{\beta {\theta _{\max }}}}{2}\!+\! \min \left\{ {\frac{{\left( {1 - \delta } \right)a}}{2 \lambda}g\left(\lambda,\gamma,\eta\right),\frac{{\beta {\theta _{\max }}}}{2}} \right\}.\label{equ:VOoptWiFiprice}
\end{align}
\end{proposition}

We can show that $p_f^*\left( \delta  \right)$ is non-increasing in $\delta$. 
When $\delta$ increases, {\emph{i.e.}}, the fraction of advertising revenue left to the VO decreases, the VO decreases its Wi-Fi price $p_f^*\left( \delta  \right)$ to attract more MUs to choose the premium access.



\section{Stage I: Ad Platform's Revenue Sharing}\label{sec:stageI}
In this section, we study the ad platform's sharing policy $\delta$ in Stage I. 
{{The ad platform decides its sharing policy by anticipating the VO's pricing strategies in Stage II and MUs' and ADs' strategies in Stage III.}}

Based on the ad platform's sharing policy $\delta\in\left[0,1-\epsilon\right]$, the ad platform and the VO obtain $\delta$ and $1-\delta$ fractions of the total advertising revenue, respectively. 
Since the VO's advertising revenue is given in (\ref{equ:combineVOad}), we compute the ad platform's revenue $\Pi^{\rm APL}\left(\delta\right)$ as
\begin{align}
\Pi^{\rm APL}\left(\delta\right) = \delta a N {\varphi _a}\left( {{p_f^*}\left(\delta\right)} \right) g\left(\lambda,\gamma,\eta\right),\label{equ:APLrevenue}
\end{align}
where $p_f^*\left( \delta  \right)$ is the VO's optimal Wi-Fi price under policy $\delta$, as computed in Proposition \ref{proposition:WiFiprice}. We formulate the ad platform's optimization problem as follows.
\begin{problem}\label{problem:platform}
The ad platform determines $\delta^*$ to maximize its revenue in (\ref{equ:APLrevenue}):
\begin{align}
&\max {~}\Pi^{\rm APL}\left(\delta\right)\\
& {\rm var.~~~} 0\le\delta\le1-\epsilon.
\end{align}
\end{problem}

In order to compute the optimal $\delta^*$, we introduce an \emph{equilibrium indicator} $\Omega$, which affects the function form of $\delta^*$. 
We define $\Omega$ as
\begin{align}
\Omega \triangleq  \frac{{\lambda \beta {\theta _{\max }}}}{{ag\left( {\lambda ,\gamma,\eta } \right)}}.\label{equ:Omega:define}
\end{align}
The intuition of $\Omega$ can be interpreted as follows. Based on (\ref{equ:varphi}) and (\ref{equ:VO:Rf}), the VO's revenue in providing the premium access can be written as
\begin{align}
\Pi_f^{\rm VO}\left(p_f\right)=\lambda \beta \theta_{\max} N{\varphi _f}\left( {{p_f}} \right){\varphi _a}\left( {{p_f}} \right).\label{equ:indicator:a}
\end{align}
Based on (\ref{equ:combineVOad}), the VO's revenue in providing the advertising sponsored access is
\begin{align}
{\Pi_a^{\rm VO}}\left( {{p_f},p_a^{\infty},\delta } \right)=a  g\left(\lambda,\gamma,\eta\right) N\left( {1 - \delta } \right) {\varphi _a}\left( {{p_f}} \right) .\label{equ:indicator:b}
\end{align}
Next we focus on the system parameters in (\ref{equ:indicator:a}) and (\ref{equ:indicator:b}). We observe that the terms $\lambda \beta \theta_{\max} N$ and $a g\left(\lambda,\gamma,\eta\right) N$ act as the coefficients for (\ref{equ:indicator:a}) and (\ref{equ:indicator:b}), respectively. 
Therefore, intuitively, the indicator $\Omega $ in (\ref{equ:Omega:define}) describes \emph{the VO's relative benefit in providing the premium access over the advertising sponsored access}.

Based on the indicator $\Omega$, we summarize the solution to Problem \ref{problem:platform} as follows.
\begin{proposition}[Revenue sharing policy]\label{proposition:delta}
The ad platform's unique optimal advertising revenue sharing policy $\delta^*$ is given by
\begin{align}
{\delta ^*} = \left\{ {\begin{array}{*{20}{l}}
{1-\epsilon,}&{{\rm if~} \Omega \in\left(0,\epsilon\right],}\\
{1-\Omega,}&{{\rm if~} \Omega \in\left(\epsilon,\frac{1}{3}\right],}\\
{\frac{1}{2} + \frac{\Omega}{{2}},}&{{\rm if~} \Omega \in\left(\frac{1}{3},1-2\epsilon\right),}\\
{1-\epsilon,}&{{\rm if~}\Omega\in\left[1-2\epsilon,\infty\right)}.
\end{array}} \right.\label{equ:APLoptdelta}
\end{align}
\end{proposition}
We can see that $\delta^*\ge \frac{2}{3}$ for all $\Omega\in\left(0,\infty\right)$. That is to say, the ad platform always takes away at least two thirds of the total advertising revenue. In particular, when $\Omega\in\left(0,\epsilon\right]$ or $\Omega\in\left[1-2\epsilon,\infty\right)$, the ad platform chooses the highest sharing ratio, \emph{i.e.}, $\delta^*=1-\epsilon$. Based on our early discussion of $\Omega$, the VO's relative benefits in providing the premium access over the advertising sponsored access under these cases are either very small or very large. Therefore, even if the ad platform decreases its sharing ratio $\delta^*$, the VO's interest in providing the advertising sponsored access will not significantly increase in these two cases. As a result, the ad platform chooses the highest sharing ratio $\delta^*$ to extract most of the advertising revenue. 

Based on Proposition \ref{proposition:delta}, we obtain the VO's Wi-Fi price at the equilibrium by plugging $\delta^*$ into the expression of $p_f^*\left(\delta\right)$ in (\ref{equ:VOoptWiFiprice}), and summarize it in the following proposition.
\begin{proposition}[Wi-Fi price at the equilibrium]\label{proposition:equWiFiprice}
The VO's unique Wi-Fi price at the equilibrium is given by
\begin{align}
{p_f^*\left(\delta^*\right)} = \left\{ {\begin{array}{*{20}{l}}
{{\beta\theta_{\max}},}&{{\rm if~} \Omega \in\left(0,\frac{1}{3}\right],}\\
{\frac{{\beta {\theta _{\max }}}}{4}+\frac{{ag\left( {\lambda ,\gamma,\eta } \right)}}{{4\lambda }},}&{{\rm if~} \Omega \in\left(\frac{1}{3},1-2\epsilon\right),}\\
{\frac{\beta\theta_{\max}}{2}+\frac{{ag\left( {\lambda ,\gamma,\eta } \right)\epsilon}}{{2\lambda }},}&{{\rm if~}\Omega\in\left[1-2\epsilon,\infty\right)}.
\end{array}} \right.\label{equ:equpf}
\end{align}
\end{proposition}
According to (\ref{equ:varphi}) and Proposition \ref{proposition:equWiFiprice}, we can compute $\varphi_a\left({p_f^*\left(\delta^*\right)}\right)$, \emph{i.e.}, the proportion of MUs choosing the advertising sponsored access at the equilibrium. 
We can show that $\varphi_a\left({p_f^*\left(\delta^*\right)}\right)\ge \frac{1}{2}$ for all $\Omega\in\left(0,\infty\right)$. Hence, at least half of the MUs choose the advertising sponsored access. In particular, when $\Omega\in\left(0,\frac{1}{3}\right]$, we have $\varphi_a\left({p_f^*\left(\delta^*\right)}\right)=1$, \emph{i.e.}, all MUs choose the advertising sponsored access. In this case, the VO has a very small relative benefit in providing the premium access, and hence it charges the highest Wi-Fi price $p_f^*\left(\delta^*\right)=\beta\theta_{\max}$ to push all MUs to choose the advertising sponsored access.{\footnote{We would like to emphasize that in order to derive clean engineering insights, our model unavoidably involves some simplifications of the much more complicated reality. It is hence most useful to focus on the engineering insights behind the results such as $\delta^* \ge \frac{2}{3}$ and $\varphi_a\left({p_f^*\left(\delta^*\right)}\right)\ge \frac{1}{2}$, instead of taking the numbers of $\frac{2}{3}$ and $\frac{1}{2}$ literally.}} 

\section{Social Welfare Analysis}\label{sec:social}
In this section, we study the social welfare ($\rm SW$) of the whole system at the equilibrium, which consists of the ad platform's revenue, the VO's total revenue, the MUs' total payoff, and the ADs' total payoff. 
{{The social welfare analysis is important for understanding how much the entire system benefits from the Wi-Fi monetization framework, and how it is affected by different system parameters.}} 
Specifically, we compute $\rm SW$ as:
\begin{align}
\nonumber
& {\rm SW}={\Pi^{\rm APL}\left(\delta^*\right)} + {\Pi ^{\rm VO}}\left( {{p_f^*\left(\delta^*\right)},\delta^* } \right)\\
\nonumber
& + \lambda N \int_{0}^{\theta_{\max}} \frac{1}{\theta_{\max}}{ {\Pi^{\rm MU}\left(\theta,d^*\left(\theta,p_f^*\left(\delta^*\right)\right),p_f^*\left(\delta^*\right)\right)}  d\theta}\\
& \!+\!M \!\int_{0}^{\sigma_{\max}}\!\! \!\!{\frac{1}{\sigma_{\max}}\! \Pi^{\rm AD}\!\!\left(\sigma,m^*\left(\sigma,p_a^{\infty},p_f^*\left(\delta^*\right)\right),p_f^*\left(\delta^*\right),p_a^\infty\right) d\sigma}.\label{equ:longSW}
\end{align}
Here, (i) the first term is the ad platform's revenue at the equilibrium, where ${\Pi^{\rm APL}\left(\delta\right)}$ is given in (\ref{equ:APLrevenue}) and $\delta^*$ is given in Proposition \ref{proposition:delta}; 
(ii) the second term is the VO's total revenue at the equilibrium, where ${\Pi ^{\rm VO}}\left( {{p_f},\delta } \right)$ is given in (\ref{equ:VOtotalrevenue}) and $p_f^*\left(\delta^*\right)$ is given in Proposition \ref{proposition:equWiFiprice}; 
(iii) the third term is the MUs' total payoff at the equilibrium, where ${\Pi^{\rm MU}\left(\theta,d,p_f\right)}$ is a type-$\theta$ MU's payoff for \emph{one} time segment given in (\ref{MUpayoff}) and $d^*\left(\theta,p_f\right)$ is given in (\ref{equ:MUoptimald}); 
(iv) the last term is the ADs' total payoff at the equilibrium, where $\Pi^{\rm AD}\left(\sigma,m,p_f,p_a\right)$ is given in (\ref{equ:ADpayoff}), $m^*\left(\sigma,p_a,p_f\right)$ is given in (\ref{equ:ADoptsolution}), and $p_a^\infty$ is given in Proposition \ref{proposition:adprice}. 

Note that the ADs' payments for displaying advertisements are transferred to the ad platform and the VO, and the MUs' payments for the premium access are collected by the VO. 
Therefore, these payments cancel out in (\ref{equ:longSW}). 
As a result, $\rm SW$ equals the total utility of all MUs and ADs. 
We show the value of $\rm SW$ in the following proposition.
\begin{proposition}[Social welfare]\label{proposition:socialwelfare}
The social welfare at the equilibrium is
\begin{align}
\nonumber
& {\rm SW}= \frac{1}{2}\lambda N\theta_{\max}-\frac{1}{2} \lambda N {p_f^*}\left(\delta^*\right) \varphi_a\left({p_f^*}\left(\delta^*\right)\right)\\
& + \eta N \varphi_a\left({p_f^*}\left(\delta^*\right)\right) \left(a-\frac{p_a^\infty}{\gamma}\left(1+\ln\left(\frac{a\gamma}{p_a^\infty}\right)\right)\right),\label{equ:SW}
\end{align}
where $p_f^*\left(\delta^*\right)$ and $p_a^\infty$ are the VO's Wi-Fi price given in Proposition \ref{proposition:equWiFiprice} and the VO's advertising price given in Proposition \ref{proposition:adprice}, respectively. 
\end{proposition}
In (\ref{equ:SW}), the first two terms correspond to the total utility of MUs, and the last term corresponds to the total utility of ADs. 
In Section \ref{subsec:simu:socialwelfare}, we will investigate the impacts of parameters $\gamma$ and $\lambda$ on the social welfare through {{numerical experiments}}. {{The {{numerical}} results show that the social welfare is always non-decreasing in $\gamma$, and is increasing in $\lambda$ for most parameter settings.}}





\section{Impact of System Parameters}\label{sec:impact}
To understand the Wi-Fi monetization at venues with different features, we analyze the impacts of the advertising concentration level $\gamma$ and visiting frequency $\lambda$ on the equilibrium outcomes. 
Compared with other parameters, 
these two parameters can be dramatically different across venues and hence better reflect the features of the venues.

\begin{figure*}[t]
  \begin{minipage}[t]{.32\linewidth}
  \centering
  \includegraphics[scale=0.33]{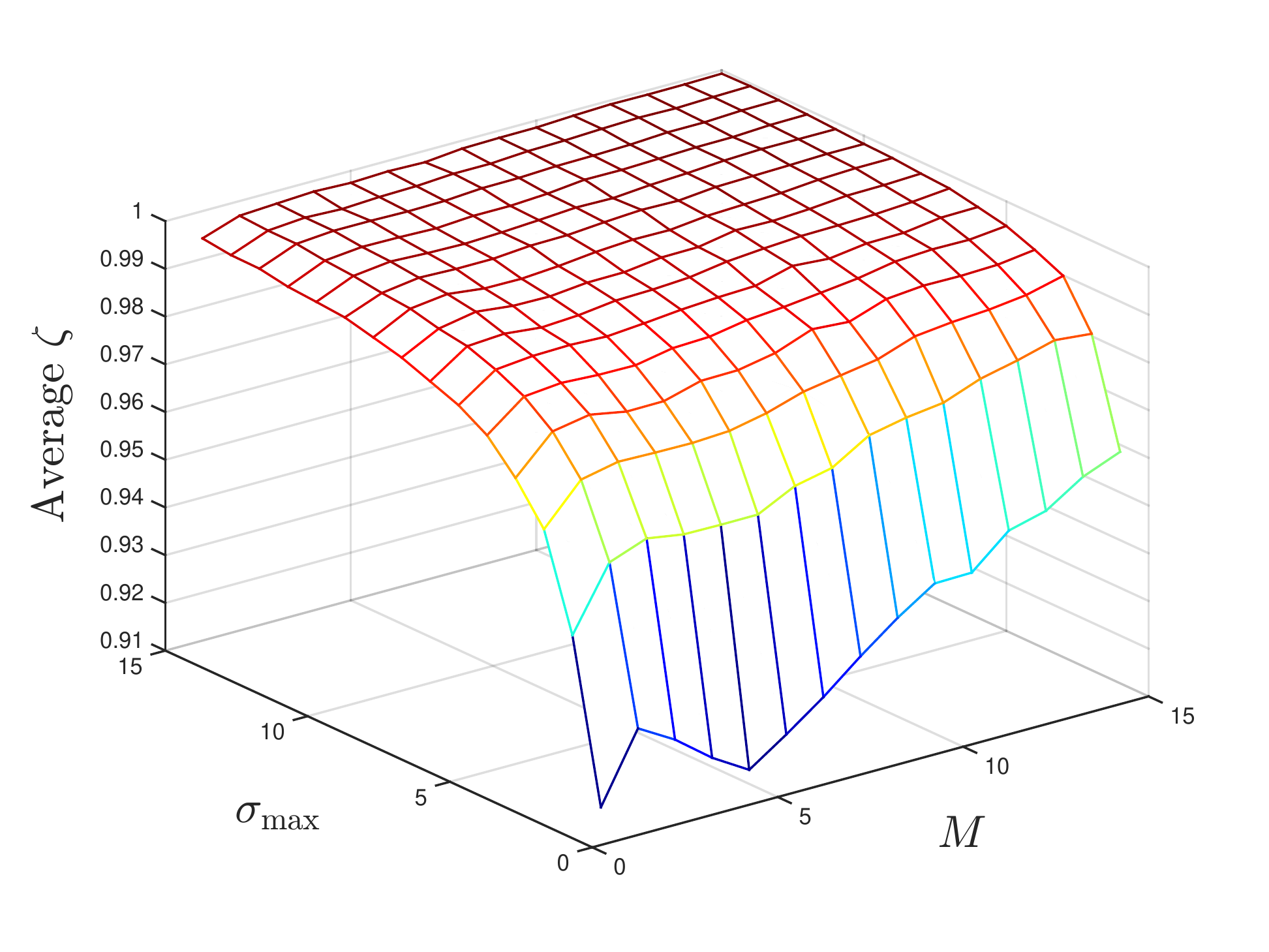}
  \caption{Optimality of $p_a^{\infty}$ without Assumption \ref{assumption:infinite}.}
  \label{fig:NEW1}
  \end{minipage}
  \begin{minipage}[t]{.32\linewidth}
  \centering
  \includegraphics[scale=0.32]{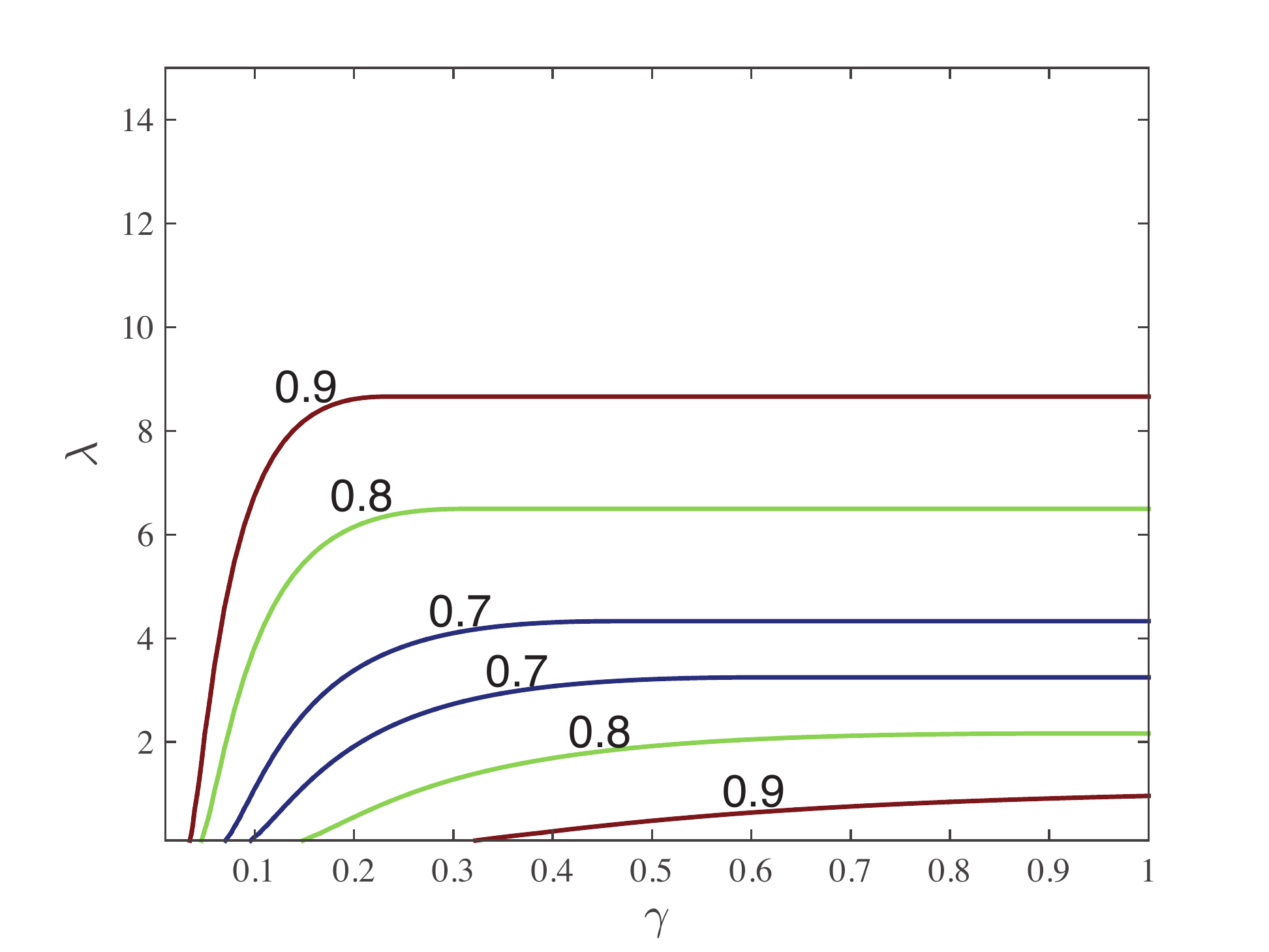}
  \caption{Ad Revenue Sharing Policy $\delta^*$.}
  \label{fig:NEW3}
  \end{minipage}
  \begin{minipage}[t]{.32\linewidth}
  \centering
  \includegraphics[scale=0.32]{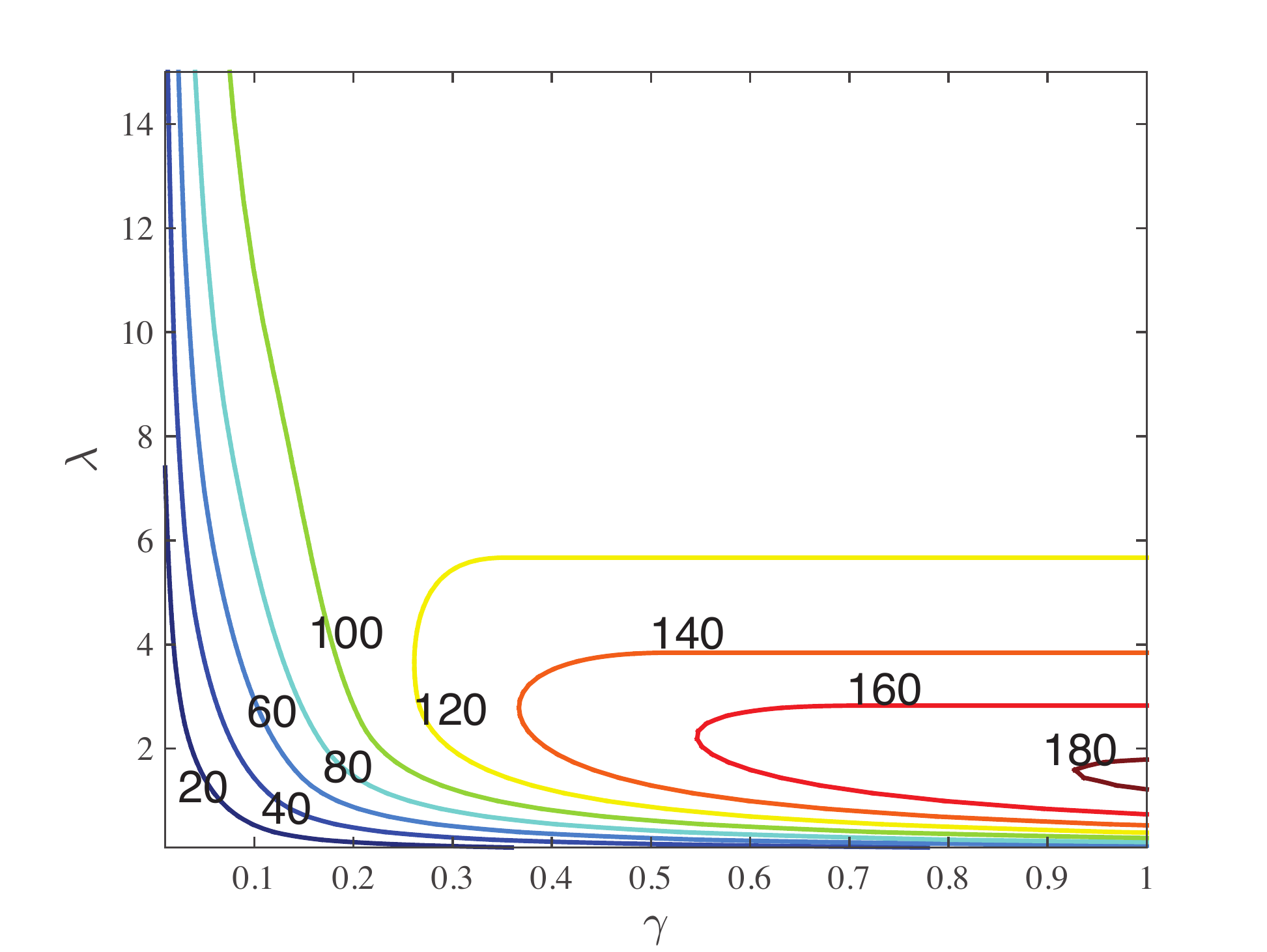}
  \caption{Ad Platform's Revenue $\Pi^{\rm APL}$.}
  \label{fig:NEW2}
  \end{minipage}
\end{figure*}

\begin{proposition}[Advertising concentration level $\gamma$]\label{proposition:gamma}
We {{show}} the following results regarding the influence of $\gamma$:

(i) The VO's advertising price $p_a^{\infty}$ in (\ref{equ:pa:infinite}) is increasing in $\gamma$;

(ii) The expected number of active ADs $\rho\left(p_a^{\infty}\right)$ in (\ref{equ:smalllambda:a}) and (\ref{equ:largelambda:a}) is decreasing in $\gamma$;

(iii) The VO's Wi-Fi price $p_f^*\left(\delta^*\right)$ in (\ref{equ:equpf}) is non-decreasing in $\gamma$;

(iv) The proportion of MUs that choose the premium access $\varphi_f\left(p_f^*\left(\delta^*\right)\right)$ is non-increasing in $\gamma$.
\end{proposition}

Items (i) and (ii) of Proposition \ref{proposition:gamma} describe the advertising sponsored access. A high concentration level $\gamma$ implies that the ADs with small $\sigma$ have much higher popularities than other ADs. Hence, when $\gamma$ increases, the ADs with small $\sigma$ have larger demand in displaying their advertisements. As a result, the VO increases $p_a^{\infty}$ to obtain more advertising revenue. On the other hand, the ADs with large $\sigma$ have smaller demand in advertising, so the expected number of active ADs decreases.~~~~

Items (iii) and (iv) of Proposition \ref{proposition:gamma} describe the premium access. A larger $\gamma$ corresponds to a smaller equilibrium indicator $\Omega$. Based on the previous discussion in Section \ref{sec:stageI}, this means providing the advertising sponsored access is more beneficial to the VO. Hence, under a larger $\gamma$, the VO charges a higher $p_f^*\left(\delta^*\right)$ to push MUs to choose the advertising sponsored access, which reduces the proportion of MUs choosing the premium access.

\begin{proposition}[Visiting frequency $\lambda$]\label{proposition:lambda}
We {{show}} the following results regarding the influence of $\lambda$:

(i) The VO's advertising price $p_a^{\infty}$ in (\ref{equ:pa:infinite}) is non-increasing in $\lambda$;

(ii) The expected number of active ADs $\rho\left(p_a^{\infty}\right)$ in (\ref{equ:smalllambda:a}) and (\ref{equ:largelambda:a}) is non-decreasing in $\lambda$;

(iii) The VO's Wi-Fi price $p_f^*\left(\delta^*\right)$ in (\ref{equ:equpf}) is non-increasing in $\lambda$;

(iv) The proportion of MUs that choose the premium access $\varphi_f\left(p_f^*\left(\delta^*\right)\right)$ is non-decreasing in $\lambda$.
\end{proposition}
Items (i) and (ii) of Proposition \ref{proposition:lambda} are related to the advertising sponsored access. According to the discussion in Section \ref{sec:stageII}, a larger $\lambda$ means the VO has more ad spaces to sell. Hence, when $\lambda$ is larger, the VO chooses a smaller $p_a^{\infty}$ to attract more ADs.

Items (iii) and (iv) of Proposition \ref{proposition:lambda} are related to the premium access. 
We can show that the equilibrium indicator $\Omega$ increases with $\lambda$. Based on the previous discussion in Section \ref{sec:stageI}, a larger indicator means providing the premium access is more beneficial to the VO. Therefore, with a larger $\lambda$, the VO charges a lower $p_f^*\left(\delta^*\right)$ to attract MUs to choose the premium access, which increases the proportion of MUs choosing the premium access. 

{{According to Propositions \ref{proposition:gamma} and \ref{proposition:lambda}, we can observe that parameters $\gamma$ and $\lambda$ have exactly the opposite impacts on the equilibrium outcomes.}}

\section{Numerical Results}\label{sec:simulation}
In this section, we provide numerical results. 
First, we study the optimality of advertising price $p_a^{\infty}$ in (\ref{equ:pa:infinite}) without Assumption \ref{assumption:infinite}. 
Then we compare the ad platform's revenue, the VO's revenue, the ADs' payoffs, and the social welfare at venues with different values of $\gamma$ and $\lambda$. {{Finally, since the ad platform may set a uniform sharing policy for multiple VOs in the practical implementation due to the fairness consideration, we investigate the uniform revenue sharing case and compare it with the VO-specific revenue sharing case studied above.}}

\begin{figure*}[t]
  \vspace{-2mm}
  \begin{minipage}[t]{.32\linewidth}
  \centering
  \includegraphics[scale=0.32]{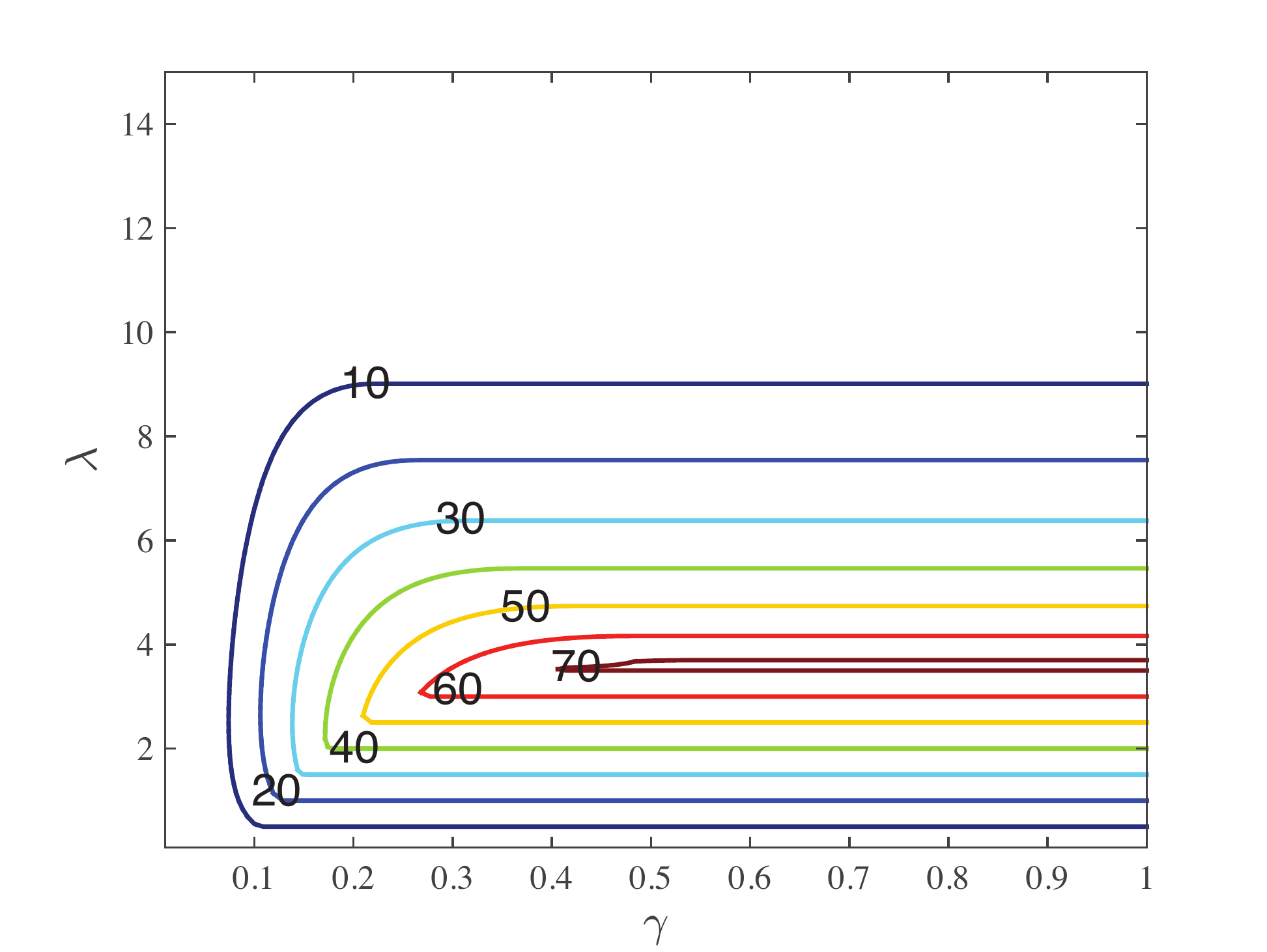}
  \caption{VO's Revenue from Advertising $\Pi_{a}^{\rm VO}$.}
  \label{fig:NEW10}
  \vspace{-2mm}
  \end{minipage}
  \begin{minipage}[t]{.32\linewidth}
  \centering
  \includegraphics[scale=0.32]{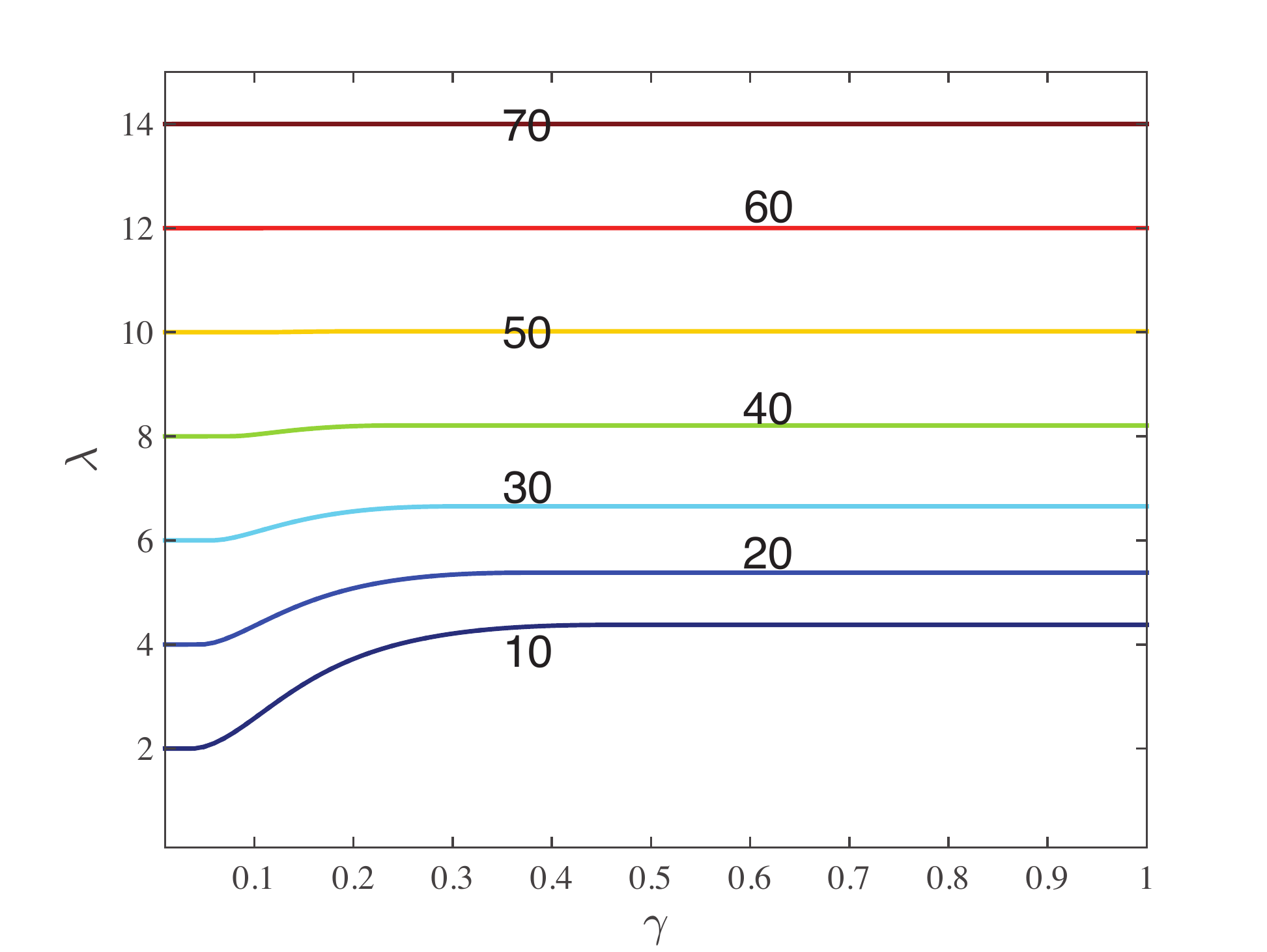}
  \caption{VO's Revenue from Premium Access $\Pi_{f}^{\rm VO}$.}
  \label{fig:NEW11}
  \vspace{-2mm}
  \end{minipage}
  \begin{minipage}[t]{.32\linewidth}
  \centering
  \includegraphics[scale=0.32]{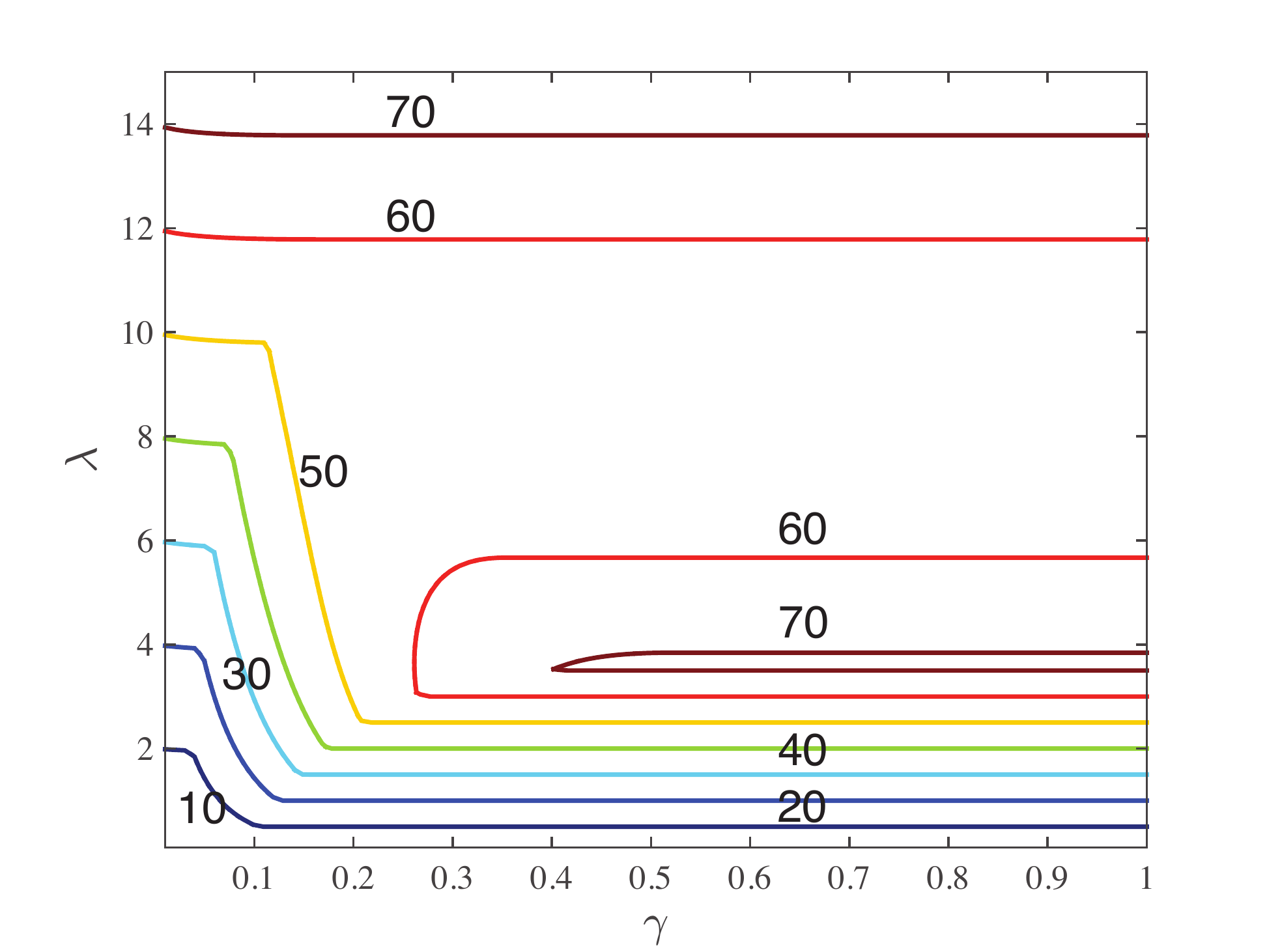}
  \caption{VO's Total Revenue $\Pi^{\rm VO}$.}
  \label{fig:NEW4}
  \vspace{-2mm}
  \end{minipage}
\end{figure*}

\subsection{Optimality of $p_a^{\infty}$ without Assumption \ref{assumption:infinite}}
In Proposition \ref{proposition:adprice}, we have shown that $p_a^\infty$ in (\ref{equ:pa:infinite}) is the optimal solution of Problem \ref{problem:VOadprice}, assuming both $M$ and $\sigma_{\max}$ going to $\infty$ (Assumption \ref{assumption:infinite}). Now we numerically demonstrate that price $p_a^{\infty}$ in (\ref{equ:pa:infinite}) {{generates a close-to-optimal advertising revenue}} to Problem \ref{problem:VOadprice} for most finite values of $M$ and $\sigma_{\max}$.

{{For a particular $\left(M,\sigma_{\max}\right)$-pair, we can compute $p_a^{\infty}$ by (\ref{equ:pa:infinite}),{\footnote{{For $p_a^{\infty}$ in (\ref{equ:pa:infinite}), $\eta$ is defined as $\lim_{M,{\sigma _{\max }} \to \infty } \frac{M}{{{\sigma _{\max }}}}$. We can simply choose $\eta=\frac{M}{\sigma_{\max}}$ to compute $p_a^{\infty}$ for the finite $M$ and $\sigma_{\max}$ situation.}}} and obtain the corresponding advertising revenue ${\Pi_a^{\rm VO}}\left( {{p_f},{p_a^{\infty}},\delta} \right)$ by (\ref{equ:VO:Ra}). Moreover, we can compute $p_a^*$ by (\ref{equ:generaladprice}), and obtain the optimal advertising revenue ${\Pi_a^{\rm VO}}\left( {{p_f},{p_a^*},\delta} \right)$ by (\ref{equ:VO:Ra}). We define \begin{align}
\zeta\triangleq \frac{{\Pi_a^{\rm VO}}\left( {{p_f},{p_a^{\infty}},\delta} \right)}{{\Pi_a^{\rm VO}}\left( {{p_f},{p_a^*},\delta} \right)}.
\end{align}
We can show that $\zeta\in\left[0,1\right]$,{\footnote{{Specifically, we can show that $p_a^{\infty}$ in (\ref{equ:pa:infinite}) is feasible to Problem \ref{problem:VOadprice}. Hence, the advertising revenue under price $p_a^{\infty}$, ${\Pi_a^{\rm VO}}\left( {{p_f},{p_a^{\infty}},\delta} \right)$, is no greater than that under the optimal price $p_a^*$, ${\Pi_a^{\rm VO}}\left( {{p_f},{p_a^*},\delta} \right)$.}}} and $\zeta$ characterizes the optimality of $p_a^\infty$ without Assumption \ref{assumption:infinite}. In particular, $\zeta=1$ implies that price $p_a^{\infty}$ generates the optimal advertising revenue.}}


We choose $\gamma\sim {\cal U}\left[0.01,1\right]$, $\lambda\sim {\cal U}\left[0.1,5\right]$, and $a\sim {\cal U}\left[1,3\right]$, where ${\cal U}$ denotes the uniform distribution.\footnote{{Since we can show that $\zeta$ is independent of $p_f$, $\delta$, and $N$, we do not specify the numerical settings for their values.}} 
We change $M$ and $\sigma_{\max}$ from $1$ to $15$. For each $\left(M,\sigma_{\max}\right)$-pair, we run the experiment $10,000$ times, {{and obtain the average $\zeta$.}}

{{In Fig. \ref{fig:NEW1}, we plot the average $\zeta$ against $M$ and $\sigma_{\max}$. We observe that the average $\zeta$ is always above $0.99$ when $M\ge 6$ and $\sigma_{\max}\ge6$. That is to say, we have ${{\Pi_a^{\rm VO}}\left( {{p_f},{p_a^{\infty}},\delta} \right)}\ge 0.99 {{\Pi_a^{\rm VO}}\left( {{p_f},{p_a^*},\delta} \right)}$ when $M\ge 6$ and $\sigma_{\max}\ge6$.}} Hence, we summarize the following observation.
\begin{observation}
{Without Assumption \ref{assumption:infinite}, the advertising price computed based on $p_a^{\infty}$ in (\ref{equ:pa:infinite}) can still generate a close-to-optimal advertising revenue for the VO.}
\end{observation}

\subsection{Ad Platform's $\delta^*$ and Revenue with Different $\left(\gamma,\lambda\right)$}\label{subsec:simulationB}

Next we compare the ad platform's revenue sharing policy and revenue for venues with different values of advertising concentration level $\gamma$ and MU visiting frequency $\lambda$. We choose $N=200$, $\theta_{\max}=1$, $\beta=0.1$, $\eta=1$, $a=4$, and $\epsilon=0.01$. We will apply the same settings for the remaining {{experiments}} in Section \ref{sec:simulation}.

Fig. \ref{fig:NEW3} is a contour plot illustrating the ad platform's revenue sharing ratio. 
The horizontal axis corresponds to parameter $\gamma$, and the vertical axis corresponds to parameter $\lambda$. The values on the contour curves are the ad platform's revenue sharing ratios, $\delta^*$, computed for venues with different $\left(\gamma,\lambda\right)$ pairs. 
The ad platform needs to strike a proper balance when choosing $\delta$ to maximize its revenue: 
(a) reduce $\delta$ can motivate the VO to push more MUs towards the advertising sponsored access, at the expense of a smaller ad platform's revenue per ad display; (b) increase $\delta$ can improve the ad platform's revenue per ad display, at the expense of making the advertising sponsored access less attractive to the VO. In Fig. \ref{fig:NEW3}, the revenue sharing ratio $\delta^*$ first decreases with $\lambda$, then increases with $\lambda$, which means approach (a) is more effective when $\lambda$ is small and approach (b) is more effective when $\lambda$ is large. This is because a large $\lambda$ leads to a large indicator $\Omega$, which means that the VO prefers the premium access, even if the ad platform leaves a large proportion of the advertising revenue to the VO. 
Hence, when $\lambda$ is large, it is optimal for the ad platform to set a large $\delta^*$ to take a large fraction of the advertising revenue. 

Fig. \ref{fig:NEW2} is a contour plot illustrating the ad platform's revenue. We observe that the ad platform obtains a large $\Pi^{\rm APL}$ from the venue when $\gamma$ is large ($\gamma>0.9$) and $\lambda$ is small ($1.2<\lambda<1.8$). 
This parameter combination corresponds to a venue with a small equilibrium indicator $\Omega$. According to Proposition \ref{proposition:equWiFiprice}, in this case, the VO chooses the highest Wi-Fi price, \emph{i.e.}, $p_f^*\left(\delta^*\right)=\beta\theta_{\max}$, and hence all MUs choose the advertising sponsored access. As a result, the total advertising revenue is large. Furthermore, based on Fig. \ref{fig:NEW3}, the ad platform sets a large sharing ratio ($\delta^*>0.8$) in this case to extract most of the advertising revenue.

We summarize the observations in Fig. \ref{fig:NEW3} and \ref{fig:NEW2} as follows.
\begin{observation}\label{observation:adplatform}
The ad platform's optimal revenue sharing ratio $\delta^*$ first decreases and then increases with $\lambda$. Furthermore, it obtains a large $\Pi^{\rm APL}$ at the venue with both a large $\gamma$ and a small $\lambda$.
\end{observation}

\vspace{-0.5cm}
\subsection{VO's Revenue with Different $\left(\gamma,\lambda\right)$}\label{subsec:simu:VOrevenue}
We investigate the VO's revenue from the advertising sponsored access $\Pi_{a}^{\rm VO}$, its revenue from the premium access $\Pi_{f}^{\rm VO}$, and its total revenue $\Pi^{\rm VO}=\Pi_{a}^{\rm VO}+\Pi_{f}^{\rm VO}$ at venues with different $\left(\gamma,\lambda\right)$ pairs.

In Fig. \ref{fig:NEW10}, we show the contour plot of $\Pi_{a}^{\rm VO}$. We observe that a VO with $\gamma>0.4$ and $3.5<\lambda<3.7$ has a large $\Pi_{a}^{\rm VO}$. Based on (\ref{equ:combineVOad}) and Proposition \ref{proposition:equWiFiprice}, the total advertising revenue at the equilibrium is $a N {\varphi _a}\left( {{p_f^*\left(\delta^*\right)}} \right) g\left(\lambda,\gamma,\eta\right)$. 
From Proposition \ref{proposition:gamma} (iv) and (\ref{equ:gfunction}), we can show that both $\varphi_a\left(p_f^*\left(\delta^*\right)\right)$ and $g\left(\lambda,\gamma,\eta\right)$ are non-decreasing in $\gamma$. 
Therefore, the total advertising revenue at the equilibrium is non-decreasing in $\gamma$. Moreover, from Fig. \ref{fig:NEW3}, the ad platform chooses a relatively small $\delta^*$ (\emph{i.e.}, $\delta^*<0.7$) at the venue {{with $\gamma>0.4$ and $3.5<\lambda<3.7$}}, and hence the VO obtains a large proportion of the total advertising revenue. 

In Fig. \ref{fig:NEW11}, we show the contour plot of $\Pi_{f}^{\rm VO}$. We find that $\Pi_{f}^{\rm VO}$ is non-decreasing in $\lambda$. The reasons are twofold. First, as $\lambda$ increases, the MUs visit the venue more frequently, and the expected number of time segments requested by the MUs increases. Second, according to Proposition \ref{proposition:lambda} (iv), the proportion of MUs choosing the premium access is non-decreasing in $\lambda$. 

In Fig. \ref{fig:NEW4}, we show the contour plot of $\Pi^{\rm VO}$, which is the summation of $\Pi_{a}^{\rm VO}$ in Fig. \ref{fig:NEW10} and $\Pi_{f}^{\rm VO}$ in Fig. \ref{fig:NEW11}. We find that the VO with both a large $\gamma$ ($\gamma>0.4$) and a medium $\lambda$ ($3.5<\lambda<3.9$) and the VO with a large $\lambda$ have large $\Pi^{\rm VO}$. According to Fig. \ref{fig:NEW10}, the former VO mainly generates its revenue from the advertising sponsored access. According to Fig. \ref{fig:NEW11}, the latter VO mainly generates its revenue from the premium access.

We summarize the key observations in Fig. \ref{fig:NEW10}, \ref{fig:NEW11}, and \ref{fig:NEW4} as follows.
\begin{observation}\label{observation:VO}
{{The VO with both a large $\gamma$ and a medium $\lambda$ has a large total revenue, which is mainly generated from the advertising sponsored access. 
The VO with a large $\lambda$ also has a large total revenue, which is mainly generated from the premium access.}}
\end{observation}

\begin{figure*}[t]
\vspace{-2mm}
  \begin{minipage}[t]{.32\linewidth}
  \includegraphics[scale=0.32]{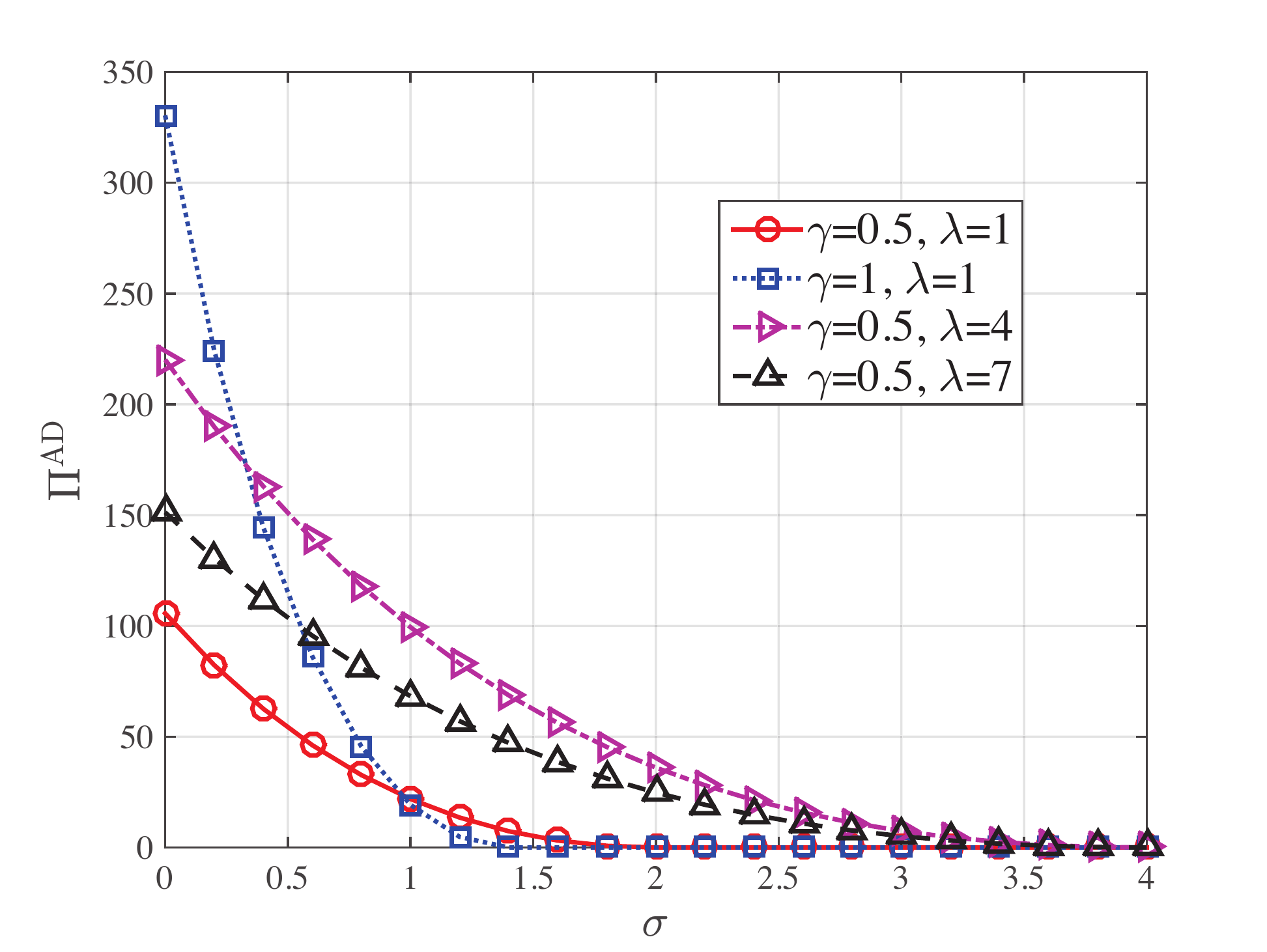}
  \centering
  \caption{AD's Payoff $\Pi^{\rm AD}$.}
  \label{fig:NEW5}
  \vspace{-4mm}
  \end{minipage}
  \begin{minipage}[t]{.32\linewidth}
  \includegraphics[scale=0.32]{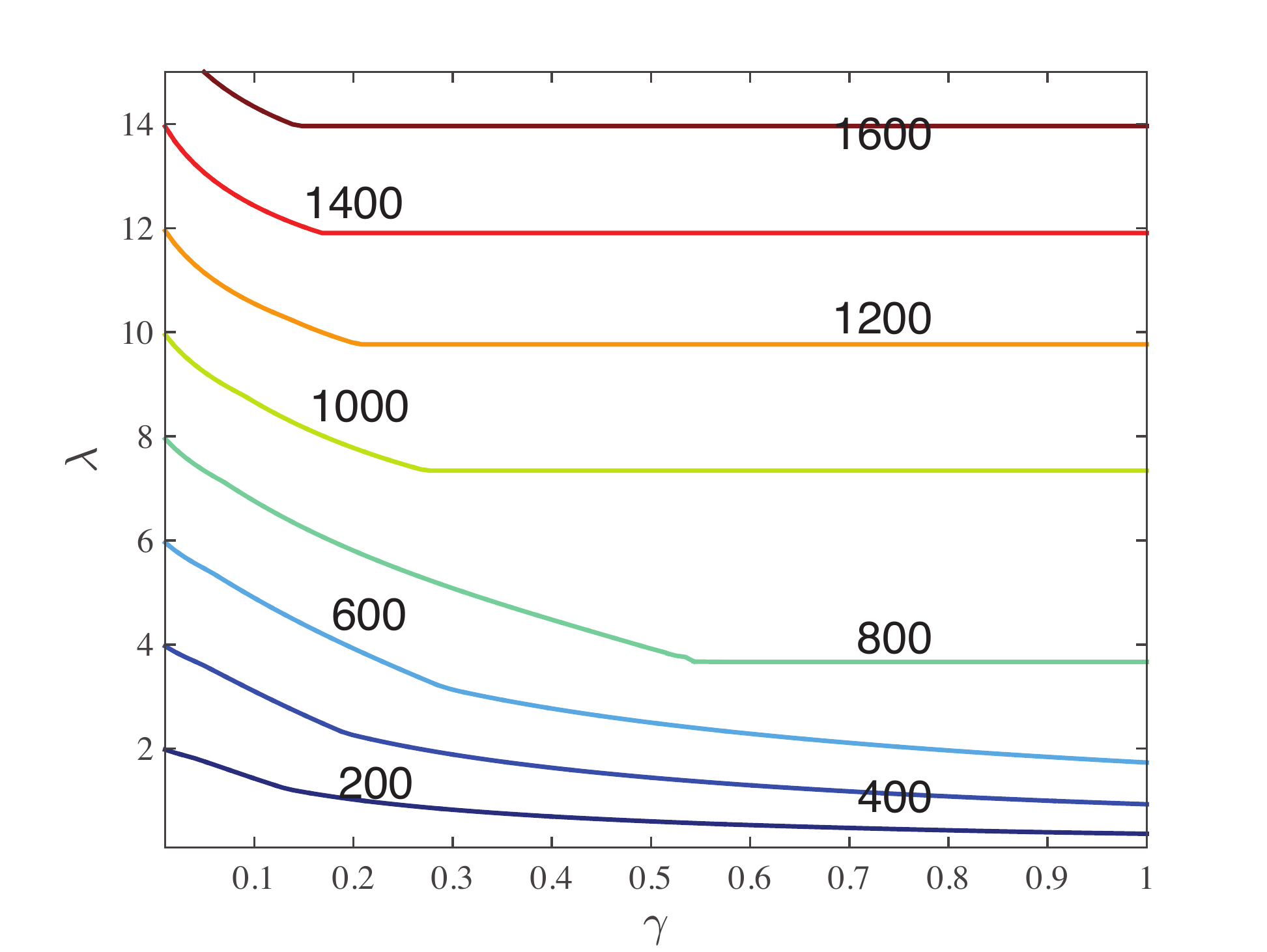}
  \centering
  \caption{Social Welfare.}
  \label{fig:NEW7}
  \vspace{-4mm}
  \end{minipage}
  \begin{minipage}[t]{.32\linewidth}
  \includegraphics[scale=0.32]{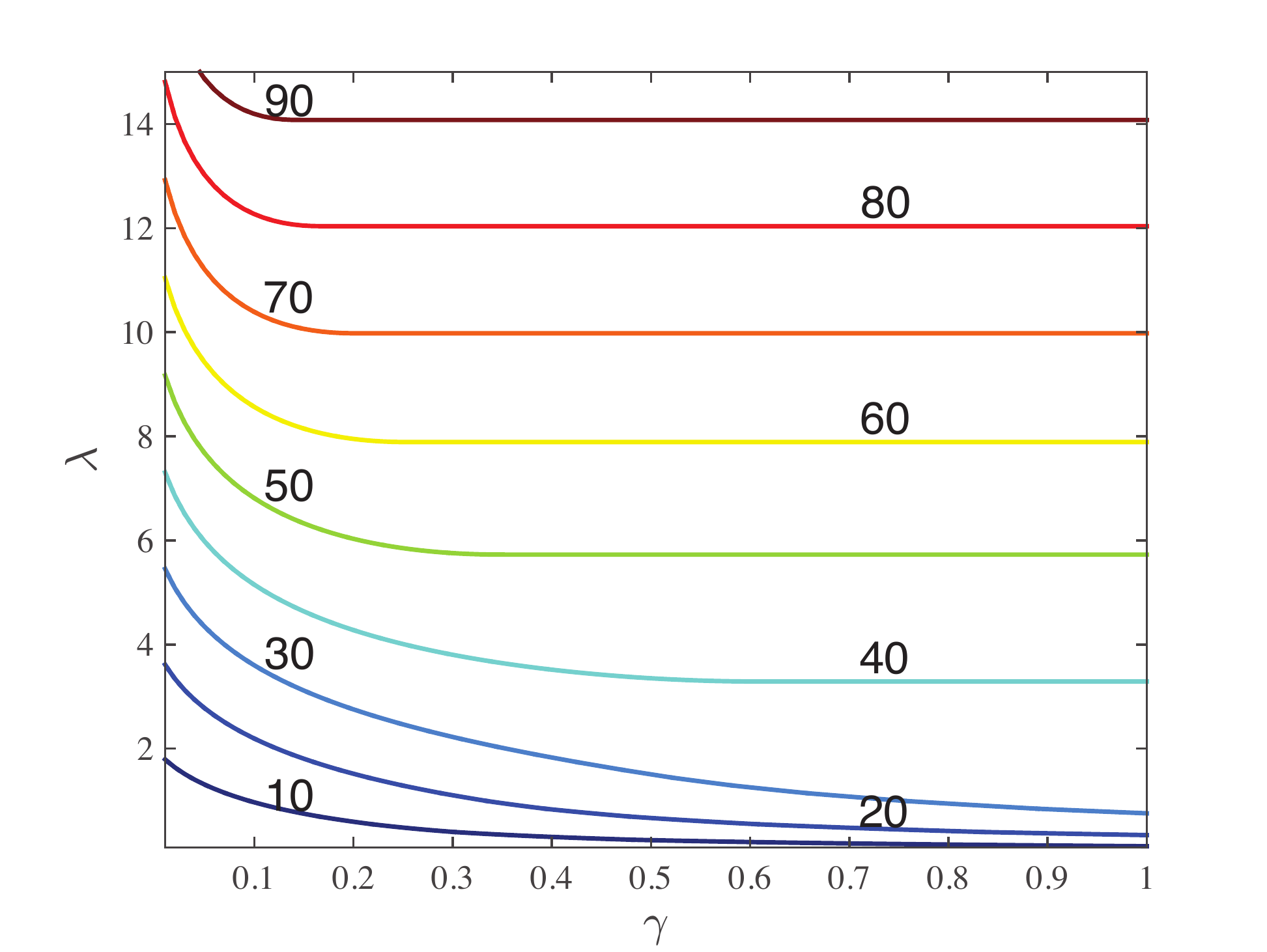}
  \centering
  \caption{Uniform Ad Revenue Sharing Policy: \\VO's Total Revenue $\Pi^{\rm VO}$.}
  \label{fig:NEW6}
  \vspace{-4mm}
  \end{minipage}
\end{figure*}

\subsection{ADs' Payoffs with Different $\left(\gamma,\lambda\right)$}\label{subsec:simu:ADpayoff}
We investigate the ADs' payoffs at venues with different $\left(\gamma,\lambda\right)$ pairs. In Fig. \ref{fig:NEW5}, we plot the ADs' payoffs $\Pi^{\rm AD}$ against the AD type $\sigma$ under different values of $\gamma$ and $\lambda$. We can observe that the ADs with higher popularities (\emph{i.e.}, smaller $\sigma$) have higher payoffs. When comparing curves with the same $\lambda=1$ and different values of $\gamma$ (0.5 and 1), we find that the increase of the concentration level $\gamma$ makes ADs with small values of $\sigma$ even more popular, and hence increases their payoffs. ADs with large values of $\sigma$ will have smaller payoffs accordingly. When comparing curves with the same $\gamma=0.5$ and different values of $\lambda$ ($1$, $4$, and $7$), we observe that ADs' payoffs first increase and then decrease with $\lambda$. According to (\ref{equ:ADpayoff}), the increase of visiting frequency $\lambda$ affects $\Pi^{\rm AD}$ in two aspects: (a) from (\ref{equ:pa:infinite}), the advertising price $p_a^{\infty}$ becomes cheaper, which {{encourages the ADs to buy more advertising spaces and hence}} potentially increases $\Pi^{\rm AD}$; (b) the VO decreases the Wi-Fi price $p_f^*\left(\delta^*\right)$ to attract MUs to the premium access, hence the proportion of MUs choosing the advertising sponsored access, \emph{i.e.}, $\varphi_a\left(p_f^*\left(\delta^*\right)\right)$, becomes smaller, which potentially decreases $\Pi^{\rm AD}$. In Fig. \ref{fig:NEW5}, impact (a) dominates when $\lambda$ increases from $1$ to $4$, and impact (b) dominates when $\lambda$ increases from $4$ to $7$.

We summarize the key observation in Fig. \ref{fig:NEW5} as follows.
\begin{observation}
ADs obtain large payoffs $\Pi^{\rm AD}$ at the venue with a medium $\lambda$, and their payoffs decrease with the index $\sigma$.
\end{observation}

\vspace{-0.4cm}
\subsection{Social Welfare with Different $\left(\gamma,\lambda\right)$}\label{subsec:simu:socialwelfare}
We study the impacts of parameters $\gamma$ and $\lambda$ on the social welfare, and show the contour plot of the social welfare in Fig. \ref{fig:NEW7}. 

First, we observe that the social welfare is non-decreasing in the advertising concentration level $\gamma$. From (\ref{equ:SW}), we can prove that the social welfare is independent of $\gamma$ for $\gamma\ge\frac{2\eta}{\lambda}$, which is consistent with the observation here. 

Second, we discuss the influence of the MU visiting frequency $\lambda$. According to (\ref{equ:SW}), the increase of parameter $\lambda$ has the following three impacts on the social welfare. First, each MU requires more time segments for the Wi-Fi connection, which increases the MUs' total utility. 
Second, as shown in Proposition \ref{proposition:lambda} (iv), more MUs choose the premium access. In this case, less MUs need to watch the advertisements (\emph{i.e.}, $\varphi_a\left(p_f^*\left(\delta^*\right)\right)$ decreases), which increases the MUs' total utility.
Third, since more MUs choose the premium access instead of the advertising sponsored access, the ADs' total utility decreases. 
For most parameter settings, the first two impacts play the dominant roles. 
In Fig. \ref{fig:NEW7}, we can observe that the social welfare always increases with $\lambda$. 
However, under a few extreme parameter settings (\emph{e.g.}, large unit advertising profit $a$ and utility reduction factor $\beta$), the third impact plays the dominant role, and the social welfare may decrease with $\lambda$ in the medium $\lambda$ regime. 
We provide a related example {{in our technical report \cite{haoran2016TON}}}.

We summarize the key observation in Fig. \ref{fig:NEW7} as follows.
\begin{observation}
The social welfare is always non-decreasing in $\gamma$. 
{{Moreover, the social welfare is increasing in $\lambda$, excluding the medium $\lambda$ regime.}}
\end{observation}

\subsection{Uniform Advertising Revenue Sharing Policy $\delta_U$}\label{subsec:simu:uniform}
In Section \ref{subsec:APL}, we assumed that the ad platform can set different advertising revenue sharing ratios for different VOs. This, however, may not be desirable in practice due to the fairness consideration. In Fig. \ref{fig:NEW6}, we consider a more practical case, where the ad platform chooses a uniform advertising revenue sharing ratio $\delta_U\in\left[0,1-\epsilon\right]$ for all VOs.

We assume that VOs have uniformly distributed $\gamma$ and $\lambda$ ($\gamma\sim {\cal U}\left[0.01,1\right]$, $\lambda\sim {\cal U}\left[0.1,15\right]$), and are identical in other parameters. We formulate the ad platform's problem as follows. 
\begin{problem}\label{problem:uniformdelta}
The ad platform decides $\delta_U^*$ by solving{\footnote{We obtain the objective function in (\ref{equ:uniformdelta:a}) by taking the expectation of the ad platform's revenue $\Pi^{\rm APL}\left(\delta\right)$ in (\ref{equ:APLrevenue}) with respect to $\gamma$ and $\lambda$.}}
\begin{align}
& \max {~}{\mathbb{E}_{\gamma,\lambda}}\left\{{\delta_U aN \varphi_a\left({p_{f}^*\left( \delta_U  \right)}\right) g\left( {\lambda ,\gamma,\eta } \right)}\right\} \label{equ:uniformdelta:a}\\
& {\rm{var.~~~}} 0\le\delta_U\le1-\epsilon,
\end{align}
where $p_{f}^*\left(\delta_U\right)$ is the VO's optimal Wi-Fi pricing response under revenue sharing ratio $\delta_U$, and is given in (\ref{equ:VOoptWiFiprice}).
\end{problem}

We consider $10,000$ VOs. By solving Problem \ref{problem:uniformdelta} numerically, we obtain the optimal $\delta_U^*=0.81$. 
Fig. \ref{fig:NEW6} is a contour figure illustrating the VO's total revenue $\Pi^{\rm VO}$ with different values of $\gamma$ and $\lambda$ under $\delta_U^*=0.81$.{\footnote{Here, we only show the impact of $\delta_U^*$ on the VO's revenue. This is because it is obvious that the ad platform's revenue under $\delta_U^*=0.81$ is not greater than its revenue under $\delta^*$ in the VO-specific revenue sharing case. Furthermore, as {{shown}} in Section \ref{subsec:adprice}, the advertising price is independent of the ad platform's sharing policy. Therefore, the uniform advertising revenue sharing policy does not affect the ADs' payoffs.}} {{Next we compare the results in Fig. \ref{fig:NEW6} (the uniform revenue sharing case) with those in Fig. \ref{fig:NEW4} (the VO-specific revenue sharing case).}} 

{{First, we find that the VO's total revenue in Fig. \ref{fig:NEW6} always increases with $\lambda$, while the VO's total revenue in Fig. \ref{fig:NEW4} decreases with $\lambda$ in some cases (\emph{e.g.}, when $\gamma>0.4$ and $3.9<\lambda<5.6$).}} 
{{This is because a larger $\lambda$ implies that the MUs request more time segments of Wi-Fi connection and there are more advertising spaces. 
In the uniform revenue sharing case, the ad platform chooses the same sharing ratio, $\delta_U^*$, for all venues. Hence, in Fig. \ref{fig:NEW6}, the VO's total revenue always increases with $\lambda$. 
In the VO-specific revenue sharing case, the ad platform's sharing ratio $\delta^*$ increases with $\lambda$ for some $\lambda$ (as shown in Fig. \ref{fig:NEW3}). 
In this situation, the proportion of advertising revenue {received} by the VO decreases with $\lambda$, and hence the VO's total revenue in Fig. \ref{fig:NEW4} may decrease with $\lambda$.}}

{{Second, a VO with a medium $\lambda$ in Fig. \ref{fig:NEW6} has a smaller total revenue than that in Fig. \ref{fig:NEW4}. 
Moreover, a VO with a large $\lambda$ in Fig. \ref{fig:NEW6} has a larger total revenue than that in Fig. \ref{fig:NEW4}. 
These are consistent with the comparison between $\delta_U^*$ here (the uniform revenue sharing case) and $\delta^*$ in Fig. \ref{fig:NEW3} (the VO-specific revenue sharing case).}} 
For those VOs with $ \delta_U^*>\delta^* $, they obtain smaller proportions of the advertising revenue in the uniform revenue sharing case, so their revenues decrease. Otherwise, they obtain larger proportions of the advertising revenue in the uniform revenue sharing case, which increases their revenue. 

We summarize the key observations in Fig. \ref{fig:NEW6} as follows.
\begin{observation}
The VO's revenue under the uniform revenue sharing policy increases with $\lambda$. 
Compared with the VO-specific revenue sharing policy, the uniform revenue sharing policy increases the revenue of the VO with a large $\lambda$, and decreases the revenue of the VO with a medium $\lambda$.
\end{observation}



\section{Conclusion}

In this work, we studied the public Wi-Fi monetization problem, and analyzed the economic interactions among the ad platform, VOs, MUs, and ADs through a three-stage Stackelberg game. 
Our analysis led to several important observations: 
(a) the ad platform's advertising revenue sharing policy affects the {{VOs'}} Wi-Fi {{prices}} but not the {{VOs'}} advertising {{prices}}; 
(b) the advertising concentration level $\gamma$ and the MU visiting frequency $\lambda$ have the opposite impacts on equilibrium outcomes; 
(c) the ad platform obtains large {{revenues}} at the {{venues}} with both large $\!\gamma$ and small $\lambda$; 
and (d) the {{VOs}} with both large concentration level and medium MU visiting frequency and the {{VOs}} with large MU visiting frequency obtain large revenues. 

{{In our future work, we plan to relax the assumptions of the uniformly distributed MU types and AD types, and also consider the MUs and ADs with multi-dimensional heterogeneity. 
For example, the MUs can have heterogeneous utility reduction factors $\beta$, besides the heterogeneous Wi-Fi access valuations $\theta$. The ADs can have heterogeneous unit advertising profits $a$, besides the heterogeneous popularity indexes $\sigma$. According to \cite{laffont1987optimal}, the optimal pricing problem for the multi-dimensional heterogeneous buyers is generally much more challenging than that for the single-dimensional heterogeneous buyers.}} 
{Moreover, the VOs can organize auctions and let the ADs bid for the ad spaces. In this situation, the VOs should allocate the ad spaces to the ADs based on the ADs' bids and advertising budgets. We are interested in applying the auction-based framework (instead of the pricing-based framework in this paper) to study the trading of the ad spaces, and investigating the corresponding influence on the equilibrium outcomes.}




\bibliographystyle{IEEEtran}
\bibliography{bare_conf}

\vspace{-1.2cm}

\begin{IEEEbiography}
[{\includegraphics[width=1in,height=1.25in,clip,keepaspectratio]{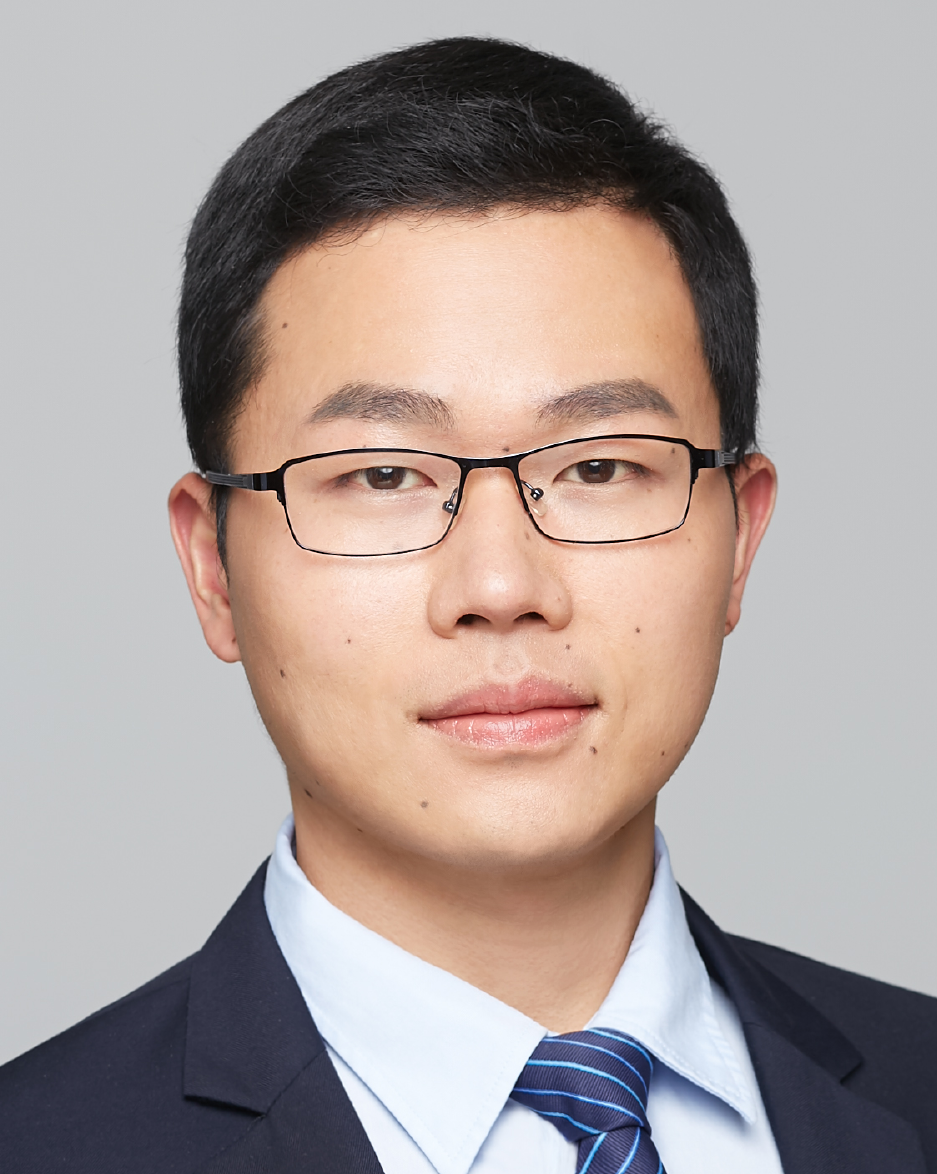}}]
{Haoran Yu} received the Ph.D. degree from the Chinese University of Hong Kong in 2016. He was a Visiting Student in the Yale Institute for Network Science and the Department of Electrical Engineering at Yale University during 2015-2016. He is currently a Post-Doctoral Fellow in the Department of Information Engineering at the Chinese University of Hong Kong. His research interests lie in the field of wireless communications and network economics, with current emphasis on cellular/Wi-Fi integration, LTE in unlicensed spectrum, economics of Wi-Fi networks, and location-based services. He was awarded the Global Scholarship Programme for Research Excellence by the Chinese University of Hong Kong. His paper in {\it IEEE INFOCOM 2016} was selected as a Best Paper Award finalist and one of top 5 papers from over 1600 submissions.
\end{IEEEbiography}

\vspace{-1.2cm}

\begin{IEEEbiography}
[{\includegraphics[width=1in,height=1.25in,clip,keepaspectratio]{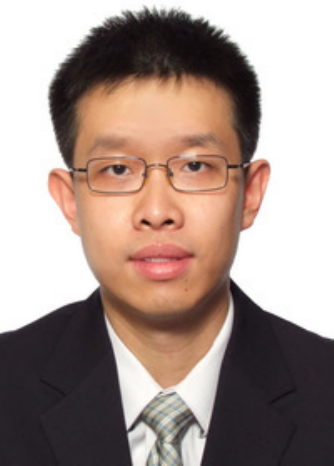}}]
{Man Hon Cheung} received the B.Eng. and M.Phil. degrees in Information Engineering from the Chinese University of Hong Kong (CUHK) in 2005 and 2007, respectively, and the Ph.D. degree in Electrical and Computer Engineering from the University of British Columbia (UBC) in 2012. Currently, he is a postdoctoral fellow in the Department of Information Engineering in CUHK. He received the IEEE Student Travel Grant for attending {\it IEEE ICC 2009}. He was awarded the Graduate Student International Research Mobility Award by UBC, and the Global Scholarship Programme for Research Excellence by CUHK. He serves as a Technical Program Committee member in {\it IEEE ICC}, {\it Globecom}, {\it WCNC}, and {\it WiOpt}. His research interests include the design and analysis of wireless network protocols using optimization theory, game theory, and dynamic programming, with current focus on mobile data offloading, mobile crowdsensing, and network economics.
\end{IEEEbiography}

\vspace{-1.2cm}

\begin{IEEEbiography}
[{\includegraphics[width=1in,height=1.25in,clip,keepaspectratio]{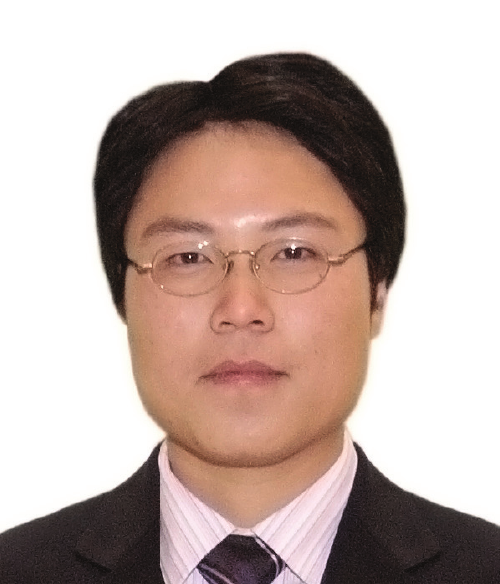}}]
{Lin Gao} (S'08-M'10-SM'16) is an Associate Professor at Harbin Institute of Technology, Shenzhen, China. He received M.S. and Ph.D. degrees in Electronic Engineering from Shanghai Jiao Tong University in 2006 and 2010, respectively. He was a Postdoc Research Fellow in the Network Communications and Economics Lab at The Chinese University of Hong Kong from 2010 to 2015. He received the IEEE ComSoc Asia-Pacific Outstanding Young Researcher Award in 2016. His research interests are in the interdisciplinary area combining telecommunications and microeconomics, with a particular focus on the game-theoretic and economic analysis for various communication and network scenarios, including cognitive radio networks, TV white space networks, cooperative communications, 5G communications, mobile crowd sensing, and Internet-of-Things. 
\end{IEEEbiography}

\vspace{-1.2cm}

\begin{IEEEbiography}
[{\includegraphics[width=1in,height=1.25in,clip,keepaspectratio]{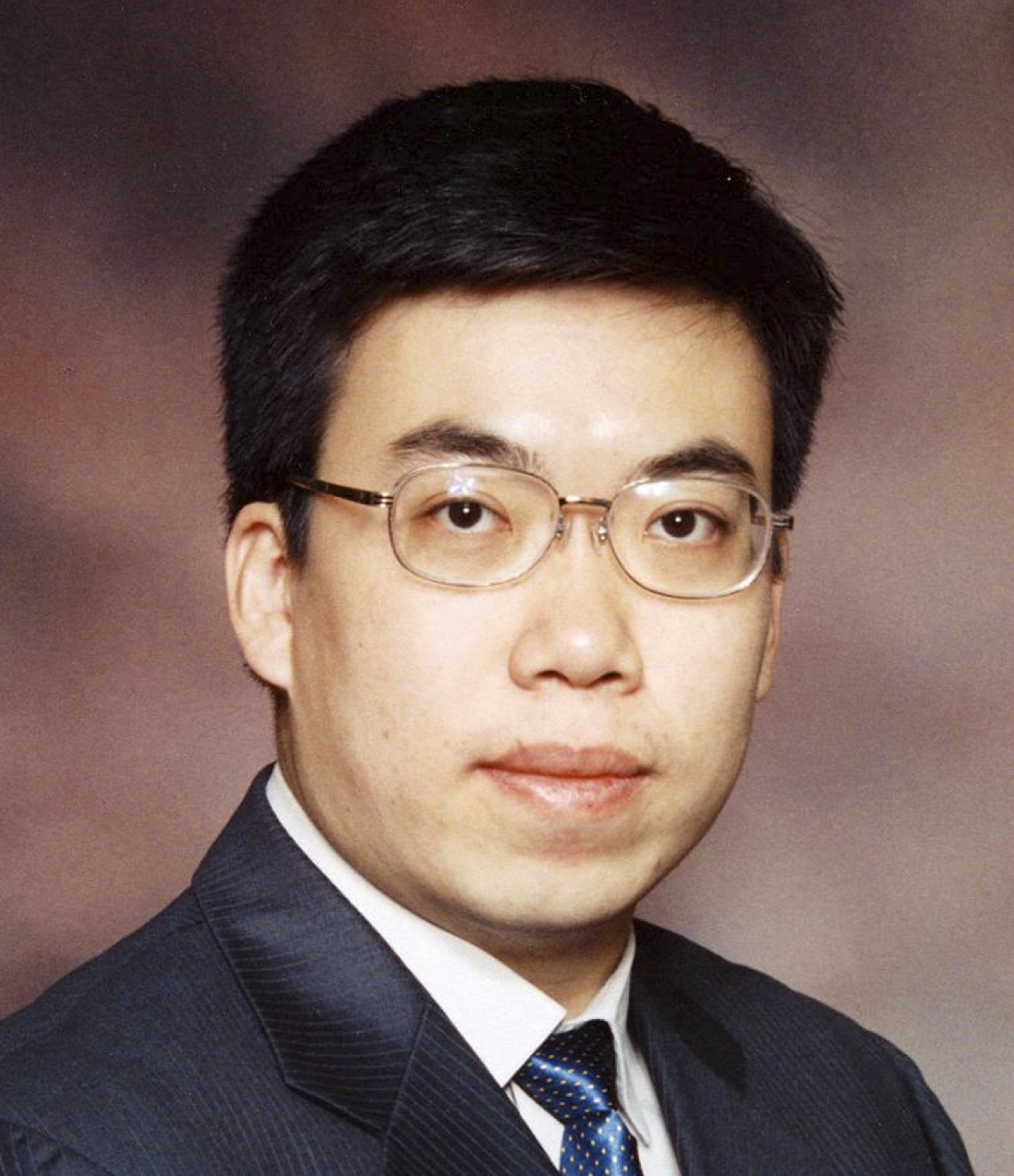}}]
{Jianwei Huang} (F'16) is an IEEE Fellow, a Distinguished Lecturer of IEEE Communications Society, and a Thomson Reuters Highly Cited Researcher in Computer Science. He is an Associate Professor and Director of the Network Communications and Economics Lab (ncel.ie.cuhk.edu.hk), in the Department of Information Engineering at the Chinese University of Hong Kong. He received the Ph.D. degree from Northwestern University in 2005, and worked as a Postdoc Research Associate at Princeton University during 2005-2007. Dr. Huang is the co-recipient of 8 Best Paper Awards, including IEEE Marconi Prize Paper Award in Wireless Communications in 2011. He has co-authored six books, including the textbook on ``Wireless Network Pricing''. He received the CUHK Young Researcher Award in 2014 and IEEE ComSoc Asia-Pacific Outstanding Young Researcher Award in 2009. He has served as an Associate Editor of IEEE/ACM Transactions on Networking, IEEE Transactions on Wireless Communications,  IEEE Journal on Selected Areas in Communications - Cognitive Radio Series, and IEEE Transactions on Cognitive Communications and Networking. He has served as the Chair of IEEE ComSoc Cognitive Network Technical Committee and Multimedia Communications Technical Committee. 
\end{IEEEbiography}

\appendix


{{
\subsection{Randomized Implementation of $m^*\left( {{\sigma},{p_a},{p_f}} \right)$}

First, we explain the implementation of a non-integer $m^*\left( {{\sigma},{p_a},{p_f}} \right)$. When $m^*\left( {{\sigma},{p_a},{p_f}} \right)$ is not an integer, the type-$\sigma$ AD purchases the ad spaces in a randomized manner such that the {expected} number of purchased ad spaces equals $m^*\left( {{\sigma},{p_a},{p_f}} \right)$. Specifically, we define 
\begin{align}
& m_{\rm floor}\triangleq \lfloor m^*\left( {{\sigma},{p_a},{p_f}} \right)  \rfloor,\\
& m_{\rm ceil}\triangleq \lceil m^*\left( {{\sigma},{p_a},{p_f}} \right)  \rceil,\\
& \kappa \triangleq m^*\left( {{\sigma},{p_a},{p_f}} \right) -\lfloor m^*\left( {{\sigma},{p_a},{p_f}} \right)  \rfloor.
\end{align}    
Here, $m_{\rm floor}$ is the largest integer no greater than $m^*\left( {{\sigma},{p_a},{p_f}} \right) $, $m_{\rm ceil}$ is the smallest integer no smaller than $m^*\left( {{\sigma},{p_a},{p_f}} \right) $, and $\kappa\in\left[0,1\right)$ is the fractional part of $m^*\left( {{\sigma},{p_a},{p_f}} \right) $. Under the randomized purchasing strategy, the AD purchases $m_{\rm floor}$ and $m_{\rm ceil}$ ad spaces with the probabilities $1-\kappa$ and $\kappa$, respectively. In this case, the number of purchased ad spaces is always an integer (either $m_{\rm floor}$ or $m_{\rm ceil}$), and the {expected} number of purchased ad spaces equals $m^*\left( {{\sigma},{p_a},{p_f}} \right) $.

Second, we show that letting the ADs implement the non-integer $m^*\left( {{\sigma},{p_a},{p_f}} \right)$ in a randomized manner does not affect our analysis for the ad platform's, VO's, and MUs' equilibrium strategies. Under the randomized implementation, a type-$\sigma$ AD's {expected} number of purchased ad spaces is always $m^*\left( {{\sigma},{p_a},{p_f}} \right)$. According to equation (\ref{equ:Qpa}), the randomized implementation does not affect the expected total number of ad spaces sold to all ADs $Q\left(p_a,p_f\right)$. Therefore, the randomized implementation does not change Problem \ref{problem:VOadprice}'s analysis, hence the randomized implementation does not change any of the later analysis for Stage II and Stage I either

Third, we show that the randomized implementation may reduce the ADs' payoffs, but such an influence is minor. According to (\ref{equ:ADpayoff}), a type-$\sigma$ AD's payoff function $\Pi^{\rm AD}\left({\sigma},m,p_f,p_a\right)$ is concave in $m$. We use $\Pi_{\rm Rand}^{\rm AD}\left({\sigma},m^*\left( {{\sigma},p_a,p_f} \right),p_f,p_a\right)$ to denote the AD's expected payoff under the randomized implementation, which can be computed as follows:
\begin{align}
\nonumber
& \Pi_{\rm Rand}^{\rm AD}\left({\sigma},m^*\left( {{\sigma},p_a,p_f} \right),p_f,p_a\right)=\\
& \left(1-\kappa\right)\Pi^{\rm AD}\left({\sigma},m_{\rm floor},p_f,p_a\right)+\kappa \Pi^{\rm AD}\left({\sigma},m_{\rm ceil},p_f,p_a\right).
\end{align}
Due to the concavity of $\Pi^{\rm AD}\left({\sigma},m,p_f,p_a\right)$, we have 
\begin{multline}
\Pi_{\rm Rand}^{\rm AD}\left({\sigma},m^*\left( {{\sigma},p_a,p_f} \right),p_f,p_a\right) \le \\\Pi^{\rm AD}\left({\sigma},m^*\left( {{\sigma},{p_a},{p_f}} \right),p_f,p_a\right).
\end{multline}
That is to say, the randomized implementation may reduce the AD's payoff. However, we show that such an influence is minor. We define
\begin{align}
\tau\left(\sigma\right) \triangleq  1-\frac{\!\Pi_{\rm Rand}^{\rm AD}\!\left({\sigma},m^*\!\!\left( {{\sigma},{p_a^\infty},{p_f^*\left(\delta^*\right)}} \right)\!,{p_f^*\left(\delta^*\right)},{p_a^\infty}\right)}{\Pi^{\rm AD}\left({\sigma},m^*\left( {{\sigma},{p_a^\infty},{p_f^*\left(\delta^*\right)}} \right),{p_f^*\left(\delta^*\right)},{p_a^\infty}\right)},
\end{align}
which characterizes the relative reduction of a type-$\sigma$ AD's payoff due to the randomized implementation. Next we show the value of $\tau\left(\sigma\right)$ through numerical experiments. 

We consider a period of one week, and assume that $N=1000$ and $\lambda=4$. Hence, there are $1000$ MUs, and on average each MU visits the venue four times during the week. For the remaining parameters, we choose $\theta_{\max}=1$, $\beta=0.1$, $\gamma=0.5$, $\eta=1$, $a=4$, and $\epsilon=0.01$. We first compute the equilibrium advertising price $p_a^\infty$ from (\ref{equ:pa:infinite}) and the equilibrium Wi-Fi price $p_f^*\left(\delta^*\right)$ from (\ref{equ:equpf}). Based on (\ref{equ:sigmaT}){}, the threshold AD type is $\sigma_T\left(p_a^\infty\right)=4$, which implies that only the ADs with $\sigma\in\left[0,4\right)$ obtain positive payoffs. 

In Fig. \ref{appendix:fig:randomized1}, we plot $\tau\left(\sigma\right)$ against $\sigma\in\left[0,4\right]$. We observe that $\tau\left(\sigma\right)$ is very small for most values of $\sigma$ (except when $\sigma$ is very close to $4$). For example, we have $\tau\left(\sigma\right)<10^{-4}$ for $\sigma\in\left[0,3.9\right]$ and $\tau\left(\sigma\right)<1.1\times 10^{-2}$ for $\sigma\in\left[0,3.99\right]$. Therefore, the influence of the randomized implementation on the ADs' payoffs is minor. 
To understand this, we plot $m^*\left( {{\sigma},{p_a^\infty},{p_f^*\left(\delta^*\right)}} \right)$ against $\sigma\in\left[0,4\right]$ in Fig. \ref{appendix:fig:randomized2}. We can observe that $m^*\left( {{\sigma},{p_a^\infty},{p_f^*\left(\delta^*\right)}} \right)$ is a large number for most values of $\sigma$ (except when $\sigma$ is very close to $4$). In this case, the randomized implementation of a non-integer $m^*\left( {{\sigma},{p_a^\infty},{p_f^*\left(\delta^*\right)}} \right)$ does not significantly change the AD's payoff, and hence the value of $\tau\left(\sigma\right)$ is small.

\begin{figure*}[t]
\begin{minipage}{0.48\linewidth}
\centering
\includegraphics[scale=0.36]{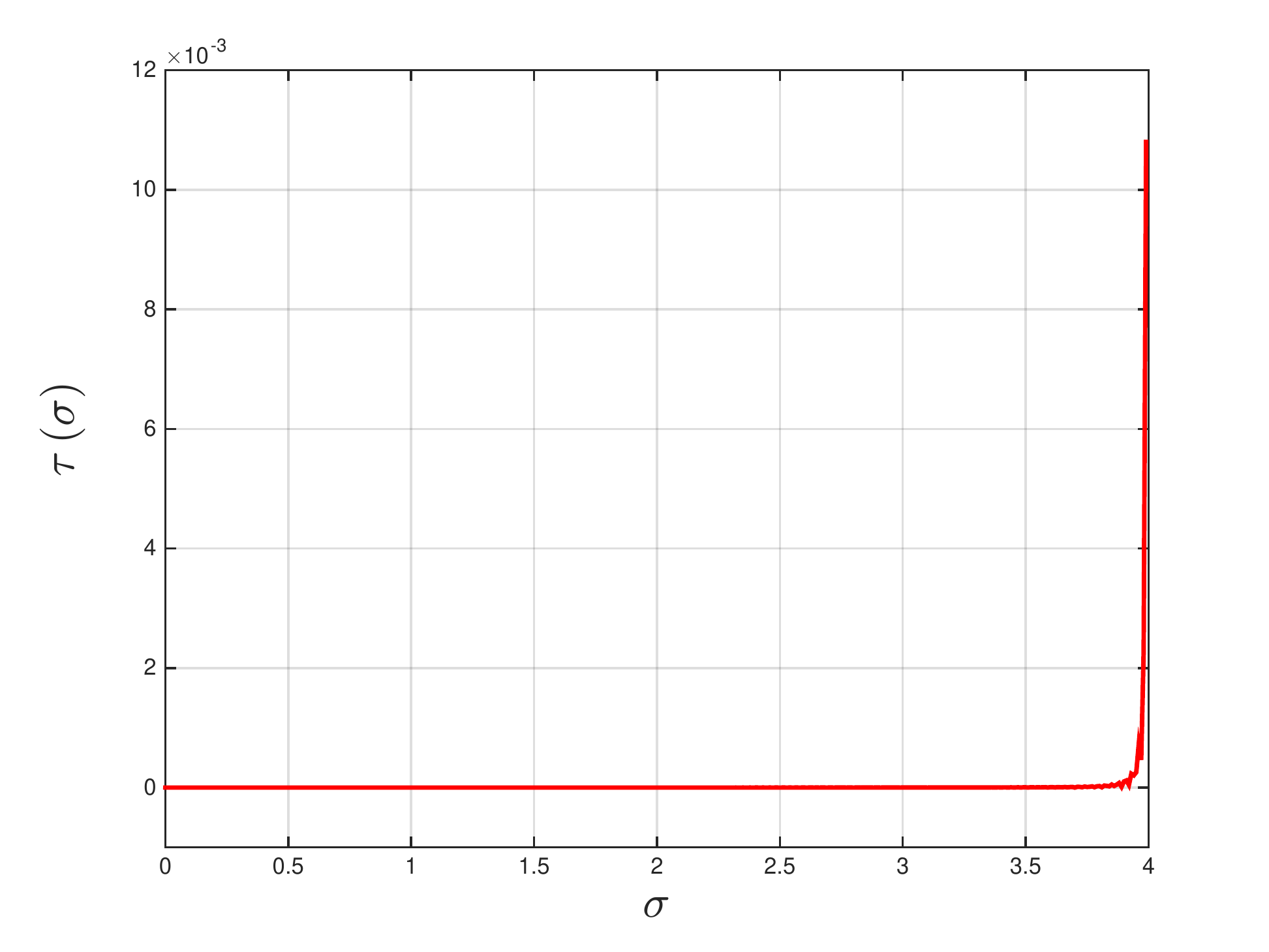}\\
\caption{{ADs' Relative Payoff Reduction.}}
\label{appendix:fig:randomized1}
\end{minipage}
\begin{minipage}{0.48\linewidth}
\centering
\includegraphics[scale=0.36]{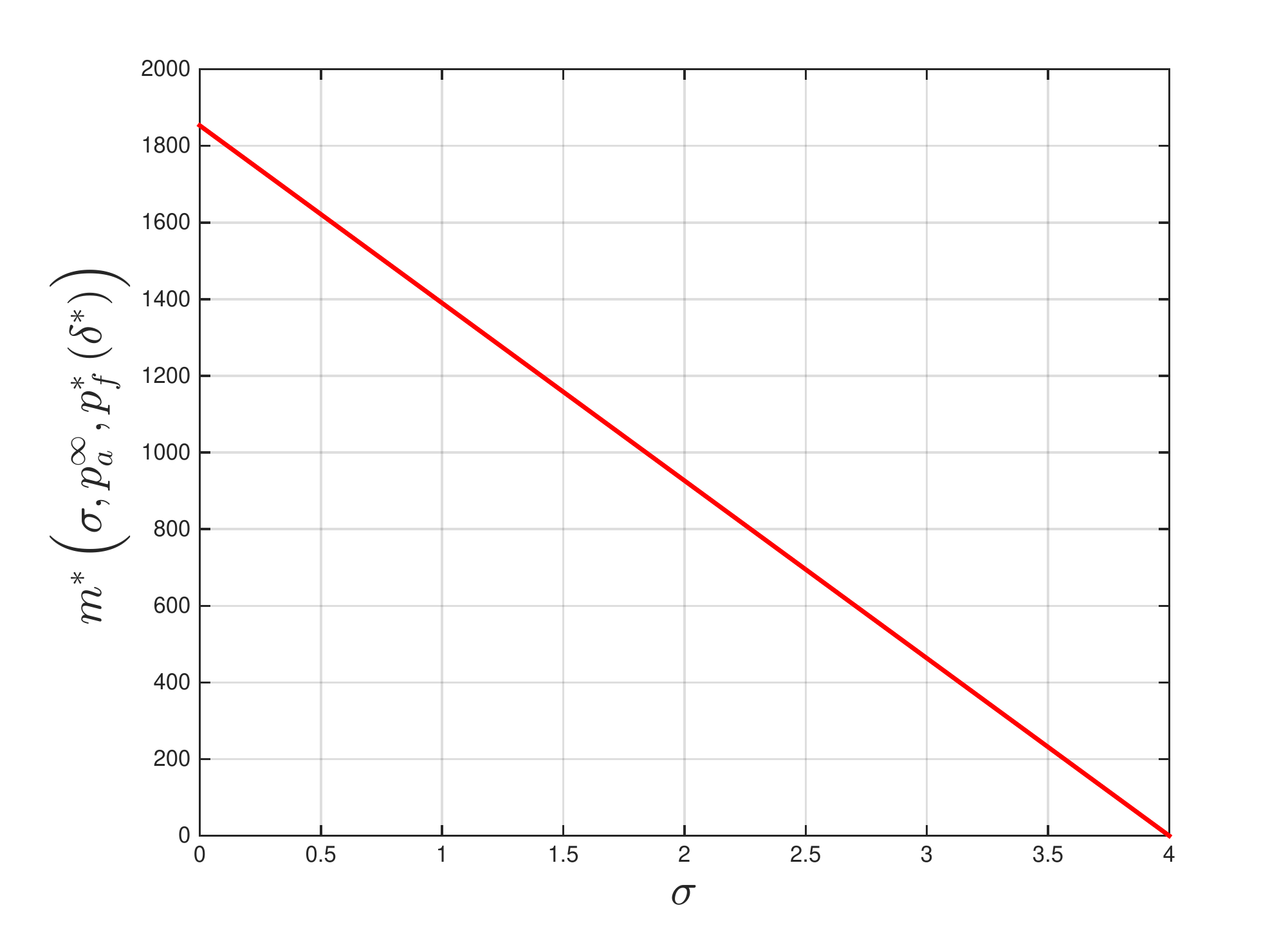}\\
\caption{{ADs' Optimal Purchasing Strategies.}}
\label{appendix:fig:randomized2}
\end{minipage}
\end{figure*}
}}

\subsection{Proof of Proposition \ref{proposition:finite}}
\begin{proof}
We analyze the solution to Problem \ref{problem:VOadprice} by considering the case $\frac{{{\sigma _{\max }}\gamma }}{2} \ge \frac{\lambda }{M}$ and the case $\frac{{{\sigma _{\max }}\gamma }}{2} < \frac{\lambda }{M}$ separately.

\subsubsection{{\textbf{Case A:}} $\frac{{{\sigma _{\max }}\gamma }}{2} \ge \frac{\lambda }{M}$}
First, we prove $\frac{1}{\gamma }\ln \left( {\frac{{a\gamma }}{{{p_a}}}} \right)\le \sigma_{\max}$ by contradiction. 
Suppose $\frac{1}{\gamma }\ln \left( {\frac{{a\gamma }}{{{p_a}}}} \right)> \sigma_{\max}$. From the definition of ${\sigma _T}\left( {{p_a}} \right)$ in (\ref{equ:sigmaT}), ${\sigma _T}\left( {{p_a}} \right) = {\sigma _{\max }}$. 
Hence, we can rewrite $Q\left( {{p_a,p_f}} \right)$ in (\ref{equ:Qpa}) as
\begin{align}
Q\left( {{p_a,p_f}} \right) = \frac{{MN{\varphi _a}\left( {{p_f}} \right)}}{{{\sigma _{\max }}}}\left( {\ln \left( {\frac{{a\gamma }}{{{p_a}}}} \right){\sigma _{\max }} - \frac{\gamma }{2}\sigma _{\max }^2} \right).
\end{align}
By using the assumption $\frac{1}{\gamma }\ln \left( {\frac{{a\gamma }}{{{p_a}}}} \right)> \sigma_{\max}$, we have the following relation:
\begin{align}
Q\left( {{p_a,p_f}} \right) > \frac{\gamma }{2}{\sigma _{\max }}MN{\varphi _a}\left( {{p_f}} \right).\label{equ:appendix:A:Qin}
\end{align}
In Case A, we have the relation $\frac{{{\sigma _{\max }}\gamma }}{2} \ge \frac{\lambda }{M}$. Hence, we can derive the following relation from (\ref{equ:appendix:A:Qin}):
\begin{align}
Q\left( {{p_a,p_f}} \right) > \lambda N{\varphi _a}\left( {{p_f}} \right).
\end{align}
This contradicts with constraint (\ref{equ:Raopt:b}) in Problem \ref{problem:VOadprice}. Therefore, we have proved that $\frac{1}{\gamma }\ln \left( {\frac{{a\gamma }}{{{p_a}}}} \right)\le \sigma_{\max}$. 

Based on $\frac{1}{\gamma }\ln \left( {\frac{{a\gamma }}{{{p_a}}}} \right)\le \sigma_{\max}$ and ${\sigma _T}\left( {{p_a}} \right)$'s definition in (\ref{equ:sigmaT}), we have ${\sigma _T}\left( {{p_a}} \right) = \frac{1}{\gamma }\ln \left( {\frac{{a\gamma }}{{{p_a}}}} \right)$.  
By plugging ${\sigma _T}\left( {{p_a}} \right) = \frac{1}{\gamma }\ln \left( {\frac{{a\gamma }}{{{p_a}}}} \right)$ into the expression of $Q\left( {{p_a,p_f}} \right)$ in (\ref{equ:Qpa}), we can rearrange (\ref{equ:Raopt:a})-(\ref{equ:Raopt:c}) as follows:
\begin{align}
& \max {~~~} \left({1 - \delta } \right){p_a}\frac{MN{\varphi _a\left( {{p_f}} \right)}}{2\sigma_{\max}\gamma
}{\left(\ln \left( {\frac{{a\gamma }}{{{p_a}}}} \right)\right)^2} \label{appendix:equ:Raopt:a}\\
& {\rm s.t.~~~~~} p_a\ge a\gamma {e^{ - \sqrt {\frac{{2\lambda \gamma {\sigma _{\max }}}}{M}} }}, \label{appendix:equ:Raopt:b}\\
& {\rm var.~~~~~} 0\le p_a\le a \gamma.\label{appendix:equ:Raopt:c}
\end{align}
We can prove that the objective function (\ref{appendix:equ:Raopt:a}) is unimodal: it increases with ${p_a}$ for ${p_a} \in \left[ {0,a\gamma {e^{ - 2}}} \right]$ and decreases with ${p_a}$ for ${p_a} \in \left( {a\gamma {e^{ - 2}},a\gamma} \right]$. 
By considering constraint (\ref{appendix:equ:Raopt:b}), we can show the solution to Problem \ref{problem:VOadprice} as follows:
\begin{align}
p_a^* = \left\{ {\begin{array}{*{20}{l}}
{a\gamma {e^{ - \sqrt {\frac{{2\lambda \gamma {\sigma _{\max }}}}{M}} }},}&{{\rm if~}
\frac{\lambda }{M} \le \frac{2}{\gamma \sigma_{\max}},}\\
{a\gamma {e^{ - 2}},}&{{\rm if~}  \frac{\lambda }{M} > \frac{2}{{\gamma {\sigma _{\max }}}}.}
\end{array}} \right.\label{equ:appendix:caseA:pre}
\end{align}
Because the condition $\frac{{{\sigma _{\max }}\gamma }}{2} \ge \frac{\lambda }{M}$  holds under Case A and $\min\left\{\frac{{\gamma {\sigma _{\max }}}}{2},\frac{2}{{\gamma {\sigma _{\max }}}}\right\}\le 1$, we can rewrite (\ref{equ:appendix:caseA:pre}) as
\begin{align}
p_a^* = \left\{ {\begin{array}{*{20}{l}}
{a\gamma {e^{ - \sqrt {\frac{{2\lambda \gamma {\sigma _{\max }}}}{M}} }},}&{{\rm if~}\frac{\lambda }{M} \le \min\left\{\frac{{\gamma {\sigma _{\max }}}}{2},1,\frac{2}{{\gamma {\sigma _{\max }}}}\right\},}\\
{a\gamma {e^{ - 2}},}&{{\rm if~} \frac{{\gamma {\sigma _{\max }}}}{2} \ge \frac{\lambda }{M} > \frac{2}{{\gamma {\sigma _{\max }}}}.}
\end{array}} \right.\label{equ:appendix:caseA}
\end{align}

\subsubsection{{\textbf{Case B:}} $\frac{{{\sigma _{\max }}\gamma }}{2} < \frac{\lambda }{M}$}
In this case, $\sigma_T\left(p_a\right)$ has different expressions for the regime $p_a\in\left[ {0,a\gamma {e^{ - \gamma {\sigma _{\max }}}}} \right]$ and regime $p_a\in\left( {a\gamma {e^{ - \gamma {\sigma _{\max }}}},a\gamma } \right]$. Next we discuss the VO's optimal pricing in these two regimens separately.

\emph{\textbf{Regime 1}: $p_a\in\left[ {0,a\gamma {e^{ - \gamma {\sigma _{\max }}}}} \right]$}: In this regime, we can show that $\frac{1}{\gamma }\ln \left( {\frac{{a\gamma }}{{{p_a}}}} \right)\ge \sigma_{\max}$. From (\ref{equ:sigmaT}), $\sigma_T\left(p_a\right)=\sigma_{\max}$. 
By plugging $\sigma_T\left(p_a\right)=\sigma_{\max}$ into the expression of $Q\left( {{p_a,p_f}} \right)$ in (\ref{equ:Qpa}), we can rearrange (\ref{equ:Raopt:a})-(\ref{equ:Raopt:c}) as follows:
\begin{align}
& \max {~} \left({1 - \delta } \right){p_a} \frac{{MN{\varphi _a}\left( {{p_f}} \right)}}{{{\sigma _{\max }}}}\left( {\ln \left( {\frac{{a\gamma }}{{{p_a}}}} \right){\sigma _{\max }} - \frac{\gamma }{2}\sigma _{\max }^2} \right) \label{appendix2:equ:Raopt:a}\\
& {\rm s.t.~~~} p_a \ge a\gamma {e^{ - \left( {\frac{{\gamma {\sigma _{\max }}}}{2} + \frac{\lambda }{M}} \right)}}, \label{appendix2:equ:Raopt:b}\\
& {\rm var.~~~} 0\le p_a\le a\gamma {e^{ - \gamma {\sigma _{\max }}}} \label{appendix2:equ:Raopt:c}.
\end{align}
We can prove that the objective function (\ref{appendix2:equ:Raopt:a}) is unimodal: it increases for ${p_a} \in \left[ {0,{a\gamma {e^{ - \left( {\frac{{\gamma {\sigma _{\max }}}}{2} + 1} \right)}}}} \right]$ and decreases for ${p_a} \in \left({a\gamma {e^{ - \left( {\frac{{\gamma {\sigma _{\max }}}}{2} + 1} \right)}} ,a\gamma^{-\gamma\sigma_{\max}}} \right]$. 
By considering (\ref{appendix2:equ:Raopt:b}) and (\ref{appendix2:equ:Raopt:c}), we show the optimal $p_a^*$ and the corresponding revenue $\Pi_a^{\rm VO}\left( {{p_f},{p_a^*},\delta } \right)$ in Regime 1 as follows:

(a) If $1 \ge \frac{\lambda }{M} > \frac{{\gamma {\sigma _{\max }}}}{2}$, we have $p_a^*=a\gamma {e^{ - \left( {\frac{{\gamma {\sigma _{\max }}}}{2} + \frac{\lambda }{M}} \right)}}$ and
$ \Pi_a^{\rm VO}\left( {{p_f},{p_a^*},\delta } \right)=\left( {1 - \delta } \right)a\gamma {e^{ - \left( {\frac{{\gamma {\sigma _{\max }}}}{2} + \frac{\lambda }{M}} \right)}}N{\varphi _a}\left( {{p_f}} \right)\lambda $;

(b) If $\frac{\lambda }{M}> \frac{{\gamma {\sigma _{\max }}}}{2} \ge 1$, we have $p_a^*={a\gamma {e^{ - \gamma {\sigma _{\max }}}}}$ and $\Pi_a^{\rm VO}\left( {{p_f},{p_a^*},\delta } \right)=\left( {1 - \delta } \right)\frac{{a{\gamma ^2}{e^{ - \gamma {\sigma _{\max }}}}{\sigma _{\max }}}}{2}MN{\varphi _a}\left( {{p_f}} \right)$;

(c) If $\frac{\lambda }{M} > 1 > \frac{{\gamma {\sigma _{\max }}}}{2}$, we have $p_a^*=a\gamma {e^{ - \left( {\frac{{\gamma {\sigma _{\max }}}}{2} + 1} \right)}}$ and $ \Pi_a^{\rm VO}\left( {{p_f},{p_a^*},\delta } \right)=\left( {1 - \delta } \right)a\gamma {e^{ - \left( {\frac{{\gamma {\sigma _{\max }}}}{2} + 1} \right)}}MN{\varphi _a}\left( {{p_f}} \right)$.

\emph{\textbf{Regime 2}: $p_a\in\left( {a\gamma {e^{ - \gamma {\sigma _{\max }}}},a\gamma} \right]$}: In this regime, we have $\frac{1}{\gamma }\ln \left( {\frac{{a\gamma }}{{{p_a}}}} \right)< \sigma_{\max}$. From (\ref{equ:sigmaT}), $\sigma_T\left(p_a\right)=\frac{1}{\gamma} {\ln \left( {\frac{{a\gamma }}{{{p_a}}}} \right)}$. By plugging it into the expression of $Q\left( {{p_a,p_f}} \right)$ in (\ref{equ:Qpa}), we can rearrange (\ref{equ:Raopt:a})-(\ref{equ:Raopt:c}) as follows:
\begin{align}
& \max {~~~} \left( {1 - \delta } \right) p_a \frac{{MN{\varphi _a}\left( {{p_f}} \right)}}{{2 {\sigma _{\max }}\gamma}}{}{\left( {\ln \left( {\frac{{a\gamma }}{{{p_a}}}} \right)} \right)^2} \label{appendix3:equ:Raopt:a}\\
& {\rm s.t.~~~~~} p_a\ge a\gamma {e^{ - \sqrt {\frac{{2\lambda \gamma {\sigma _{\max }}}}{M}} }}, \label{appendix3:equ:Raopt:b}\\
& {\rm var.~~~~~} a\gamma {e^{ - \gamma {\sigma _{\max }}}} \le p_a\le a \gamma. \label{appendix3:equ:Raopt:c}
\end{align}
The objective function (\ref{appendix3:equ:Raopt:a}) is the same as (\ref{appendix:equ:Raopt:a}), which increases with ${p_a}$ for ${p_a} \in \left[ {0,a\gamma {e^{ - 2}}} \right]$ and decreases with ${p_a}$ for ${p_a} \in \left( {a\gamma {e^{ - 2}},a\gamma} \right]$. 
By considering (\ref{appendix3:equ:Raopt:b}) and (\ref{appendix3:equ:Raopt:c}), we show the optimal $p_a^*$ and the corresponding revenue $\Pi_a^{\rm VO}\left( {{p_f},{p_a^*},\delta } \right)$ in Regime 2 as follows:

(a) If $\frac{\lambda}{M}> \frac{{\gamma {\sigma _{\max }}}}{2}\ge1$, we have $p_a^*={a\gamma {e^{ - 2}}}$ and $\Pi_a^{\rm VO}\left( {{p_f},{p_a^*},\delta } \right)=\frac{{2\left( {1 - \delta } \right)MN{\varphi _a}\left( {{p_f}} \right)a{e^{ - 2}}}}{{{\sigma _{\max }}}}$;

(b) If $\min\left\{\frac{\lambda}{M},1 \right\}> \frac{{\gamma {\sigma _{\max }}}}{2}$, we have $p_a^*=a\gamma {e^{ - \gamma {\sigma _{\max }}}}$ and $\Pi_a^{\rm VO}\left( {{p_f},{p_a^*},\delta } \right)=\frac{{\left( {1 - \delta } \right)MN{\varphi _a}\left( {{p_f}} \right)a{e^{ - \gamma {\sigma _{\max }}}}{\gamma ^2}{\sigma _{\max }}}}{2}$.

\emph{\textbf{Combination of Regime 1 and Regime 2}:} Next we combine the optimal solutions in Regime 1 and Regime 2. Based on the comparison of $\Pi_a^{\rm VO}\left( {{p_f},{p_a^*},\delta } \right)$ in Regime 1 and Regime 2, we can show that:
\begin{align}
p_a^* = \left\{ {\begin{array}{*{20}{l}}
{a\gamma {e^{ - \left( {\frac{{\gamma {\sigma _{\max }}}}{2} + \frac{\lambda }{M}} \right)}},}&{{\rm if~}1 \ge \frac{\lambda }{M} > \frac{{\gamma {\sigma _{\max }}}}{2},}\\
{a\gamma e^{-2},}&{{\rm if~} \frac{\lambda }{M} > \frac{{\gamma {\sigma _{\max }}}}{2} \ge 1,}\\
{a\gamma {e^{ - \left( {\frac{{\gamma {\sigma _{\max }}}}{2} + 1} \right)}}}&{{\rm if~\frac{\lambda }{M} > 1 > \frac{{\gamma {\sigma _{\max }}}}{2}}.}
\end{array}} \right.\label{equ:appendix:caseB}
\end{align}

\subsubsection{Combination of Case A and Case B}
Finally, we show the solution based on the analysis in Case A and Case B. From (\ref{equ:appendix:caseA}) and (\ref{equ:appendix:caseB}), we can obtain
\begin{align}
\!\!{p_a^* \!\!= \!\left\{ {\begin{array}{*{20}{l}}
{\!\!a\gamma {e^{ -\! \sqrt {\frac{{2\lambda \gamma {\sigma _{\max }}}}{M}} }},}&{{\rm \!\!\!\!if~\!}\frac{\lambda }{M} \!\le \!\min\! \left\{ \!{\frac{{\gamma {\sigma _{\max }}}}{2},\!1,\!\frac{2}{{\gamma {\sigma _{\max }}}}} \!\right\}\!\!,}\\
{\!\!a\gamma {e^{ - \left( {\frac{{\gamma {\sigma _{\max }}}}{2} + \frac{\lambda }{M}} \right)}},}&{{\rm \!\!\!\!if~}\frac{{\gamma {\sigma _{\max }}}}{2} < \frac{\lambda }{M} \le 1,}\\
{\!\!a\gamma {e^{ - \left( {\frac{{\gamma {\sigma _{\max }}}}{2} + 1} \right)}},}&{{\rm \!\!\!\!if}\frac{{\gamma {\sigma _{\max }}}}{2} < 1 < \frac{\lambda }{M},}\\
{\!\!a\gamma {e^{ - 2}},}&{{\rm \!\!\!\!other~cases},}
\end{array}} \right.}
\end{align}
which completes the proof of Proposition \ref{proposition:finite}.
\end{proof}

\subsection{Proof of Proposition \ref{proposition:adprice}}
\begin{proof}
The VO's optimal advertising price under general $\sigma_{\max}$ and $M$ is given in (\ref{equ:generaladprice}). With Assumption \ref{assumption:infinite}, we have $\sigma_{\max}\rightarrow \infty$. Hence, the conditions $\frac{{\gamma {\sigma _{\max }}}}{2} < \frac{\lambda }{M} \le 1$ and $\frac{{\gamma {\sigma _{\max }}}}{2} < 1 < \frac{\lambda }{M}$ do not hold. Furthermore, the condition $\frac{\lambda }{M} \!\le \!\min\! \left\{ \!{\frac{{\gamma {\sigma _{\max }}}}{2},\!1,\!\frac{2}{{\gamma {\sigma _{\max }}}}} \!\right\}$ can be simplified as $\frac{\lambda }{M} \le \frac{2}{{\gamma {\sigma _{\max }}}}$. Using $\eta=\lim_{M,{\sigma _{\max }} \to \infty } \frac{M}{{{\sigma _{\max }}}}$, we can further rewrite this condition as $\lambda  \le \frac{{2\eta }}{\gamma }$. Therefore, the optimal advertising price under Assumption \ref{assumption:infinite} is simply
\begin{align}
p_a^{\infty} = \left\{ {\begin{array}{*{20}{l}}
{a\gamma {e^{ - \sqrt {\frac{2\lambda \gamma}{\eta} } }},}&{{\rm if~}0< \lambda   \le \frac{2\eta}{\gamma},}\\
{a\gamma {e^{ - 2}},}&{{\rm if~}\lambda   > \frac{2\eta}{\gamma}.}
\end{array}} \right.
\end{align}
This completes the proof of Proposition \ref{proposition:adprice}.
\end{proof}

\subsection{Proof of Proposition \ref{proposition:WiFiprice}}
\begin{proof}
The objective function (\ref{equ:formula:WiFiprice:a}) is a quadratic function of $p_f$. We can show that the objective function increases with $p_f$ for ${p_f} \in \left[ {0,\frac{{\beta {\theta _{\max }}}}{2} + \left( {1 - \delta } \right)\frac{{ag\left( {\lambda ,\gamma ,\eta } \right)}}{2 \lambda }} \right]$ and decreases with $p_f$ for ${p_f} \in \left( {\frac{{\beta {\theta _{\max }}}}{2} + \left( {1 - \delta } \right)\frac{{ag\left( {\lambda ,\gamma ,\eta } \right)}}{2 \lambda },\infty } \right)$. 
From (\ref{equ:formula:WiFiprice:b}), $p_f\in\left[ {0,\beta {\theta _{\max }}} \right]$. Hence, we can compute the optimal Wi-Fi price under the sharing policy $\delta$ as
\begin{align}
p_f^*\left( \delta  \right) = \min \left\{ \beta \theta_{\max}, \frac{{\beta {\theta _{\max }}}}{2} \!+ \frac{{\!\left( {1\! -\! \delta } \right)\!a}}{2 \lambda }g\left( {\lambda ,\gamma ,\eta } \right)\right\}.\label{equ:appendix:pffunction}
\end{align}
We can rearrange (\ref{equ:appendix:pffunction}) as
\begin{align}
p_f^*\left( \delta  \right)\! =\! \frac{{\beta {\theta _{\max }}}}{2}\!+\! \min \left\{ {\frac{{\left( {1 - \delta } \right)a}}{2 \lambda}g\left(\lambda,\gamma,\eta\right),\frac{{\beta {\theta _{\max }}}}{2}} \right\},
\end{align}
which completes the proof of Proposition \ref{proposition:WiFiprice}.
\end{proof}

\subsection{Proofs of Proposition \ref{proposition:delta} and Proposition \ref{proposition:equWiFiprice}}
\begin{proof}
We discuss the ad platform's sharing policy and the VO's Wi-Fi price at the equilibrium in the following four cases: $\Omega\in\left(0,\epsilon\right]$, $\Omega\in\left(\epsilon,\frac{1}{3}\right]$, $\Omega\in\left(\frac{1}{3},1-2\epsilon\right)$, and $\Omega\in\left[1-2\epsilon,\infty\right)$.

\subsubsection{\textbf{Case A:} $\Omega\in\left(0,\epsilon\right]$} 
In this case, $1-\epsilon\le 1-\Omega$. Since $\delta\in\left[0,1-\epsilon\right]$, we have $\delta\le 1-\Omega$. 
By using $\Omega= \frac{{\lambda \beta {\theta _{\max }}}}{{ag\left( {\lambda ,\gamma,\eta } \right)}}$, we can rearrange $\delta\le 1-\Omega$ as $\left( {1 - \delta } \right)\frac{{a g\left(\lambda,\gamma,\eta\right)}}{2\lambda}\ge \frac{\beta\theta_{\max}}{2}$. 
Based on (\ref{equ:VOoptWiFiprice}), this implies that $p_f^*\left(\delta\right)=\beta\theta_{\max}$.
Then we can rewrite Problem \ref{problem:platform} as
\begin{align}
&\max {~} \delta aNg\left( {\lambda ,\gamma,\eta } \right)\\
& {\rm var.~~~} 0\le\delta\le1-\epsilon.
\end{align}
We can easily show that at the equilibrium the optimal sharing policy is $\delta^*=1-\epsilon$, and the VO's Wi-Fi price is $p_f^*\left(\delta^*\right)=\beta\theta_{\max}$.

\subsubsection{\textbf{Case B:} $\Omega\in\left(\epsilon,\frac{1}{3}\right]$} In this case, we discuss $\delta\in\left[0,1-\Omega\right]$ and $\delta\in\left(1-\Omega,1-\epsilon\right]$ separately.

(a) When $\delta\in\left[0,1-\Omega\right]$, we can show $p_f^*\left(\delta\right)=\beta\theta_{\max}$ based on (\ref{equ:VOoptWiFiprice}). From (\ref{equ:APLrevenue}), the ad platform's revenue is $\Pi^{\rm APL}\left(\delta\right)=\delta aNg\left( {\lambda ,\gamma ,\eta } \right)$. 
Hence, we can easily obtain the optimal sharing policy $\delta^*=1-\Omega$, and compute the corresponding revenue as ${\Pi ^{{\rm{APL}}}}\left(\delta^*\right) = ag\left( {\lambda ,\gamma ,\eta } \right)N - \lambda \beta {\theta _{\max }}N$.

(b) When $\delta\in\left(1-\Omega,1-\epsilon\right]$, we can show $p_f^*\left( \delta  \right) = \frac{{\beta {\theta _{\max }}}}{2}+  {\frac{{\left( {1 - \delta } \right) \beta\theta_{\max}}}{2 \Omega}}$ based on (\ref{equ:VOoptWiFiprice}). According to (\ref{equ:APLrevenue}), we can compute the ad platform's revenue as
\begin{align}
\Pi^{\rm APL}\left(\delta\right)=\frac{{aNg\left( {\lambda ,\gamma ,\eta } \right)}}{2}\delta \left( {1 + \frac{{1 - \delta }}{\Omega }} \right).\label{equ:appendix:APLrevenue}
\end{align}
This is a quadratic function of $\delta$, and we can prove that the ad platform's revenue is always below $ag\left( {\lambda ,\gamma ,\eta } \right)N - \lambda \beta {\theta _{\max }}N$ for all $\delta\in\left(1-\Omega,1-\epsilon\right]$.

Summarizing (a) and (b), we show that in Case B, the optimal sharing policy is $\delta^*=1-\Omega$, and the VO's Wi-Fi price is $p_f^*\left(\delta^*\right)=\beta\theta_{\max}$.

\subsubsection{\textbf{Case C:} $\Omega\in\left(\frac{1}{3},1-2\epsilon\right)$} In this case, we discuss $\delta\in\left[0,1-\Omega\right]$ and $\delta\in\left(1-\Omega,1-\epsilon\right]$ separately.

(a) When $\delta\in\left[0,1-\Omega\right]$, the analysis is the same as item (a) of Case B. The ad platform's optimal sharing policy is $\delta^*=1-\Omega$, and the corresponding revenue is ${\Pi ^{{\rm{APL}}}}\left(\delta^*\right) = ag\left( {\lambda ,\gamma ,\eta } \right)N - \lambda \beta {\theta _{\max }}N$.

(b) When $\delta\in\left(1-\Omega,1-\epsilon\right]$, the ad platform's revenue function $\Pi^{\rm APL}\left(\delta\right)$ is the same as (\ref{equ:appendix:APLrevenue}). 
We can easily show that function $\Pi^{\rm APL}\left(\delta\right)$ achieves the maximum value at point $\delta=\frac{1+\Omega}{2}$. 
Furthermore, from $\Omega\in\left(\frac{1}{3},1-2\epsilon\right)$ (condition of Case C), we can prove that $\frac{1+\Omega}{2}\in\left(1-\Omega,1-\epsilon\right)$. Therefore, the ad platform's optimal sharing policy is $\delta^*  = \frac{{1+\Omega}}{2}$, and the corresponding revenue is ${\Pi ^{{\rm{APL}}}}\left(\delta^*\right) =\frac{{ag\left( {\lambda ,\gamma ,\eta } \right)N}}{2}\frac{{{{\left( {\Omega  + 1} \right)}^2}}}{{4\Omega }}$.

Next we summarize (a) and (b). 
We can show that the value of ${\Pi ^{{\rm{APL}}}}\left(\delta^*\right)$ in (b) is greater than that in (a). As a result, in Case C, the optimal sharing policy is $\delta^*  = \frac{{1+\Omega}}{2}$, and the corresponding Wi-Fi price can be computed as $p_f^*\left(\delta^*\right)=\frac{{\beta {\theta _{\max }}}}{4}+\frac{{ag\left( {\lambda ,\gamma,\eta } \right)}}{{4\lambda }}$.

\subsubsection{\textbf{Case D:} $\Omega\in\left[1-2\epsilon,\infty\right)$} In this case, we discuss $\delta\in\left[0,1-\Omega\right]$ and $\delta\in\left(1-\Omega,1-\epsilon\right]$ separately.

(a) When $\delta\in\left[0,1-\Omega\right]$, the analysis is the same as item (a) of Case B. The ad platform's optimal sharing policy is $\delta^*=1-\Omega$, and the corresponding revenue is ${\Pi ^{{\rm{APL}}}}\left(\delta^*\right) = ag\left( {\lambda ,\gamma ,\eta } \right)N - \lambda \beta {\theta _{\max }}N$.

(b) When $\delta\in\left(1-\Omega,1-\epsilon\right]$, the ad platform's revenue function $\Pi^{\rm APL}\left(\delta\right)$ is the same as (\ref{equ:appendix:APLrevenue}). We can easily prove that $\Pi^{\rm APL}\left(\delta\right)$ increases with $\delta$ for $\delta\in\left(1-\Omega,1-\epsilon\right]$. 
Hence, the optimal sharing policy is $\delta^*=1-\epsilon$, and the corresponding revenue is ${\Pi ^{{\rm{APL}}}}\left(\delta^*\right)= \frac{{aNg\left( {\lambda ,\gamma ,\eta } \right)}}{2}\left(1-\epsilon\right)\left(1+\frac{\epsilon}{\Omega}\right)$.

Next we summarize (a) and (b). We can show that the value of ${\Pi ^{{\rm{APL}}}}\left(\delta^*\right)$ in (b) is greater than that in (a). As a result, in Case D, the ad platform's optimal sharing policy is $\delta^*  =1-\epsilon$, and the corresponding Wi-Fi price can be computed as $p_f^*\left(\delta^*\right)=\frac{{\beta {\theta _{\max }}}}{2}+\frac{{ag\left( {\lambda ,\gamma,\eta } \right)}\epsilon}{{2\lambda }}$.

Summarizing Case A, Case B, Case C, and Case D, we complete the proofs of Proposition \ref{proposition:delta} and Proposition \ref{proposition:equWiFiprice}.
\end{proof}


\subsection{Proof of Proposition \ref{proposition:socialwelfare}}

\begin{proof}
First, we compute the total utility of MUs. If a type-$\theta$ MU chooses the premium access, its utility for connecting Wi-Fi for one time segment is $\theta$; otherwise, its utility for connecting Wi-Fi for one time segment is $\left(1-\beta\right)\theta$. Under the VO's Wi-Fi price $p_f^*\left(\delta^*\right)$, the MUs with types in $\left[0,{\theta}_{T}\left(p_f^*\left(\delta^*\right)\right)\right)$ choose the advertising sponsored access, and the MUs with types in $\left[{\theta}_{T}\left(p_f^*\left(\delta^*\right)\right),\theta_{\max}\right]$ choose the premium access. 
Therefore, we can compute the MUs' total utility as  
\begin{align}
\nonumber
& \lambda N \int_{0}^{{\theta}_{T}\left(p_f^*\left(\delta^*\right)\right)} \left(1-\beta\right)\frac{\theta}{\theta_{\max}}d\theta + \lambda N \int_{{\theta}_{T}\left(p_f^*\left(\delta^*\right)\right)}^{\theta_{\max}} \frac{\theta}{\theta_{\max}}d\theta
\\
& =\frac{1}{2}\lambda N\theta_{\max}-\frac{1}{2} \lambda N {p_f^*}\left(\delta^*\right) \varphi_a\left({p_f^*}\left(\delta^*\right)\right).\label{equ:appendix:SW1}
\end{align}

Second, we compute the total utility of ADs. Under the VO's advertising price $p_a^\infty$, only the ADs with types in $\left[0,\sigma_T\left(p_a^\infty\right)\right]$ purchase the ad spaces, and the amounts of purchased ad spaces are given in (\ref{equ:ADoptsolution}). Hence, we can compute the ADs' total utility as
\begin{align}
\nonumber
& a N \varphi_a\left(p_f^*\left(\delta^*\right)\right) M   \int_0^{\sigma_{T}\left(p_a^\infty\right)} \frac{\gamma e^{-\gamma\sigma}}{\sigma_{\max}} \left(1-e^{-\ln\left(\frac{a\gamma}{p_a^\infty}\right)+\gamma\sigma}\right)d\sigma \\
& = \eta N \varphi_a\left({p_f^*}\left(\delta^*\right)\right) \left(a-\frac{p_a^\infty}{\gamma}\left(1+\ln\left(\frac{a\gamma}{p_a^\infty}\right)\right)\right).\label{equ:appendix:SW2}
\end{align}

Since the social welfare equals the total utility of MUs and ADs, we obtain the social welfare by combining the MUs' total utility in (\ref{equ:appendix:SW1}) and the AD's total utility in (\ref{equ:appendix:SW2}). This completes the proof.
\end{proof}



\subsection{Proof of Proposition \ref{proposition:gamma}}

\subsubsection{Proof of Item (i)}
The VO's advertising price $p_a^{\infty}$ is given in (\ref{equ:pa:infinite}). 
We find that $p_a^{\infty}$ is continuous in $\gamma$ for $\gamma\in\left(0,\infty\right)$. 
Furthermore, by checking $\frac{{\partial p_a^\infty }}{{\partial \gamma }}$, we can show that $p_a^{\infty}$ is increasing in $\gamma$ for $\gamma\in\left(0,\frac{2\eta}{\lambda}\right)$ and $\gamma\in\left(\frac{2\eta}{\lambda},\infty\right)$. 
Hence, the VO's advertising price $p_a^{\infty}$ is increasing in $\gamma$ for $\gamma\in\left(0,\infty\right)$.

\subsubsection{Proof of Item (ii)}

According to (\ref{equ:smalllambda:a}) and (\ref{equ:largelambda:a}), 
$\rho\left(p_a^\infty\right)$ is continuous at point $\gamma=\frac{2\eta}{\lambda}$. Furthermore, we can prove that $\rho\left(p_a^\infty\right)$ is decreasing in $\gamma$ for $\gamma\in\left(0,\frac{2\eta}{\lambda}\right)$ and $\gamma\in\left(\frac{2\eta}{\lambda},\infty\right)$. 
Hence, we can show that $\rho\left(p_a^\infty\right)$ is decreasing in $\gamma$ for $\gamma\in\left(0,\infty\right)$.

\subsubsection{Proof of Item (iii)}
First, we show some properties for $\frac{\lambda\beta\theta_{\max}}{a g\left(\lambda,\gamma,\eta\right)}$. 
Based on (\ref{equ:gfunction}), we can show that $\frac{\lambda\beta\theta_{\max}}{a g\left(\lambda,\gamma,\eta\right)}$ is continuous in $\gamma$ for $\gamma\in\left(0,\infty\right)$. Furthermore, we can show that $\frac{\lambda\beta\theta_{\max}}{a g\left(\lambda,\gamma,\eta\right)}$ is strictly decreasing in $\gamma$ for $\gamma\in\left(0,\frac{2\eta}{\lambda}\right)$, and does not change with $\gamma$ for $\gamma\in\left(\frac{2\eta}{\lambda},\infty\right)$. We can also prove that $\lim_{\gamma\rightarrow 0^{+}}\frac{\lambda \beta \theta_{\max}}{a g\left(\lambda,\gamma,\eta\right)}=\infty$ (\emph{i.e.}, for any $V>0$, there exists a $\xi>0$ such that $\frac{\lambda \beta \theta_{\max}}{a g\left(\lambda,\gamma,\eta\right)}>V$ for all $\gamma\in\left(0,\xi\right)$) and $\frac{\lambda \beta \theta_{\max}}{a g\left(\lambda,\gamma,\eta\right)}=\frac{\lambda\beta\theta_{\max}}{a2\eta e^{-2}}$ for all $\gamma\in\left[\frac{2\eta}{\lambda},\infty\right)$. 
Therefore, for any value $W\in\left(\frac{\lambda\beta\theta_{\max}}{a2\eta e^{-2}},\infty\right)$, we can always find a unique $\gamma_0\in\left(0,\infty\right)$ such that $\frac{\lambda\beta\theta_{\max}}{a g\left(\lambda,\gamma_0,\eta\right)}=W$. 

We can prove Item (iii) by considering the following three situations separately: $\frac{\lambda\beta\theta_{\max}}{a2\eta e^{-2}}<\frac{1}{3}$, $\frac{1}{3}\le\frac{\lambda\beta\theta_{\max}}{a 2\eta e^{-2}}<1-2\epsilon$, and $\frac{\lambda\beta\theta_{\max}}{a2\eta e^{-2}}\ge1-2\epsilon$. Next we discuss the situation where $\frac{\lambda\beta\theta_{\max}}{a2\eta e^{-2}}<\frac{1}{3}$. The situations where $\frac{1}{3}\le\frac{\lambda\beta\theta_{\max}}{a 2\eta e^{-2}}<1-2\epsilon$ and $\frac{\lambda\beta\theta_{\max}}{a2\eta e^{-2}}\ge1-2\epsilon$ can be analyzed in similar approaches.

Since $\epsilon\in\left(0,\frac{1}{3}\right)$, we have $\frac{1}{3}<1-2\epsilon$. 
When $\frac{\lambda\beta\theta_{\max}}{a2\eta e^{-2}}<\frac{1}{3}$, we have $\frac{\lambda\beta\theta_{\max}}{a2\eta e^{-2}}<\frac{1}{3}<1-2\epsilon$. 
Based on the analysis of $\frac{\lambda\beta\theta_{\max}}{a g\left(\lambda,\gamma,\eta\right)}$ above, we can always find unique $\gamma_1$ and $\gamma_2$ such that $\frac{\lambda\beta\theta_{\max}}{a g\left( {\lambda ,\gamma_1 ,\eta } \right)}=1-2\epsilon$ and $\frac{\lambda\beta\theta_{\max}}{a g\left( {\lambda ,\gamma_2 ,\eta } \right)}=\frac{1}{3}$. Moreover, we have $\gamma_1<\gamma_2$. 

Based on the monotonicity of $\frac{\lambda\beta\theta_{\max}}{a g\left(\lambda,\gamma,\eta\right)}$, we can show that $\frac{\lambda\beta\theta_{\max}}{a g\left(\lambda,\gamma,\eta\right)}\in\left[1-2\epsilon,\infty\right)$ for $\gamma\in\left(0,\gamma_1\right]$, $\frac{\lambda\beta\theta_{\max}}{a g\left(\lambda,\gamma,\eta\right)}\in\left(\frac{1}{3},1-2\epsilon\right)$ for $\gamma\in\left(\gamma_1,\gamma_2\right)$, and $\frac{\lambda\beta\theta_{\max}}{a g\left(\lambda,\gamma,\eta\right)}\in\left[\frac{\lambda\beta\theta_{\max}}{a2\eta e^{-2}},\frac{1}{3}\right]$ for $\gamma\in\left[\gamma_2,\infty\right)$. 
Notice that $\Omega=\frac{\lambda\beta\theta_{\max}}{a g\left(\lambda,\gamma,\eta\right)}$. From Proposition \ref{proposition:equWiFiprice}, we can easily derive that 
\begin{align}
p_f^*\left(\delta^*\right) = \left\{ {\begin{array}{*{20}{l}}
{\frac{{\beta {\theta _{\max }}}}{2} + \frac{{ag\left( {\lambda ,\gamma ,\eta } \right)}\epsilon}{{2\lambda }},}&{{\rm if~}\gamma\in\left(0,\gamma_1\right],}\\
{\frac{{\beta {\theta _{\max }}}}{4} + \frac{{ag\left( {\lambda ,\gamma ,\eta } \right)}}{{4\lambda }},}&{{\rm if~}\gamma\in\left(\gamma_1,\gamma_2\right),}\\
{\beta\theta_{\max},}&{{\rm if~}\gamma\in\left[\gamma_2,\infty\right).}\label{equ:appendix:pflongday}
\end{array}} \right.
\end{align}
According to (\ref{equ:gfunction}), $g\left(\lambda,\gamma,\eta\right)$ is non-decreasing in $\gamma$ for $\gamma\in\left(0,\infty\right)$. 
Hence, from (\ref{equ:appendix:pflongday}), we can easily show that $p_f^*\left(\delta^*\right)$ is non-decreasing in $\gamma$ for $\gamma\in\left(0,\gamma_1\right]$, $\left(\gamma_1,\gamma_2\right)$, or $\left[\gamma_2,\infty\right)$. 
Moreover, from (\ref{equ:appendix:pflongday}), $p_f^*\left(\delta^*\right)$ is continuous at point $\gamma=\gamma_1$ and point $\gamma=\gamma_2$. 
Therefore, we show that $p_f^*\left(\delta^*\right)$ is non-decreasing in $\gamma$ for $\gamma\in\left(0,\infty\right)$.

For situations $\frac{1}{3}\le\frac{\lambda\beta\theta_{\max}}{a2\eta e^{-2}}<1-2\epsilon$ and $\frac{\lambda\beta\theta_{\max}}{a2\eta e^{-2}}\ge1-2\epsilon$, we can apply similar approaches and show that $p_f^*\left(\delta^*\right)$ is also non-decreasing in $\gamma$ for $\gamma\in\left(0,\infty\right)$.

\subsubsection{Proof of Item (iv)}

Based on (\ref{equ:varphi}), we have $\varphi_f\left(p_f\right)=1-\frac{p_f}{\beta\theta_{\max}}$ for $p_f\in\left[0,\beta\theta_{\max}\right]$. Since we have shown that $p_f^*\left(\delta^*\right)$ is non-decreasing in $\gamma$, we can show that $\varphi_f\left(p_f^*\left(\delta^*\right)\right)$ is non-increasing in $\gamma$ for $\gamma\in\left(0,\infty\right)$.

\subsection{Proof of Proposition \ref{proposition:lambda}}

\subsubsection{Proof of Item (i)}
From (\ref{equ:pa:infinite}), we can show that $p_a^\infty$ is continuous in $\lambda$ for $\lambda\in\left(0,\infty\right)$. Moreover, it is decreasing in $\lambda$ for $\lambda\in\left(0,\frac{2\eta}{\gamma}\right)$, and is independent of $\lambda$ for $\lambda\in\left(\frac{2\eta}{\gamma},\infty\right)$. Therefore, the VO's advertising price $p_a^\infty$ is non-increasing in $\lambda$ for $\lambda\in\left(0,\infty\right)$.

\subsubsection{Proof of Item (ii)}
Based on (\ref{equ:smalllambda:a}) and (\ref{equ:largelambda:a}), $\rho\left(p_a^\infty\right)$ is continuous at point $\lambda=\frac{2\eta}{\gamma}$. Furthermore, 
$\rho\left(p_a^\infty\right)$ is increasing in $\lambda$ for $\lambda\in\left(0,\frac{2\eta}{\gamma}\right)$, and is independent of $\lambda$ for $\lambda\in\left(\frac{2\eta}{\gamma},\infty\right)$. Hence, we can show that $\rho\left(p_a^\infty\right)$ is non-decreasing in $\lambda$ for $\lambda\in\left(0,\infty\right)$.

\subsubsection{Proof of Item (iii)}
Based on (\ref{equ:gfunction}), we can rewrite $\frac{\lambda\beta\theta_{\max}}{a g\left(\lambda,\gamma,\eta\right)}$ as
\begin{align}
\frac{\lambda\beta\theta_{\max}}{a g\left(\lambda,\gamma,\eta\right)} = \left\{ {\begin{array}{*{20}{l}}
{\frac{\beta\theta_{\max}}{a\gamma}e^{\sqrt{\frac{2\lambda\gamma}{\eta}}},}&{{\rm if~}0 < \lambda  \le \frac{{2\eta }}{\gamma },}\\
{\frac{\beta\theta_{\max}}{2a\eta e^{-2}}\lambda,}&{{\rm if~}\lambda  > \frac{{2\eta }}{\gamma }.}\label{equ:appendix:lastproof}
\end{array}} \right.
\end{align}
From (\ref{equ:appendix:lastproof}), we can easily show that $\frac{\lambda\beta\theta_{\max}}{a g\left(\lambda,\gamma,\eta\right)}$ is continuous and strictly increasing in $\lambda$ for $\lambda\in\left(0,\infty\right)$. Furthermore, we can find that $\lim_{\lambda\rightarrow 0^{+}}\frac{\lambda \beta \theta_{\max}}{a g\left(\lambda,\gamma,\eta\right)}=\frac{\beta\theta_{\max}}{a\gamma}$ and $\lim_{\lambda\rightarrow \infty}\frac{\lambda \beta \theta_{\max}}{a g\left(\lambda,\gamma,\eta\right)}=\infty$. 
Therefore, for any value $W\in\left(\frac{\beta\theta_{\max}}{a\gamma},\infty\right)$, we can find a unique $\lambda\in\left(0,\infty\right)$ such that $\frac{\lambda\beta\theta_{\max}}{a g\left(\lambda,\gamma,\eta\right)} =W$. 

We can prove Item (iii) by considering the following three situations separately: $\frac{\beta\theta_{\max}}{a\gamma}<\frac{1}{3}$, $\frac{1}{3}\le\frac{\beta\theta_{\max}}{a\gamma}<1-2\epsilon$, and $\frac{\beta\theta_{\max}}{a\gamma}\ge1-2\epsilon$. 
Next we discuss the situation where $\frac{\beta\theta_{\max}}{a\gamma}<\frac{1}{3}$. The situations where $\frac{1}{3}\le\frac{\beta\theta_{\max}}{a\gamma}<1-2\epsilon$ and $\frac{\beta\theta_{\max}}{a\gamma}\ge1-2\epsilon$ can be analyzed in similar approaches.

Since $\epsilon\in\left(0,\frac{1}{3}\right)$, we have $\frac{1}{3}<1-2\epsilon$. 
When $\frac{\beta\theta_{\max}}{a\gamma}<\frac{1}{3}$, we have $\frac{\beta\theta_{\max}}{a\gamma}<\frac{1}{3}<1-2\epsilon$. 
Based on the analysis above, we can always find unique $\lambda_1$ and $\lambda_2$ such that $\frac{\lambda\beta\theta_{\max}}{g\left( {\lambda_1 ,\gamma ,\eta } \right)}=\frac{1}{3}$ and $\frac{\lambda\beta\theta_{\max}}{g\left( {\lambda_2 ,\gamma ,\eta } \right)}=1-2\epsilon$. Also, we can show that $\lambda_1<\lambda_2$.

Based on the monotonicity of $\frac{\lambda\beta\theta_{\max}}{a g\left(\lambda,\gamma,\eta\right)}$, we can show that $\frac{\lambda\beta\theta_{\max}}{a g\left(\lambda,\gamma,\eta\right)}\in\left(\frac{\beta\theta_{\max}}{a\gamma},\frac{1}{3}\right]$ for $\lambda\in\left(0,\lambda_1\right]$, $\frac{\lambda\beta\theta_{\max}}{a g\left(\lambda,\gamma,\eta\right)}\in\left(\frac{1}{3},1-2\epsilon\right)$ for $\lambda\in\left(\lambda_1,\lambda_2\right)$, and $\frac{\lambda\beta\theta_{\max}}{a g\left(\lambda,\gamma,\eta\right)}\in\left[1-2\epsilon,\infty\right)$ for $\lambda\in\left[\lambda_2,\infty\right)$. 
From Proposition \ref{proposition:equWiFiprice}, we can easily derive that 
\begin{align}
{p_f^*\left(\delta^*\right)} = \left\{ {\begin{array}{*{20}{l}}
{{\beta\theta_{\max}},}&{{\rm if~} \lambda\in\left(0,\lambda_1\right],}\\
{\frac{{\beta {\theta _{\max }}}}{4}+\frac{{ag\left( {\lambda ,\gamma,\eta } \right)}}{{4\lambda }},}&{{\rm if~} \lambda\in\left(\lambda_1,\lambda_2\right),}\\
{\frac{\beta\theta_{\max}}{2}+\frac{{ag\left( {\lambda ,\gamma,\eta } \right)\epsilon}}{{2\lambda }},}&{{\rm if~}\lambda\in\left[\lambda_2,\infty\right)}.\label{equ:appendix:pf:2}
\end{array}} \right.
\end{align}
From (\ref{equ:gfunction}), we can easily show that $\frac{{g\left( {\lambda ,\gamma,\eta } \right)}}{{\lambda }}$ is decreasing in $\lambda$. 
Based on (\ref{equ:appendix:pf:2}), we can prove that $p_f^*\left(\delta^*\right)$ is non-increasing in $\lambda$ for $\lambda\in\left(0,\lambda_1\right]$, $\left(\lambda_1,\lambda_2\right)$, or $\left[\lambda_2,\infty\right)$. Moreover, we can prove that $p_f^*\left(\delta^*\right)$ is continuous at point $\lambda=\lambda_1$ and point $\lambda=\lambda_2$. Therefore, we show that $p_f^*\left(\delta^*\right)$ is non-increasing in $\lambda$ for $\lambda\in\left(0,\infty\right)$.

\subsubsection{Proof of Item (iv)}

According to (\ref{equ:varphi}), we have $\varphi_f\left(p_f\right)=1-\frac{p_f}{\beta\theta_{\max}}$ for $p_f\in\left[0,\beta\theta_{\max}\right]$. Since we have shown that $p_f^*\left(\delta^*\right)$ is non-increasing in $\lambda$, we can show that $\varphi_f\left(p_f^*\left(\delta^*\right)\right)$ is non-decreasing in $\lambda$ for $\lambda\in\left(0,\infty\right)$.

\subsection{Example of $\lambda$'s Impact on $\rm SW$}\label{appendix:example}
In this section, we show an example where the social welfare may decrease with $\lambda$ for some medium $\lambda$. We choose $N=200$, $\theta_{\max}=1$, $\beta=0.8$, $\eta=1$, $a=20$, $\epsilon=0.01$, and $\gamma=0.8$. We change $\lambda$ from $0.01$ to $15$, and plot the social welfare against $\lambda$ in Fig. \ref{fig:NEW12}. We can observe that the social welfare decreases with $\lambda$ for $2.3<\lambda<6.6$.
  \begin{figure}[h]
  \centering
  \includegraphics[scale=0.35]{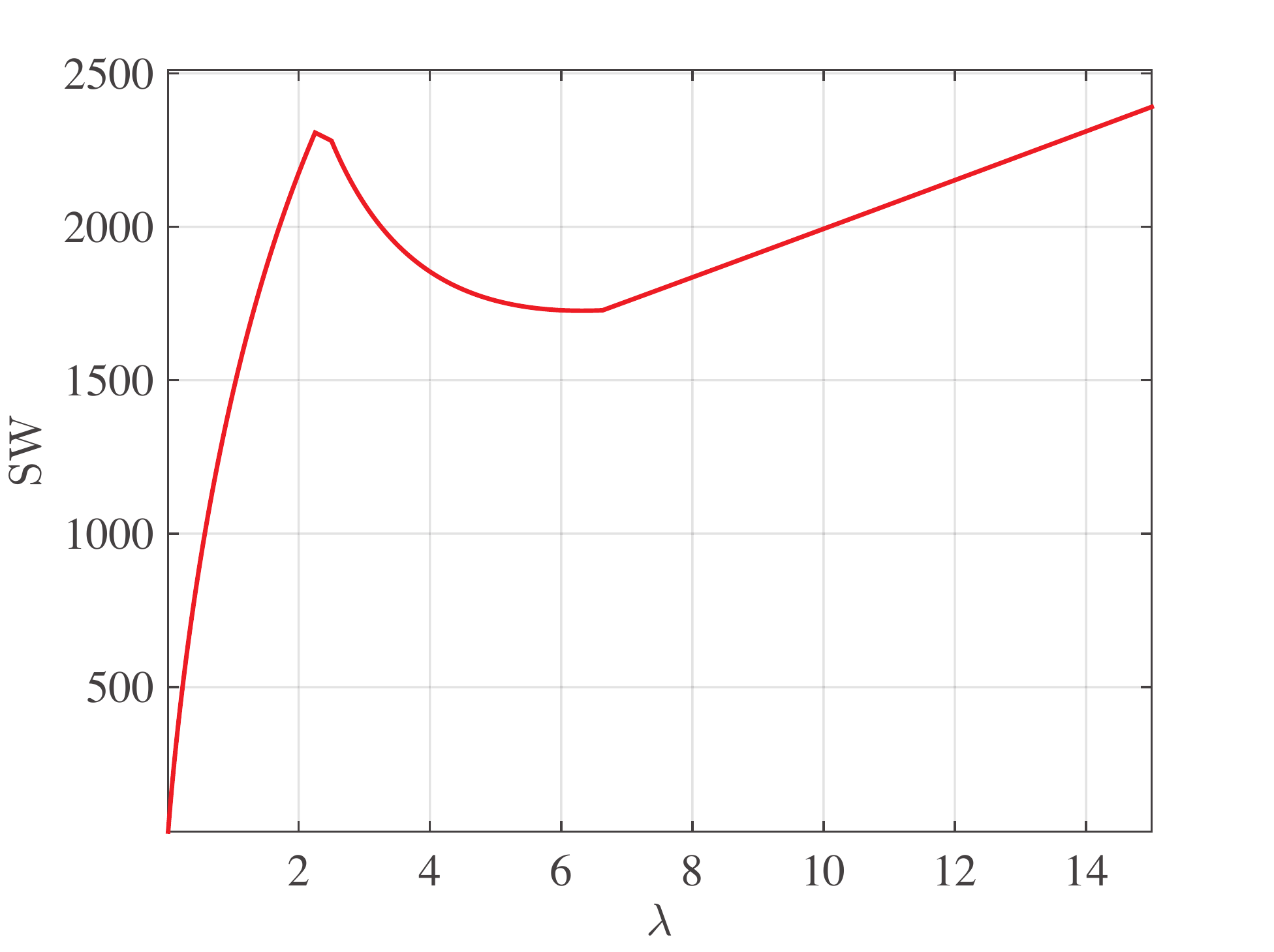}
  \caption{A Special Example of $\lambda$'s Impact on Social Welfare.}
  \label{fig:NEW12}
  \end{figure}

\end{document}